\documentclass[12pt,preprint]{aastex62}
\shorttitle{Nature vs. Nurture}
\shortauthors{Safsten et al.}

\usepackage{mathtools}
\usepackage{amsmath}
\usepackage{multirow}
\usepackage{appendix}

\begin{document}
\title{Nature vs. nurture: a Bayesian framework for assessing apparent correlations between planetary orbital properties and stellar ages}
\received{}
\revised{}
\correspondingauthor{Emily D. Safsten}
\email{eds36@psu.edu}
\author[0000-0002-3425-7803]{Emily D. Safsten}
\affiliation{Department of Astronomy \& Astrophysics, Center for Exoplanets and Habitable Worlds, The Pennsylvania State University, University Park, PA 16802, USA}
\author[0000-0001-9677-1296]{Rebekah I. Dawson}
\affiliation{Department of Astronomy \& Astrophysics, Center for Exoplanets and Habitable Worlds, The Pennsylvania State University, University Park, PA 16802, USA}
\author[0000-0003-2862-6278]{Angie Wolfgang}
\affiliation{Department of Astronomy \& Astrophysics, Center for Exoplanets and Habitable Worlds, The Pennsylvania State University, University Park, PA 16802, USA}

\begin{abstract}
Many exoplanets have orbital characteristics quite different from those seen in our own solar system, including planets locked in orbital resonances and planets on orbits that are elliptical or highly inclined from their host star's spin axis.  It is debated whether the wide variety in system architecture is primarily due to differences in formation conditions (nature) or due to evolution over time (nurture).  Identifying trends between planetary and stellar properties, including stellar age, can help distinguish between these competing theories and offer insights as to how planets form and evolve.  However, it can be challenging to determine whether observed trends between planetary properties and stellar age are driven by the age of the system -- pointing to evolution over time being an important factor -- or other parameters to which the age may be related, such as stellar mass or stellar temperature. The situation is complicated further by the possibilities of selection biases, small number statistics, uncertainties in stellar age, and orbital evolution timescales that are typically much shorter than the range of observed ages.  Here we develop a Bayesian statistical framework to assess the robustness of such observed correlations and to determine whether they are indeed due to evolutionary processes, are more likely to reflect different formation scenarios, or are merely coincidental.  We apply this framework to reported trends between stellar age and 2:1 orbital resonances, spin-orbit misalignments, and hot Jupiters' orbital eccentricities. We find strong support for the nurture hypothesis only in the final case.
\end{abstract}

\section{Introduction}
Many exoplanets have orbital characteristics quite different from those seen in our own solar system, including planets locked in orbital resonances, planets on orbits that are elliptical or highly inclined from their host star's spin axis, planets parked close to their stars, and compact multi-planet systems (see \citealt{winnfab2015} and references therein).  It is debated whether the wide variation in system architecture is primarily due to differences in formation conditions (nature) or due to evolution over time (nurture).  Identifying trends between planetary and stellar properties, including stellar age (see, e.g., \citealt{christiansenwhitepaper}), can help distinguish between these competing theories and offer insights as to how planets form and evolve. As our sample of known exoplanets have grown, some apparent correlations between planetary orbital properties and stellar age have started to emerge. However, we face two major problems in interpreting such correlations.

First, the true underlying source of such correlations can be unclear because stellar age is interrelated with other stellar properties. For example, \cite{winn2010} found that hot Jupiters orbiting stars warmer than 6250 K are preferentially misaligned with their host star's spin axis, compared to those orbiting cooler stars. However, \cite{triaud2011} argued that the trend with temperature noted by \cite{winn2010} could instead be a trend with stellar age. The difficulty in distinguishing between nature vs. nurture is complicated further by the possibilities of selection biases, small number statistics, and uncertainties in stellar age (e.g., \citealt{mamhill2008,skumanich1972}).

Second, orbital evolution timescales tend to be strongly dependent on system properties such as semimajor axis, meaning that the distribution of such timescales can easily span many orders of magnitude (e.g., \citealt{hotjupreview2018}; \citealt{socrates2012}; \citealt{winn2010}). Therefore a typical observed sample of main sequence stars typically probes only a sliver of orbital evolution timescales -- those which are comparable to the stellar ages in the sample -- casting suspicion on the nurture hypothesis as the explanation for an observed correlation. For example, \citet{kz2011} reported that systems with 2:1 period commensurabilities are younger than those without, suggesting that some 2:1 resonances are eventually disrupted. However, \citet{dd2016} argued that the fine tuning of the disruption timescale necessary to account for the trend suggests that the difference in ages could be due to chance.  In another case, \cite{quinn2014} found evidence that hot Jupiters younger than their circularization timescales are more eccentric than ones older than their circularization timescales, which supports high-eccentricity migration as the primary hot Jupiter formation mechanism.  However, the strong dependence of circularization timescale on semimajor axis, combined with the relatively small age range, raises the possibility that the eccentricity may primarily depend on semimajor axis rather than age.

Despite these challenges, disentangling the underlying sources of proposed trends is essential for understanding how these planetary properties came to be. Furthermore, as a larger sample of planets with well-characterized stellar hosts becomes available with data from Gaia, TESS, and PLATO (e.g., \citealt{veras2015}), being able to distinguish the true driving factor will be essential for interpreting the data.

To address this problem, we develop a Bayesian statistical framework to assess the robustness of observed correlations of planetary properties with age.  We define three hypotheses that represent possible sources for a correlation:
 
\begin{itemize}
 \item[]1.  Nurture: The planetary property is due to the age of the system.
 \item[]2.  Nature: The planetary property is due to an observed system parameter other than age, but which may be connected to the age.
 \item[]3.  Chance: The planetary property is independent of observed system parameters, and any apparent correlation between them is merely a coincidence.
\end{itemize}

Here we lay out an approach for testing relative strengths of these three hypotheses using the available data and apply the approach to several reported trends.  In the next section, we derive general equations for comparing the three hypotheses.  We then apply our approach to reported trends with stellar age for 2:1 orbital resonances (Section \ref{sec:resonances}), stellar obliquities (Section \ref{sec:alignment}), and hot Jupiter eccentricities (Section \ref{sec:eccentricities}). We end in Section \ref{sec:conclusions} with our conclusions and a brief discussion of future work.

\section{Generalized Hypotheses}
\label{sec:general}

Here we present a general framework for comparing Nurture, Nature, and Chance.  To compare the strengths of two hypotheses, we need to compute their odds ratio.  The odds ratio for two hypotheses of interest, $H_{\rm A}$ and $H_{\rm B}$, as explanations for an observed correlation between a planetary property and stellar age is:
\begin{equation}
\label{eqn:oddsratio}
    \frac{p(H_{\rm A}|\{X_{\rm p,i},t_{\rm {\star},i},{\bf X}_{\rm{ob},i}\})}{p(H_{\rm B}|\{X_{\rm p,i},t_{\rm {\star},i},{\bf X}_{\rm{ob},i}\})}=\frac{\int[\prod\limits_{\rm i}{p(X_{\rm p,i},t_{\rm {\star},i},{\bf X}_{\rm{ob},i}|{\bf Y},H_{\rm A})]p({\bf Y})d{\bf Y}}}{\int[\prod\limits_{\rm i}{p(X_{\rm p,i},t_{\rm {\star},i},{\bf X}_{\rm{ob},i}|{\bf Y},H_{\rm B})]p({\bf Y})d{\bf Y}}}~ \frac{p(H_{\rm A})}{p(H_{\rm B})}
\end{equation}
where $X_{\rm p}$ is the observed property of interest of the planet, $t_{\star}$ is the age of the star, and ${\bf X}_{\rm ob}$ contains other observed system quantities relevant to the problem (e.g., stellar mass, stellar temperature,  planetary semimajor axis).  $\{X_{\rm p,i},t_{\rm \star,i},{\bf X}_{\rm{ob},i}\}$ represents the set of observed properties for a collection of planetary systems and their host stars, with the $i$ subscript denoting an individual planet, star, or system.  ${\bf Y}$ represents hyperparameters, population-wide variables that describe the distributions from which the individual system properties are drawn; these hyperparameters are marginalized over outside the product of individual system likelihoods.  We give equal weight to each hypothesis a priori, so $p(H_{\rm A})/p(H_{\rm B}) = 1$. For each system and each hypothesis, we need to find $p(X_{\rm p,i},t_{\rm \star,i},{\bf X}_{\rm{ob},i}|{\bf Y})$, the probability of a system existing with a certain configuration given the hyperparameter(s).

In general, $t_{\star}$, $X_{\rm p}$, and other relevant data may depend on stellar and planetary properties that have not been observed and that will thus need to be marginalized over.  We lump these into a vector of non-observed quantities ${\bf X}_{\rm nob}$.  Additionally, from here on out, we do not explicitly indicate the dependence on a particular hypothesis with $|H$ in the equations, and we omit the $i$ subscript, with the understanding that all variables except ${\bf Y}$ denote individual system, planet, or stellar properties.  The following equation is the general form of the likelihood from which we derive equations for each hypothesis (Sections \ref{subsec:genhyp1}, \ref{subsec:genhyp2}, and \ref{subsec:genhyp3}) and, later, each specific application (Sections \ref{sec:resonances}, \ref{sec:alignment}, and \ref{sec:eccentricities}):
\begin{equation}
\label{eqn:mostgeneral}
\begin{split}
    p(X_{\rm p},t_{{\star}},{\bf X}_{\rm ob}|{\bf Y})=\int& p(X_{\rm p},t_{{\star}},{\bf X}_{\rm ob},{\bf X}_{\rm nob}|{\bf Y})d{\bf X}_{\rm nob}.
\end{split}
\end{equation}

We will first consider, in Section \ref{subsec:simpmod}, an oversimplified version of the model before deriving the full model we will actually use in the subsequent sections.

\subsection{Simplified model of the three hypotheses} \label{subsec:simpmod}
\begin{figure}[ht]
 \centering
    \includegraphics[width=3.0 in]{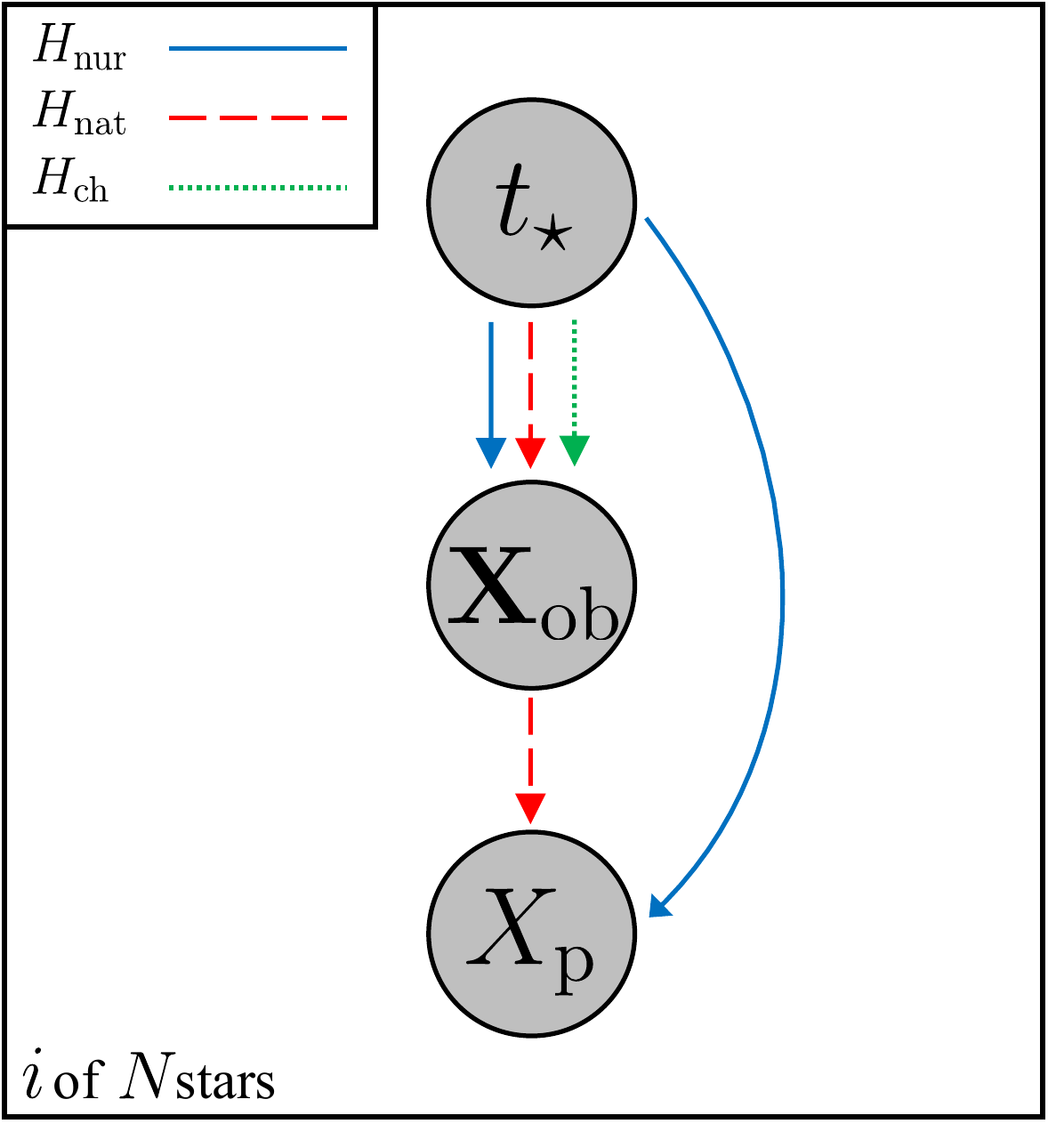}
    \caption{Graphical representation of the relationships between $X_{\rm p}$, $t_{\star}$, and ${\bf X}_{\rm ob}$ under the three hypotheses.  Relations under the Nurture hypothesis are shown with a solid blue line, those under the Nature hypothesis with a dashed red line, and those under the Chance hypothesis with a dotted green line.  The ``$i$ of $N$ stars" indicates that the plate is iterated over each of the $N$ systems in the sample.  Gray circles represent observed parameters:  the planetary property of interest ($X_{\rm p}$), the stellar age ($t_{\star}$), and other observed parameters (${\bf X}_{\rm ob}$).}
    \label{fig:basicmodel}
 \end{figure}

To further clarify what each hypothesis proposes, first we will consider more closely the relationships between $X_{\rm p}$, $t_{\star}$, and ${\bf X}_{\rm ob}$ under each hypothesis.  We will ignore ${\bf X}_{\rm nob}$ and ${\bf Y}$ for now and consider how $p(X_{\rm p},t_{\star},{\bf X}_{\rm ob})$ differs among hypotheses.  The relationships are depicted in Figure \ref{fig:basicmodel}.  Note that this is a simplification for the purposes of clarification, and does not fully represent the full model we use.

Under the Nurture hypothesis, the age of the system drives the planetary property of interest; in other words, $X_{\rm p}$ evolves over time.  Thus $X_{\rm p}$ only depends directly on $t_{\star}$:
\begin{equation}
    p(X_{\rm p},t_{\star},{\bf X}_{\rm ob})=p(X_{\rm p}|t_{\star})p({\bf X}_{\rm ob}|t_{\star})p(t_{\star}).
\end{equation}
Note that ${\bf X}_{\rm ob}$ may have some dependence on $t_{\star}$, or vice versa, (e.g., stars changing temperature as they age) so that $X_{\rm p}$ appears correlated ${\bf X}_{\rm ob}$, but ${\bf X}_{\rm ob}$ is not the fundamental driver of $X_{\rm p}$.  Such a relationship makes it challenging to distinguish whether the underlying source of observed correlations is Nature (i.e., ${\bf X}_{\rm ob}$ is the driver of $X_{\rm p}$) or Nurture (i.e., $t_\star$ is the driver of $X_{\rm p}$).

Under the Nature hypothesis, a component of ${\bf X}_{\rm ob}$ -- a system parameter other than the age -- is directly responsible for $X_{\rm p}$; in other words, $X_{\rm p}$ does not evolve with time but is instead determined by the conditions of the system's formation.  So we can write:
\begin{equation}
    p(X_{\rm p},t_{\star},{\bf X}_{\rm ob})=p(X_{\rm p}|{\bf X}_{\rm ob}(t_{\star}))p({\bf X}_{\rm ob}|t_{\star})p(t_{\star}).
\end{equation}
Note that $X_{\rm p}$ may still appear correlated with the system age because of the relationship between ${\bf X}_{\rm ob}$ and $t_{\star}$, but in the Nature hypothesis, $X_{\rm p}$ is only directly connected to ${\bf X}_{\rm ob}$.

Under the Chance hypothesis, $X_{\rm p}$ is a random process, so it is not related to ${\bf X}_{\rm ob}$ or $t_{\star}$ at all.  There still may be a relationship between ${\bf X}_{\rm ob}$ and $t_{\star}$.  So we can write:
\begin{equation}
    p(X_{\rm p},t_{\star},{\bf X}_{\rm ob})=p(X_{\rm p})p({\bf X}_{\rm ob}|t_{\star})p(t_{\star}).
\end{equation}

In this subsection we have rewritten $p(X_{\rm p},{\bf X}_{\rm ob},t_{\star})$ for each hypothesis as a chain of dependencies.  For the rest of the paper, we will continue to separate out $p(X_{\rm p})$ as appropriate for each hypothesis, but instead of expressing the rest as a series of conditional probabilities, we will use the joint probability for $t_{\star}$, ${\bf X}_{\rm ob}$, and ${\bf X}_{\rm nob}$ (which we have excluded in this subsection): $p(t_{\star},{\bf X}_{\rm ob},{\bf X}_{\rm nob})$.  Due to the definition of conditional probability, this is equivalent to a chain of dependencies like those expressed above but allows for more flexibility in how the chain is constructed (i.e. in expressing which parameters are dependent upon which).

\subsection{Hypothesis 1: Nurture (Age-driven)}
\label{subsec:genhyp1}
In the Nurture hypothesis, the observed planetary property is driven by the age of the star.  Since this hypothesis involves evolution over time, the current configuration depends on the configuration in which the system was formed, which we represent as $X_{\rm p0}$, and an evolutionary timescale $t_{\rm e}$, which in general may depend on ${\bf X}_{\rm ob}$, ${\bf X}_{\rm nob}$, and $X_{\rm p0}$.  This timescale represents the time it takes for a planetary system to attain some value of $X_{\rm p}$ that is different from $X_{\rm p0}$ (e.g., the disruption of an orbital resonance, the orbital realignment with a host star's spin axis).  $X_{\rm p0}$ and $t_{\rm e}$ are members of ${\bf X}_{\rm nob}$, but here we write them out explicitly, along with ${\bf X}_{\rm nob}$, because of their importance in this hypothesis.  We also separate out the integrand in Eqn. \ref{eqn:mostgeneral} according to the rules of conditional probability for this hypothesis, where $X_{\rm p}$ depends on $t_{\star}$ but not explicitly on ${\bf X}_{\rm ob}$ or anything in ${\bf X}_{\rm nob}$ except $t_{\rm e}$ and $X_{\rm p0}$.  Finally, we note that $t_{\star}$ may also correlate with other parameters contained within ${\bf X}_{\rm ob}$ and ${\bf X}_{\rm nob}$; such a dependence can be responsible for apparent correlations of $X_{\rm p}$ with ${\bf X}_{\rm ob}$, as explained in Section \ref{subsec:simpmod}.

In this time-dependent hypothesis, the initial configuration $X_{\rm p0}$ is independent of all parameters in ${\bf X}_{\rm ob}$ or ${\bf X}_{\rm nob}$.  There could, in theory, be a situation in which the probability of some value of $X_{\rm p0}$ depends on other properties of the system, and then that value evolves over time.  This would represent a sort of hybrid between the Nurture and Nature hypotheses, as both evolution over time and dependence on other system properties would play an important role.  For our framework, however, we are considering whether a system's $X_{\rm p}$ is \textit{primarily} driven by evolution over time or \textit{primarily} driven by other system properties.  So in this hypothesis, where time is the important factor, we can separate $X_{\rm p0}$ from other observed and non-observed parameters.  Putting this all together yields a probability
\begin{align}
\label{eqn:hyp1gen3v2}
\begin{split}
    p(X_{\rm p},t_\star,{\bf X}_{\rm ob}|{\bf Y})&=\iiint p(X_{\rm p}|t_\star,t_{\rm e},X_{\rm p0},{\bf X}_{\rm ob},{\bf X}_{\rm nob},{\bf Y}) p(t_\star,t_{\rm e},X_{\rm p0},{\bf X}_{\rm ob},{\bf X}_{\rm nob}|{\bf Y}) dt_{\rm e} dX_{\rm p0}d{\bf X}_{\rm nob}\\
    &= \iiint p(X_{\rm p}|t_\star,t_{\rm e},X_{\rm p0},{\bf Y}) p(t_{\rm e}|{\bf X}_{\rm ob},{\bf X}_{\rm nob},X_{\rm p0},{\bf Y}) \\
    & \qquad\quad \times p(t_\star,{\bf X}_{\rm ob},{\bf X}_{\rm nob}|{\bf Y})p(X_{\rm p0}|{\bf Y})dt_{\rm e} dX_{\rm p0}d{\bf X}_{\rm nob}.
\end{split}
\end{align}
\noindent where we have assumed that $t_{\rm e}$ does not depend on $t_\star$ and that $X_{\rm p0}$ is independent of $t_{\star}$, ${\bf X}_{\rm ob}$, and ${\bf X}_{\rm nob}$, and vice versa.  While $X_{\rm p}$ does not explicitly depend on ${\bf X}_{\rm ob}$ and ${\bf X}_{\rm nob}$, it is related through $t_{\rm e}$, as discussed above.

In general, $X_{\rm p}$ may be a continuous quantity spanning a range of possible values, so $p(X_{\rm p})$ and $p(X_{\rm p0})$ would be continuous distributions that may depend on general population hyperparameters, ${\bf Y}$, and/or system-specific properties, $t_{\star}$, ${\bf X}_{\rm ob}$, and ${\bf X}_{\rm nob}$.  However, in some applications of these equations, the planetary property of interest may be a binary parameter, e.g., a system either has a 2:1 resonance or does not.   When $X_{\rm p}$ is a binary parameter, $p(X_{\rm p0})$ is a Bernoulli distribution set by an overall, population-wide fraction of systems, $f_0$, that start out with said property -- e.g., the overall fraction of systems that begin with a 2:1 resonance.  This fraction will not necessarily be known a priori, so we treat it as a hyperparameter over which we marginalize outside the product of the likelihoods of individual systems.

Furthermore, the discrete nature of $X_{\rm p}$ and $X_{\rm p0}$ in this binary scenario means that we can change the integral over $X_{\rm p0}$ in Eqn. \ref{eqn:hyp1gen3v2} to a sum instead.  $X_{\rm p} = 1$ if the system possesses the property of interest and 0 if it does not, and similarly for $X_{\rm p0}$.  The initial fraction $f_0$ then represents the proportion of systems with $X_{\rm p0}=1$.  Thus, when $X_{\rm p}$ is a binary parameter, the general system likelihood equation we use for the Nurture hypothesis, modified from Eqn. \ref{eqn:hyp1gen3v2}, is:
\begin{align}
\begin{split}
\label{eqn:hyp1gen3binary2}
    p(X_{\rm p},t_{{\star}},{\bf X}_{\rm ob}|f_0,{\bf Y})=\iint\sum_{X_{\rm p0}=0}^1& p(X_{\rm p}|t_{{\star}},t_{\rm e},X_{\rm p0},{\bf Y})p(t_{\rm e}|X_{\rm p0},{\bf X}_{\rm ob},{\bf X}_{\rm nob},{\bf Y})\\&\times p(t_{{\star}},{\bf X}_{\rm ob},{\bf X}_{\rm nob}|{\bf Y})p(X_{\rm p0}|f_0)dt_{\rm e} d{\bf X}_{\rm nob}.
\end{split}
\end{align}
\noindent where $p(X_{\rm p0}|f_0) = 1-f_0$ for $X_{\rm p0}=0$ and $p(X_{\rm p0}|f_0) = f_0$ for $X_{\rm p0}=1$.

In many applications, including all applications we consider in this paper, $X_{\rm p}$ can only evolve in a single direction; for example, we can consider a host star realigning with its planet's orbital axis without subsequently becoming misaligned again.  In such cases, when both $X_{\rm p}$ and $X_{\rm p0}$ are discrete binary parameters, let the evolutionary timescale $t_{\rm e}$ represent the time it takes a system to attain $X_{\rm p}=0$ from $X_{\rm p0}=1$.  We separate out Eqn. \ref{eqn:hyp1gen3binary2} into specific instances for each possible combination of $X_{\rm p}$ and $X_{\rm p0}$.  In such a scenario, a system with $X_{\rm p0}=1$ and $X_{\rm p}=0$ must be older than its evolutionary timescale; a system with $X_{\rm p0}=1$ and $X_{\rm p}=1$ must be younger than its evolutionary timescale; and a system with $X_{\rm p0}=0$ must have $X_{\rm p}=0$ and has no evolutionary timescale.  We can thus adjust the limits of integration for $t_{\rm e}$ in Eqn. \ref{eqn:hyp1gen3binary2} for each case, yielding (see Appendix \ref{sec:appendixa} for full derivation):
\begin{align}
    \label{eqn:hyp1genbinaryfinal}
    p(X_{\rm p},t_{{\star}},{\bf X}_{\rm ob}|f_0,{\bf Y})=
    \begin{cases}
    \text{\Large$\int$} \left[1-f_0+f_0\int_0^{t_{\star}}p(t_{\rm e}|X_{\rm p0},{\bf X}_{\rm ob},{\bf X}_{\rm nob},{\bf Y})dt_{\rm e}\right] \\ \quad\times p(t_{{\star}},{\bf X}_{\rm ob},{\bf X}_{\rm nob}|{\bf Y}) d{\bf X}_{\rm nob} &,X_{\rm p}=0\\
    f_0\iint_{t_{\star}}^\infty p(t_{\rm e}|X_{\rm p0},{\bf X}_{\rm ob},{\bf X}_{\rm nob},{\bf Y})dt_{\rm e}\\ \quad\times p(t_{{\star}},{\bf X}_{\rm ob},{\bf X}_{\rm nob}|{\bf Y}) d{\bf X}_{\rm nob}&,X_{\rm p}=1.
    \end{cases}
\end{align}
Note that the above equations are only valid for a binary $X_{\rm p}$ when the system can go from $X_{\rm p0}=1$ to $X_{\rm p}=0$ but cannot go from $X_{\rm p0}=0$ to $X_{\rm p}=1$.  If the system is better described as going from $X_{\rm p0}=0$ to $X_{\rm p}=1$ but not from $X_{\rm p0}=1$ to $X_{\rm p}=0$, one only need switch the values of $X_{\rm p}$ in Eqn. \ref{eqn:hyp1genbinaryfinal} to get the correct equations.

\subsection{Hypothesis 2: Nature (Driven by inherent system properties)}
\label{subsec:genhyp2}
In this scenario, the planet property of interest does not change with time but instead depends deterministically on one or more system properties which are included in ${\bf X}_{\rm ob}$.  In this hypothesis, a relation between ${\bf X}_{\rm ob}$ and $t_{\star}$ may make $X_{\rm p}$ appear to be correlated with age, but ${\bf X}_{\rm ob}$ is the true driving force behind $X_{\rm p}$. Furthermore, because it depends directly on ${\bf X}_{\rm ob}$, we will assume $X_{\rm p}$ is also independent of ${\bf X}_{\rm nob}$.  This means the integrand in Eq. \ref{eqn:mostgeneral} can be separated into two probabilities to give the following:
\begin{align}
\label{eqn:genhyp2final}
p(X_{\rm p},t_{{\star}},{\bf X}_{\rm ob}|{\bf Y})&=\int p(X_{\rm p}|{\bf X}_{\rm ob},{\bf Y})p(t_{{\star}},{\bf X}_{\rm ob},{\bf X}_{\rm nob}|{\bf Y})d{\bf X}_{\rm nob}.
\end{align}

Note that there is no overall fraction here when $X_{\rm p}$ is a binary parameter because $X_{\rm p}$ does not evolve over time and so the fraction of the sample with $X_{\rm p}$ is set by the sample's distribution of ${\bf X}_{\rm ob}$.  However, there may be some applications in which the dependence of $X_{\rm p}$ on ${\bf X}_{\rm ob}$ is stochastic, rather than deterministic.  In such cases, a fraction may be introduced, as a hyperparameter if it is unknown, to describe the distribution of $X_{\rm p}$.

We note that stellar metallicity is an important system parameter that has a great influence on planetary system formation and is also connected to stellar age.  While several aspects of the connection between metallicity and planets have been documented (see, for example, \citealt{fischervalenti2005,dawson2013,buchhave2014,wang2015,dawson2015}), stellar metallicity has not been proposed as an explanation for the observed trends we examine in this particular work.  There may be future applications where an observed trend with age could actually be due to metallicity, or vice versa.  These could be assessed with a Nature vs. Nurture odds ratio by including the metallicity parameter in ${\bf X}_{\rm ob}$.

\subsection{Hypothesis 3: Chance}
\label{subsec:genhyp3}
Here, the planet property of interest is not related to any observed system parameters.  Therefore we can separate $X_{\rm p}$ from $t_{\star}$, ${\bf X}_{\rm ob}$, and ${\bf X}_{\rm nob}$, and Eqn. \ref{eqn:mostgeneral} becomes
\begin{align}
    \label{eqn:genhyp3cont}
    p(X_{\rm p},t_{{\star}},{\bf X}_{\rm ob}|{\bf Y})&=\int p(X_{\rm p}|{\bf Y})p(t_{{\star}},{\bf X}_{\rm ob},{\bf X}_{\rm nob}|{\bf Y})d{\bf X}_{\rm nob}.
\end{align}

In the case of a binary $X_{\rm p}$, we can again use an overall fraction $f$, representing the proportion of systems with $X_{\rm p}=1$.  Under the Chance hypothesis, this fraction is independent of ${\bf X}_{\rm ob}$ and ${\bf X}_{\rm nob}$.  The discrete nature of $X_{\rm p}$ in the binary scenario allows us to write separate expressions for the cases of $X_{\rm p}=0$ and $X_{\rm p}=1$:
\begin{align}
    \label{eqn:genhyp3final}
    p(X_{\rm p},t_{{\star}},{\bf X}_{\rm ob}|f,{\bf Y})=
    \begin{cases}
    (1-f)\int p(t_{{\star}},{\bf X}_{\rm ob},{\bf X}_{\rm nob}|{\bf Y})d{\bf X}_{\rm nob} &,X_{\rm p}=0\\
    f\int p(t_{{\star}},{\bf X}_{\rm ob},{\bf X}_{\rm nob}|{\bf Y})d{\bf X}_{\rm nob} &,X_{\rm p}=1
    \end{cases}
\end{align}
where we use $p(X_{\rm p}|f)=(1-f)$ for $X_{\rm p}=0$ and $p(X_{\rm p}|f)=f$ for $X_{\rm p}=1$.

\subsection{Odds Ratios}
\label{subsec:genoddsratios}
In the previous three subsections, we have derived equations for $p(X_{\rm p},t_{\star},{\bf X}_{\rm ob})$ for three different explanations for the observed distribution of $X_{\rm p}$.  Each quantity in the parentheses -- $X_{\rm p}$, $t_{\star}$, and ${\bf X}_{\rm ob}$ -- represents the driving force behind $X_{\rm p}$ in one hypothesis.  In the Nurture hypothesis, $t_{\star}$ is the driver: the planetary property evolves with time and is thus caused by the age of the system.  In the Nature hypothesis, ${\bf X}_{\rm ob}$ is the driver: the planetary property does not evolve with time, but instead depends on some other characteristic of its formation environment.  In the Chance hypothesis, since the planetary property does not evolve with time and does not depend on the formation environment, the only determining factor is the chance values of $X_{\rm p}$.

\begin{figure}[ht]
\centering
    \includegraphics[width=3.0 in]{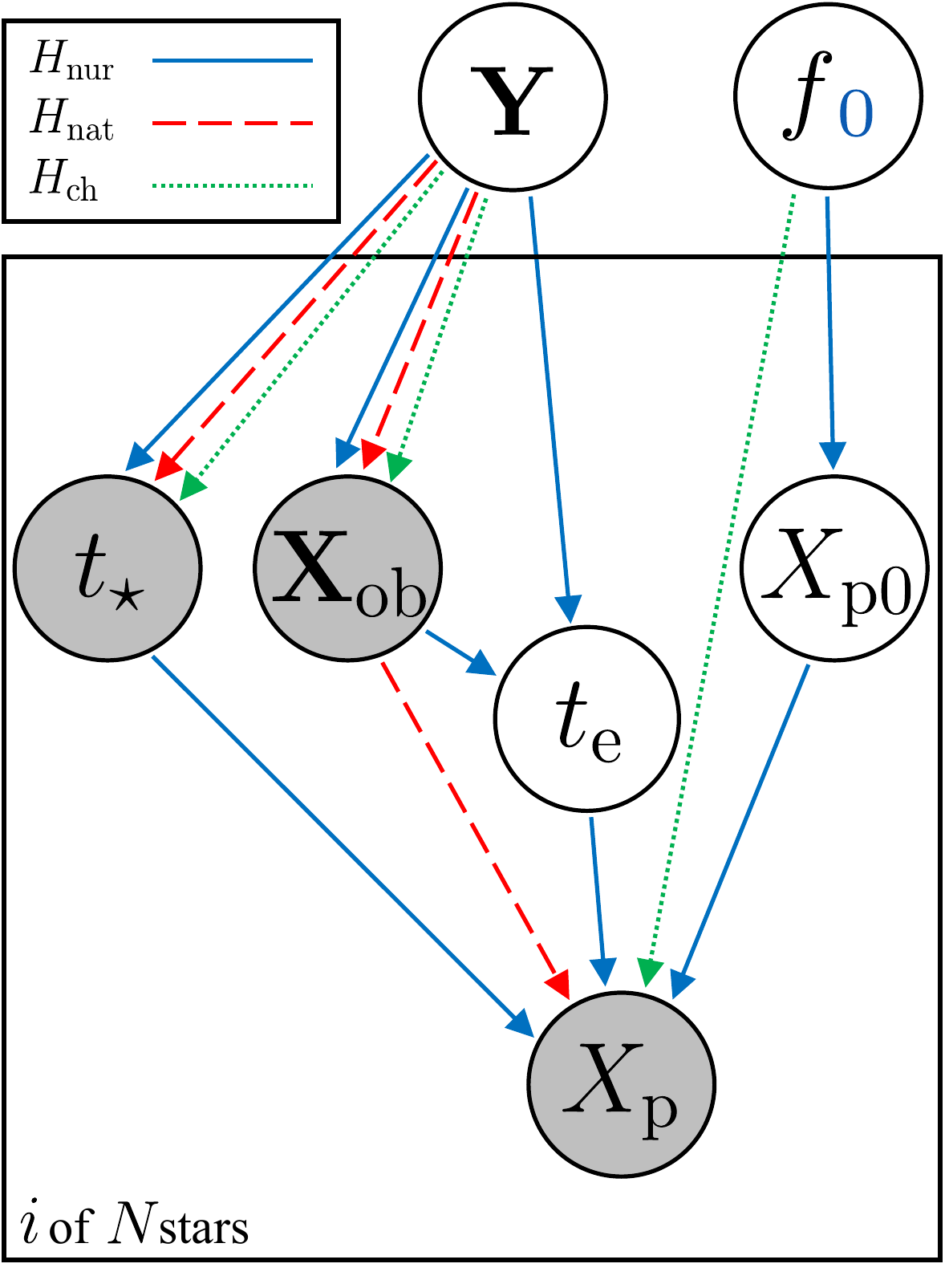}
    \caption{Graphical representation of the three hypotheses in the general model, showing possible relationships between parameters under different hypotheses.  Rather than show every possible dependence that a model could have, we focus on those that most often become relevant in the applications we examine.  Relations under the Nurture hypothesis are shown with a solid blue line, those under the Nature hypothesis with a dashed red line, and those under the Chance hypothesis with a dotted green line.  The ``$i$ of $N$ stars" indicates that the plate is iterated over each of the $N$ systems in the sample.  Gray circles represent observed parameters:  the planetary property of interest ($X_{\rm p}$), the stellar age ($t_{\star}$), and other observed parameters (${\bf X}_{\rm ob}$).  White circles are unobserved individual parameters (on the plate) -- the initial value of $X_{\rm p}$ ($X_{\rm p0}$) and the evolutionary timescale ($t_{\rm e}$) -- and hyperparameters -- the fraction of systems with a given value of $X_{\rm p}$ ($f$), the fraction of systems that start out with a given value of $X_{\rm p}$ ($f_0$) and other hyperparameters (${\bf Y}$).  The hyperparameters $f$ and $f_0$ are in the same circle because mathematically, they behave the same way in each hypothesis; they are distinguished by the blue subscript, because $f_0$ occurs in the Nurture hypothesis while the Nature and Chance hypotheses use $f$.  Note that we do not explicitly include the relevant non-observed quantities (${\bf X}_{\rm nob}$), but if we did, it would be in a white circle; it would have each type of arrow pointing from ${\bf Y}$ to it and from it to $t_\star$ and ${\bf X}_{\rm ob}$, as well as a Nurture hypothesis arrow from it to $t_{\rm e}$.}
    \label{fig:graphicalmodel}
\end{figure}

In Figure \ref{fig:graphicalmodel}, we present a graphical representation of possible relationships between parameters in Eqn. \ref{eqn:mostgeneral} under different hypotheses.  For simplicity, we do not show every possible relationship, but rather focus on those that most often become relevant in the applications we examine.

Ultimately we are interested in the odds ratios for pairs of hypotheses to determine which hypothesis is most favored by the data.  The odds ratios are obtained by applying the equations in the previous subsections to the observational data according to Eqn. \ref{eqn:oddsratio}.  For the purposes of this paper, we will consider odds ratios of $\sim1-10$ to be inconclusive, $\sim10-100$ to be moderately supportive but not decisive, and $\sim100$ and above to be strong, similar to the scale given by \cite{jeffreys1961} and \cite{kass1995}.

The calculation of the ratios may result in quantities cancelling out in some cases, reducing the need to determine suitable priors for and relationships between some parameters.  In particular, if there are no parameters of interest in ${\bf X}_{\rm nob}$ to consider and if $p(t_\star,{\bf X}_{\rm ob})$ is independent of ${\bf Y}$, then we can write, from Eqns. \ref{eqn:hyp1gen3v2}, \ref{eqn:genhyp2final}, and \ref{eqn:genhyp3cont}:
\begin{align}
    H_{\rm nur}:  p(X_{\rm p},t_\star,{\bf X}_{\rm ob}|{\bf Y})&= \iint p(X_{\rm p}|t_\star,t_{\rm e},X_{\rm p0},{\bf Y}) p(t_{\rm e}|{\bf X}_{\rm ob},X_{\rm p0},{\bf Y})\nonumber \\
    & \qquad\quad \times p(X_{\rm p0}|{\bf Y})dt_{\rm e} dX_{\rm p0}p(t_\star,{\bf X}_{\rm ob})\\
    H_{\rm nat}:  p(X_{\rm p},t_{{\star}},{\bf X}_{\rm ob}|{\bf Y})&= p(X_{\rm p}|{\bf X}_{\rm ob},{\bf Y})p(t_{{\star}},{\bf X}_{\rm ob})\\
    H_{\rm ch}:  p(X_{\rm p},t_{{\star}},{\bf X}_{\rm ob}|{\bf Y})&= p(X_{\rm p}|{\bf Y})p(t_{{\star}},{\bf X}_{\rm ob}).
\end{align}
The common term between these three equations, $p(t_{{\star}},{\bf X}_{\rm ob})$, will cancel when the odds ratios are calculated, eliminating the need to specify the exact relationship between $t_\star$ and ${\bf X}_{\rm ob}$ or the underlying distribution of these parameters.

Finally, we note that our framework does not account for measurement uncertainties on $X_{\rm p}$, $t_{\star}$, or ${\bf X}_{\rm ob}$.  This is an additional complicating factor and we recognize its omission makes our framework less general.  We plan to address this in a future paper and incorporate measurement uncertainties in our general framework as well as the applications we investigate here.

\section{Are systems with 2:1 resonances younger than those without?}
\label{sec:resonances}

Mean motion resonances are important tracers of planetary systems' formation and evolution. Dissipative processes -- including tides (e.g.,\citealt{yode79}) and planet-disk interactions (e.g., \citealt{lee02,macd18}) -- can capture planets into resonance. The initial frequency of resonances informs us about the planets' formation environment and early evolution (e.g., \citealt{malh93,gold14}), but resonances can be broken by later dynamical interactions among planets (e.g., \citealt{thommes2008,izid17}).  Investigating if and how the occurrence of orbital resonances changes with system age can help us trace back the initial fraction and also better understand the processes that subsequently destabilize resonances.

\citet{kz2011} examined the ages and resonance states of a sample of multiplanet systems.  For the sake of homogeneity, they limited their sample to 30 systems of planets discovered via radial velocity around F, G, and K stars with stellar ages derived from chromospheric activity calcium H and K lines (averaging when a target had multiple age estimates from different studies). As described in the review by \cite{soderblom2010}, measuring stellar activity levels is one of the most widely used methods of age determination, but it is subject to systematics that are not fully understood, and activity levels for stars of the same age can vary considerably.  With calcium H and K lines, the age is calculated from an index $R'_{\rm HK}$ derived from the H and K line ratios.  While individual stellar age uncertainties were not reported, several of the sources cited by \cite{kz2011} calculated stellar age with a fit for $R'_{\rm HK}$ versus age determined by \cite{mamhill2008}, who estimated a 60\% uncertainty on ages derived from their relation when taking into account calibration and observational uncertainties as well as astrophysical scatter.

To determine those systems that were in resonance, they defined a “normalized commensurability proximity” score as follows:
\begin{equation}
    \delta=2\frac{|r-r_{\rm c}|}{r+r_{\rm c}}
    \label{eqn:ncp}
\end{equation}
where $r$ is the measured period ratio and $r_{\rm c}$ is the period commensurability ratio -- 2:1 in this case -- against which $r$ is compared.  They defined a system to have a period commensurability if it had $\delta < 0.1$.  With this criteria, the dataset consists of five systems with 2:1 resonances and 25 systems without 2:1 resonances.

\begin{figure}[ht]
 \centering
    \includegraphics[width=5.0 in]{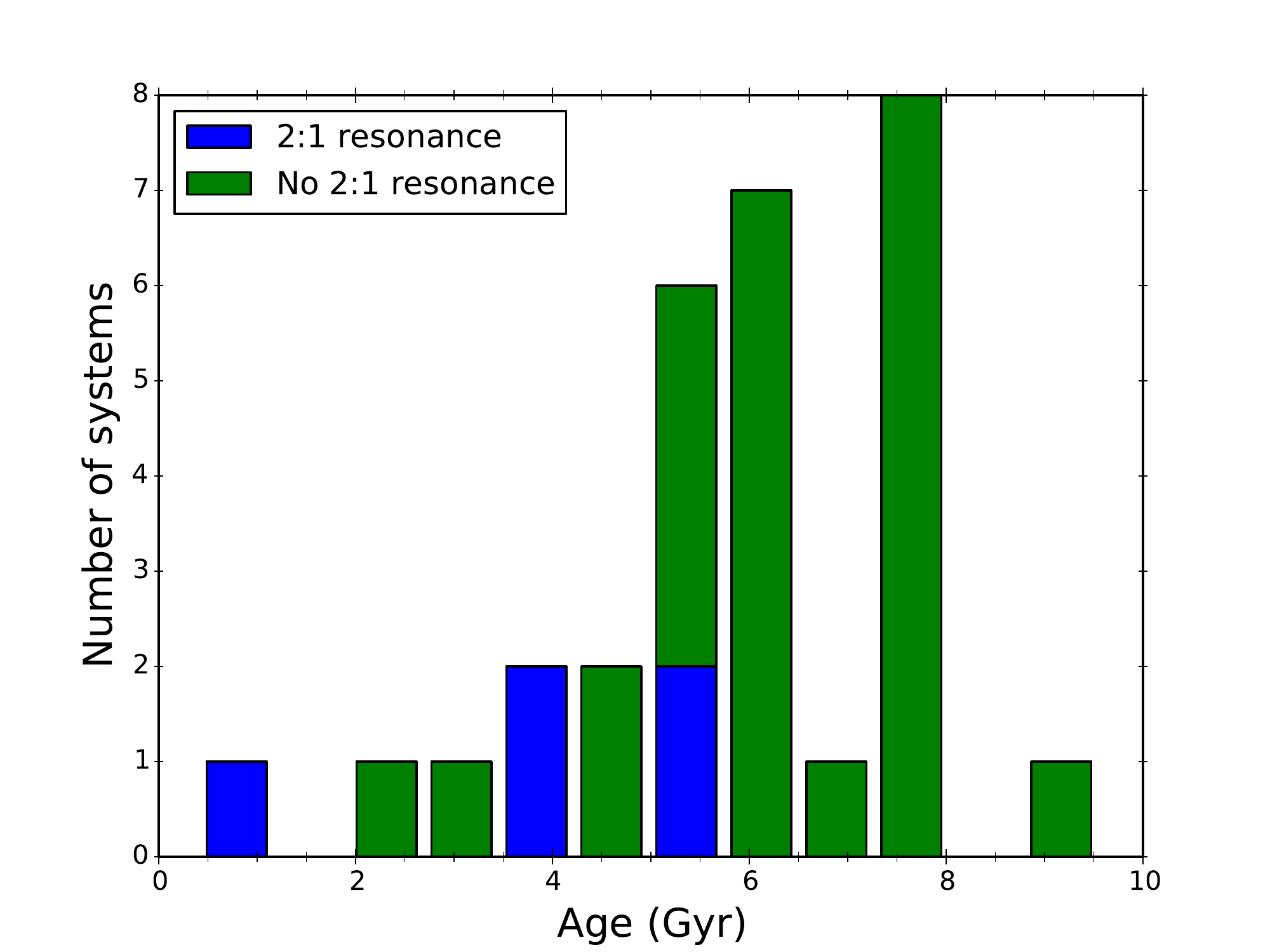}
    \caption{Histogram showing the ages of systems in the \citet{kz2011} sample with 2:1 resonances (blue) and without 2:1 resonances (green). This is the data we use in our analysis in Section \ref{sec:resonances}.}
    \label{fig:ageshist}
 \end{figure}

\citet{kz2011} found that the planetary systems in their sample that exhibit 2:1 period ratios tended to be younger than those that do not, with a median age difference of 2.15 Gyr (Figure \ref{fig:ageshist}). The authors ran a series of random permutation tests on the sample and found that only about 0.4\% of the randomized samples yielded a median age difference between systems with and without a 2:1 resonance that is greater than the measured value.  While acknowledging the small sample size and uncertainties in stellar ages, the authors suggested this finding is indicative of such commensurabilities having only a finite lifetime. 

Random permutation tests reveal the probability of a trend emerging by chance, but in the framework of Bayesian hypothesis testing, we need to compare this probability to that expected from some competing model. \citet{dd2016} took the first step toward such a comparison by exploring the initial fraction of resonant systems and resonant disruption timescale needed to account for the observed trend. They found that their model could account for the data only if all systems begin in resonance and are disrupted on a timescale very similar to the age difference between the 2:1 resonant and nonresonant samples. Their results implied that the model parameters may need to be quite fine-tuned, which would penalize the resonance disruption hypothesis when assessed within a Bayesian hypothesis testing framework.

\begin{figure}[ht]
 \centering
    \includegraphics[width=3.0 in]{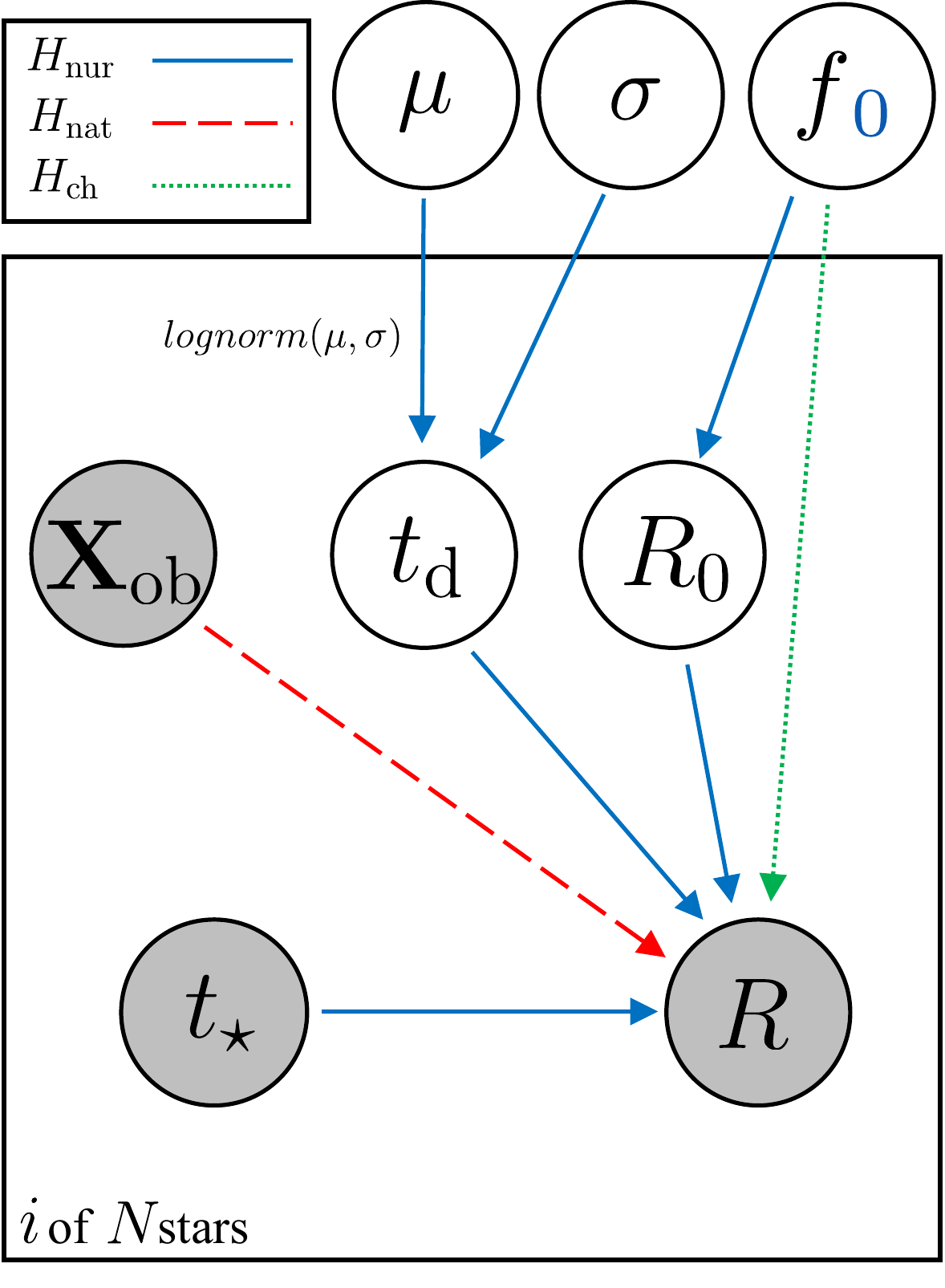}
    \caption{Graphical representation of the three hypotheses applied to the case of 2:1 resonances.  Relations under the Nurture hypothesis are shown with a solid blue line, those under the Nature hypothesis with a dashed red line, and those under the Chance hypothesis with a dotted green line.  The ``$i$ of $N$ stars" indicates that the plate is iterated over each of the $N$ systems in the sample.  Gray circles represent observed parameters: the current 2:1 resonance state of the system ($R$), the stellar age ($t_{\star}$), and other observed parameters that may be used to test the Nature hypothesis (${\bf X}_{\rm ob}$), which we do not test here. White circles are unobserved individual parameters (on the plate) -- the initial 2:1 resonance state ($R_{0}$) and the 2:1 resonance disruption timescale ($t_{\rm d}$) -- and hyperparameters -- the mean ($\mu$) and standard deviation ($\sigma$) of a log-normal prior for $t_{\rm d}$, the fraction of systems with a 2:1 resonance ($f$), and the fraction of systems that start out with a 2:1 resonance ($f_0$).  The hyperparameters $f$ and $f_0$ are in the same circle because mathematically, they behave the same way in each hypothesis; they are distinguished by the blue subscript, because $f_0$ occurs in the Nurture hypothesis while the Chance hypothesis uses $f$.}
    \label{fig:resonancemodel}
 \end{figure}

Here we use the same dataset as \cite{kz2011} with the framework from Section \ref{sec:general} to compare the evidence for the Nurture and Chance hypotheses: that 2:1 resonances are systematically disrupted (Nurture); and that there is no astrophysical correlation between 2:1 resonances and age, so that the perceived trend is merely due to chance.  We show a graphical representation of these two hypotheses in Figure \ref{fig:resonancemodel}.  Since we are unaware of any proposed correlations between the presence of 2:1 resonances and system parameters other than age, we do not test the Nature hypothesis for this case in this work, though we may investigate it in the future.  In Section \ref{subsec:reseqns}, we adapt the general equations derived in Section \ref{sec:general} to this specific case. We present the results of our analysis in Section \ref{subsec:resresults}, show and discuss contour plots of hyperparameters in Section \ref{subsec:rescontourplots}, and briefly summarize in Section \ref{subsec:ressummary}.

\subsection{Equations for each hypothesis}
\label{subsec:reseqns}
\subsubsection{Hypothesis 1: Nurture (Age-driven)}
\label{subsec:reshyp1}
Under the Nurture hypothesis applied to the 2:1 resonance case, the planetary property of interest, $X_{\rm p}$, is the resonance state of the system.  We represent this state by the binary parameter $R$, which is 1 when a system has a 2:1 resonance and 0 when it does not.  The initial state $X_{\rm p0}$ is the initial resonance state of the system, $R_0$, which is defined in the same way as $R$.  The evolutionary timescale in this case is a resonance disruption timescale, which we represent with $t_{\rm d}$.  There are no other observational data we consider, so we drop ${\bf X}_{\rm ob}$.  Substituting these variables into the general Nurture hypothesis equation for a binary parameter with one-directional evolution (Eqn. \ref{eqn:hyp1genbinaryfinal}) gives:
\begin{align}
    \label{eqn:resnurfirst}
    p(R,t_{\star}|f_0)=
    \begin{cases}
    \text{\Large$\int$} \left[(1-f_0)+f_0\int_0^{t_{\star}}p(t_{\rm d}|R_0,{\bf X}_{\rm nob})dt_{\rm d}\right] p(t_{\star},{\bf X}_{\rm nob}) d{\bf X}_{\rm nob} &,R=0\\
    f_0\iint_{t_{\star}}^\infty p(t_{\rm d}|R_0,{\bf X}_{\rm nob})p(t_{\star},{\bf X}_{\rm nob})dt_{\rm d} d{\bf X}_{\rm nob}&,R=1
    \end{cases}
\end{align}
where $f_0$ represents the fraction of systems that begin with a 2:1 resonance.

We assume the underlying distribution of disruption timescales is log-normal -- and note that a system with $R_0=0$ must have $R=0$ and has no timescale -- and that it is independent of anything in ${\bf X}_{\rm nob}$: 
\begin{equation}
    p(\text{log}_{10}(t_{\rm d})|\mu,\sigma)=\frac{1}{\sqrt{2\pi \sigma^2}} \text{exp}\left(\frac{-(\text{log}_{10}(t_{\rm d})-\mu)^2}{2\sigma^2}\right).
\end{equation}
This choice is appropriate because orbital evolution processes can happen on timescales that span many orders of magnitude, making a logarithmic form suitable.  Resonance systems close to the stability limit might break apart quickly, while others may be stable for longer than the current age of the universe.  A wide log-normal prior is an uninformative way to capture this range of behavior.

This formulation introduces two unknown parameters, namely the mean $\mu$ and standard deviation $\sigma$ of the prior of log$_{10}(t_{\rm d})$.  Both $\mu$ and $\sigma$ are population-wide variables; thus, they are hyperparameters that are marginalized over outside the product of individual likelihoods of each system. We impose uniform hyperpriors on these variables with the following ranges, given in log$_{10}$([yr]) space:
\begin{align}
    p(\mu)&=\text{U}(6,16)\\
    p(\sigma)&=\text{U}(0,20).
\end{align}
These wide ranges are to account for our relative lack of knowledge about resonance disruption timescales and encompass timescales ranging from roughly right after planet formation -- the earliest conceivable time that a resonance could be broken, which is roughly $10^6$ years after system formation -- to many orders of magnitude beyond the current age of the universe.  We note that the uniform hyperprior on $\mu$ is uninformative, but the uniform hyperprior on $\sigma$ is not.  The true uninformative hyperprior for $\sigma$, as a scale parameter, would be $1/\sigma^2$.  The uniform hyperprior is more weighted to large values of $\sigma$.  However, as we expect resonance disruption timescales to span many orders of magnitude, such a weighting is appropriate for this case and we stick with the uniform hyperprior.

For simplicity, we will assume that $p(t_{\star})$ does not depend on anything in ${\bf X}_{\rm nob}$.  Since, then, neither $t_{\star}$ nor $t_{\rm d}$ depends explicitly on other system properties, we can remove ${\bf X}_{\rm nob}$ in Eqn. \ref{eqn:resnurfirst}, and the $p(t_{\star})$ term will cancel out when the odds ratio is calculated, as described in Section \ref{subsec:genoddsratios}.

We give $f_0$ a uniform hyperprior from 0 to 1:
\begin{equation}
    p(f_0)=\text{U}(0,1).
\end{equation}
We note that since $f_0$ is a Bernoulli distribution parameter, a uniform hyperprior is not actually uninformative.  For such a variable, the uninformative hyperprior is a Beta distribution with both parameters set to 1/2, Beta(1/2,1/2).  We will also calculate the odds ratios using the Beta distribution on $f_0$ and compare to the results from a uniform hyperprior.

With the above assumptions, Eqn. \ref{eqn:resnurfirst} becomes:
\begin{align}
    \label{eqn:hyp1resfinal}
    p(R,t_{\star}|\mu,\sigma,f_0)=
    \begin{cases}
    \left[(1-f_0)+f_0\int_0^{t_{\star}}p(t_{\rm d}|\mu,\sigma)dt_{\rm d}\right] p(t_{\star}) &,R=0\\
    f_0\int_{t_{\star}}^\infty p(t_{\rm d}|\mu,\sigma)dt_{\rm d} p(t_{\star}) &,R=1.
    \end{cases}
\end{align}

\subsubsection{Hypothesis 2: Nature (Driven by inherent system properties)}
\label{subsec:reshyp2}

We are not aware of any proposed correlations between the presence of 2:1 resonances and system parameters other than age, so we do not test this hypothesis for the resonances case at this time.  If a possible relationship does emerge, such as with stellar mass $M_{\star}$, Eqn. \ref{eqn:genhyp2final} could be applied to give
\begin{align}
\label{eqn:resnatfirst}
p(R,t_{\star},M_{\star})&=\int p(R|M_{\star})p(t_{\star},M_{\star},{\bf X}_{\rm nob})d{\bf X}_{\rm nob},
\end{align}
where we have set ${\bf X}_{\rm ob}=M_\star$. ${\bf X}_{\rm nob}$ could be, for example, a property of the initial proto-planetary disk thought to affect capture into resonance.

\subsubsection{Hypothesis 3: Chance}
\label{subsec:reshyp3}
Again, we do not consider any other observational data or unobserved parameters.  Thus under the Chance hypothesis, the general equation for a binary $X_{\rm p}$ (Eqn. \ref{eqn:genhyp3final}) applied to the resonance case gives
\begin{align}
\label{eqn:hyp3resfinal}
p(R,t_{\star}|f)=
\begin{cases}
p(t_{\star})(1-f) &, R=0\\
p(t_{\star}) f&, R=1.
\end{cases}
\end{align}
We give $f$, the fraction of systems with a 2:1 resonance, a uniform distribution from 0 to 1, but will also compare the odds ratios to those obtained when $f$ is given the uninformative hyperprior of Beta(1/2,1/2).

\subsection{Results}
\label{subsec:resresults}

To calculate the odds ratio, we compute the individual system likelihoods from Eqn. \ref{eqn:hyp1resfinal} (Nurture hypothesis) and Eqn. \ref{eqn:hyp3resfinal} (Chance hypothesis).  For each hypothesis, we multiply those individual likelihoods and then marginalize over the hyperparameters $f_0$, as well as $\mu$ and $\sigma$ for the Nurture hypothesis, according to Eqn. \ref{eqn:oddsratio}.  The complete equations for each hypothesis are
\begin{align}
    p(H_{\rm nur})&=\iiint \prod_{R=0}\left[p(t_{\star})(1-f_0) + p(t_{\star})f_0\int_0^{t_{\star}}p(t_{\rm d}|\mu,\sigma)dt_{\rm d}\right] \nonumber\\&\qquad\quad\times\prod_{R=1}\left[p(t_{\star})f_0\int_{t_{\star}}^\infty p(t_{\rm d}|\mu,\sigma) dt_{\rm d}\right]p(\mu)p(\sigma)p(f_0)d\mu d\sigma df_0\\
    p(H_{\rm ch})&=\int \prod_{R=0}\left[p(t_{\star})(1-f)\right]\prod_{R=1}\left[p(t_{\star}) f\right]p(f)df.
\end{align}
The priors of relevant parameters are summarized in Table \ref{tab:respriors}, where we have written out the complete equations for the Nurture and Chance hypotheses, where the ranges for $\mu$ and $\sigma$ are given in log$_{10}$([yr]).

\begin{table}
	\centering
	\begin{tabular}{|l|l|}
		\hline
		Parameter & Prior \\
		\hline
        Disruption timescale (log$_{10}(t_{\rm d})|\mu,\sigma$) & $\frac{1}{\sqrt{2\pi \sigma^2}} \text{exp}\left(\frac{-(\text{log}_{10}(t_{\rm d})-\mu)^2}{2\sigma^2}\right)$\\
        \hline
        Mean of log$_{10}(t_{\rm d})$ ($\mu$) & U(6,16)\\ 
        \hline
        Standard deviation of log$_{10}(t_{\rm d})$ ($\sigma$) & U(0,20)\\
        \hline
        Initial 2:1 resonance fraction ($f_0$) & U(0,1)\\
        \hline
        2:1 resonance fraction ($f$) & U(0,1)\\
        \hline
	\end{tabular}
	\caption{Priors of relevant parameters for the resonance case.  The ranges for $\mu$ and $\sigma$ are given in log$_{10}$([yr]).}
	\label{tab:respriors}
\end{table}

The integrations are performed using the Python package scipy.integrate.nquad with default settings, which uses techniques from the Fortran library QUADPACK.  Specifically, it uses a Clenshaw-Curtis method with Chebyshev moments if the integration limits are finite, and it uses a Fourier integral in the case of an infinite limit. For comparison we also compute integrals using the numerical integrator NIntegrate in Mathematica, which by default uses an adaptive Genz–Malik algorithm, and find excellent agreement with the results from Python.

We find that the odds ratio of the Nurture hypothesis to the Chance hypothesis is 2.2.  This result only weakly supports the Nurture scenario, and we cannot favor one hypothesis over the other with the current information.  We can say that according to this approach, there is not enough evidence in the \cite{kz2011} dataset to support systematic resonance disruption.  To get a better idea of whether 2:1 resonances do eventually get disrupted, we would need to use a larger sample of 2:1 resonant systems, ideally with a wider age range.

We also compute the odds ratio with $f_0$ and $f$ both assigned uninformative hyperpriors of Beta(1/2,1/2).  We obtain an odds ratio of the Nurture hypothesis to the Chance hypothesis of 2.0.  This is not a significant change from before and still does not allow us to favor one hypothesis over the other.

In Appendix \ref{sec:appendixb}, we perform additional variations in our calculations by removing systems with highly-uncertain ages, bootstrapping the data, and marginalizing over the threshold for a system having a 2:1 resonance.  In each of these treatments, we obtain odds ratios of Nurture vs. Chance close to our original value of 2.2, and thus still cannot favor one hypothesis over the other.

\subsection{Contour plots}
\label{subsec:rescontourplots}
\begin{figure}[ht]
    \includegraphics[width=3.5 in]{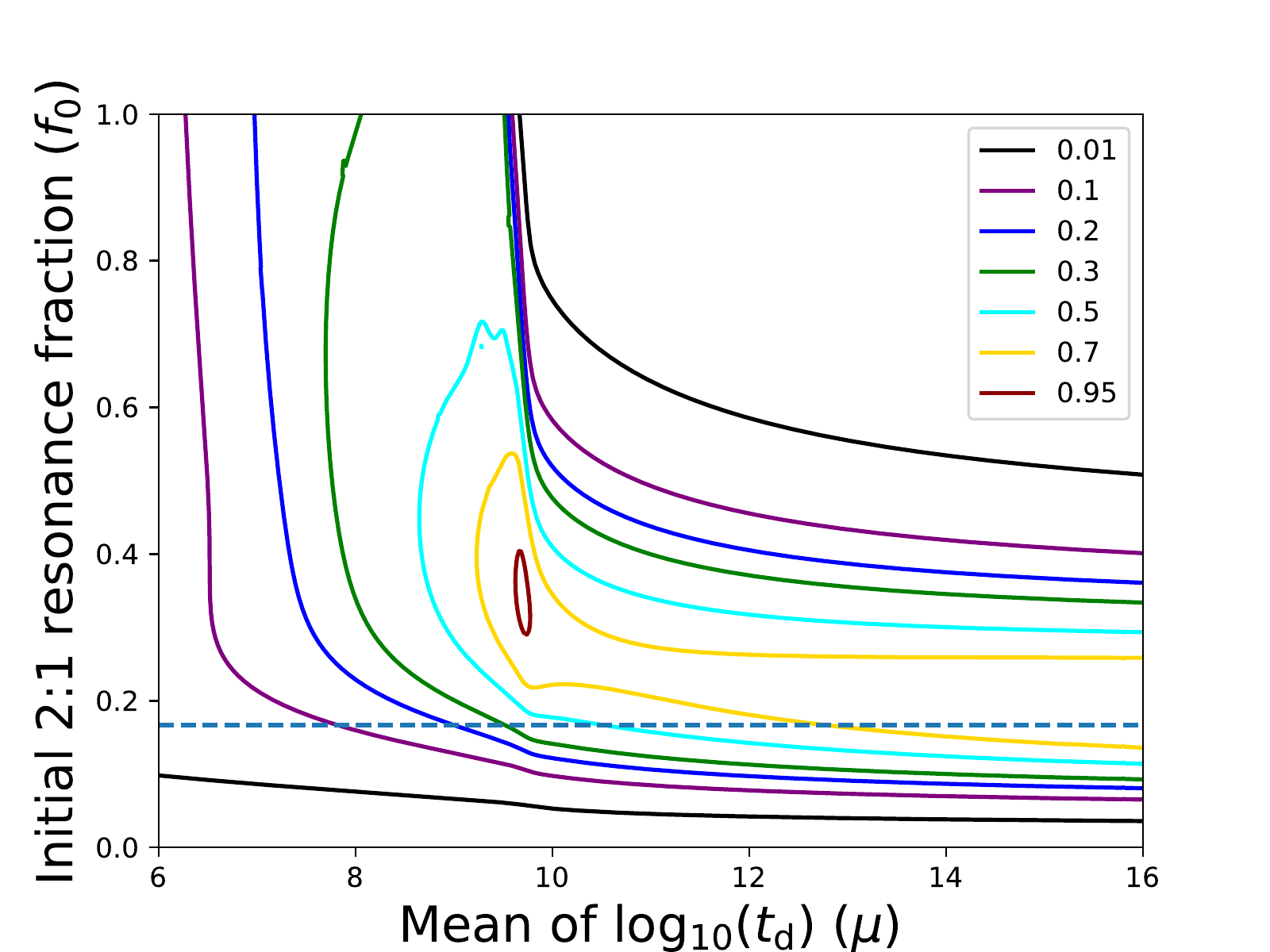}
    \includegraphics[width=3.5 in]{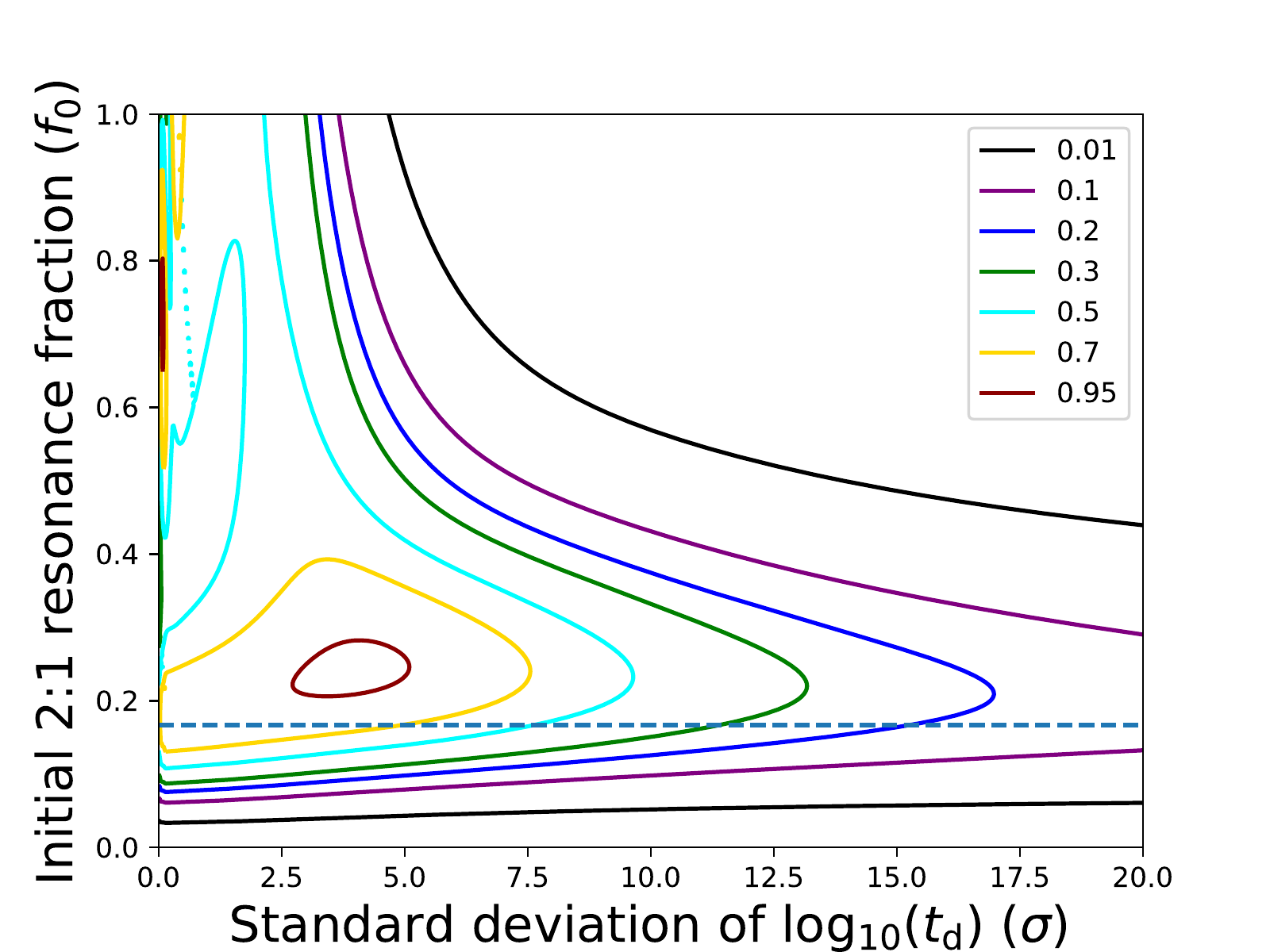}
    \includegraphics[width=3.5 in]{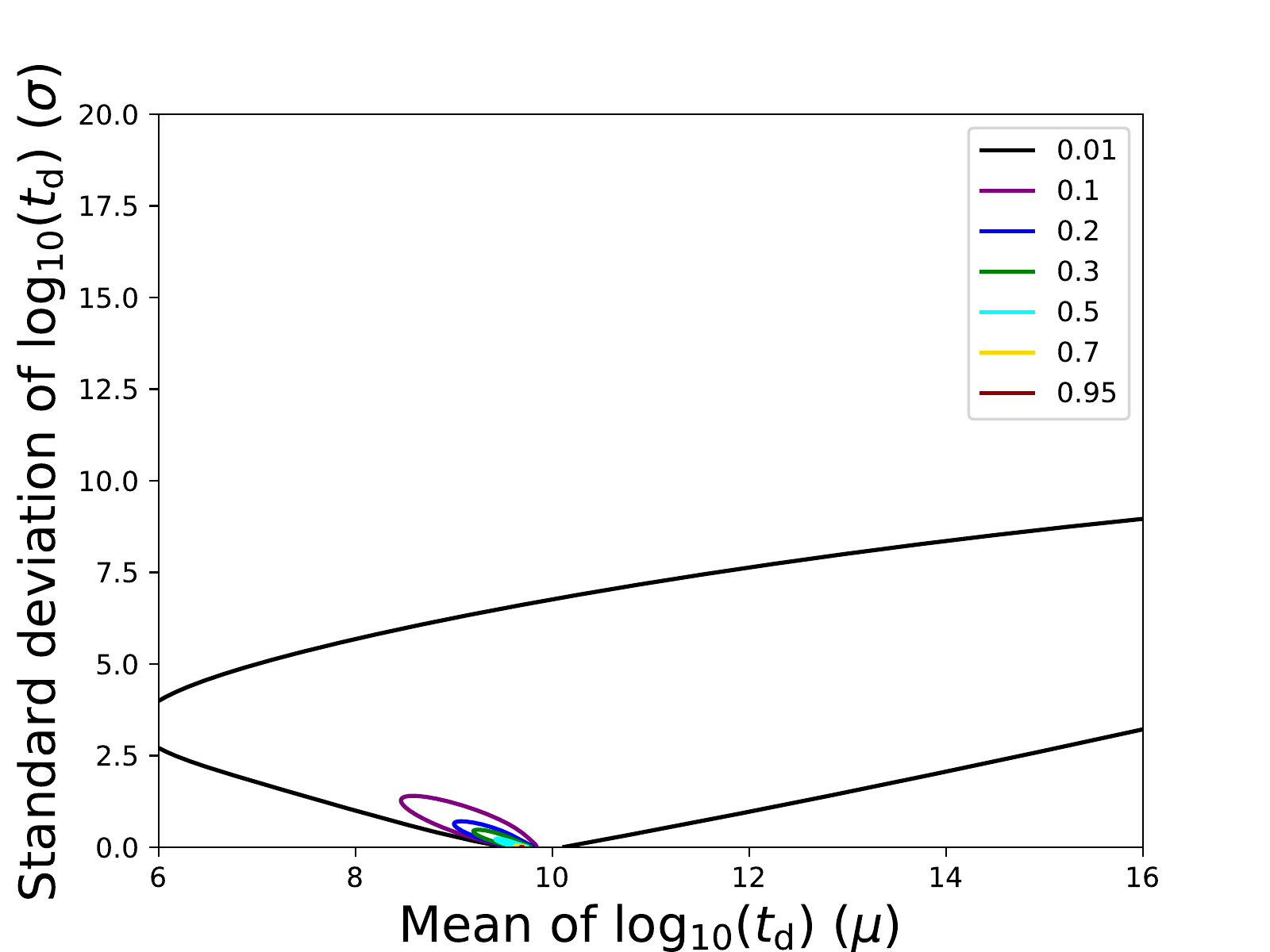}
    \includegraphics[width=3.5 in]{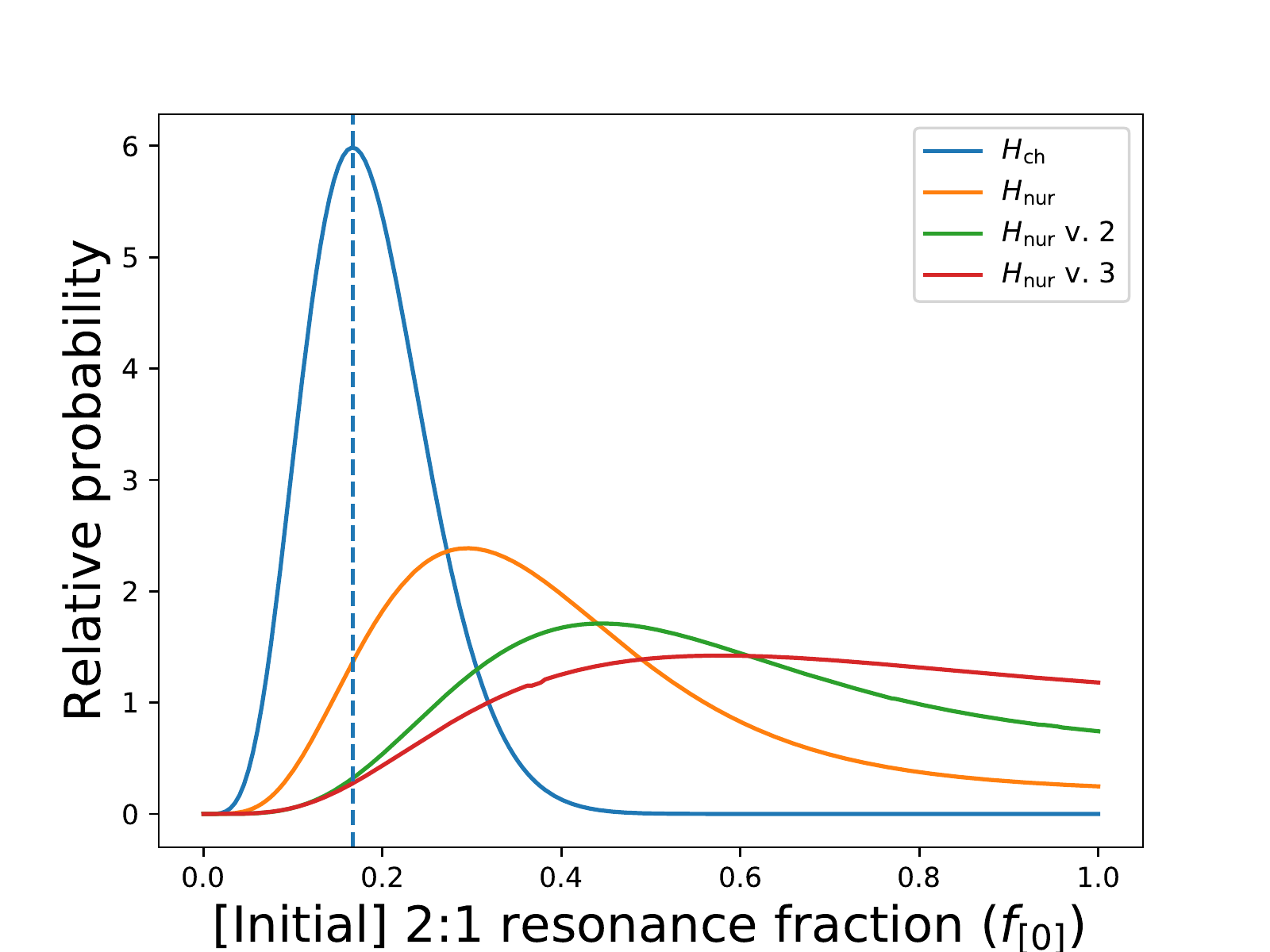}
    \caption{Top row and bottom left: Probability contour plots for the Nurture hypothesis for the 2:1 resonances application, showing $f_0$ vs. $\mu$ (top left), $f_0$ vs. $\sigma$ (top right), and $\sigma$ vs. $\mu$ (bottom left).  Contour levels are 0.01, 0.05, 0.2, 0.5, 0.6, 0.7, 0.8, and 0.95.  The observed fraction of systems with a 2:1 resonance (5/30) is shown with a dashed line.  Bottom right: Probability histogram of the 2:1 resonance fraction ($f$) under the Chance hypothesis (blue), and the initial 2:1 resonance fraction ($f_0$) under the Nurture hypothesis with original hyperprior ranges on $\mu$ and $\sigma$ (orange), with the first reduction of $\mu$ and $\sigma$ hyperprior ranges (green, labeled as ``$H_{\rm nur}$ v. 2"), and with the second reduction of $\mu$ and $\sigma$ hyperprior ranges (red, labeled as ``$H_{\rm nur}$ v. 3").  The curves have been normalized to have the same area.  The observed 2:1 resonance fraction (5/30) is again shown with a dashed line.  Note that because resonances are broken over time in the Nurture hypothesis, we do not expect $f_0$ to match up with the observed 2:1 resonance fraction.}
    \label{fig:rescontourplots}
 \end{figure}

In Figure \ref{fig:rescontourplots} we display probability contour plots of hyperparameters $f_0$ vs. $\mu$ (top left), $\sigma$ vs. $\mu$ (top right), and $f_0$ vs. $\sigma$ (bottom left) from the equations for the Nurture hypothesis.  Note that these are not plots of the posterior or the likelihood; rather, they are plots of the part of the posterior that does not get cancelled in the odds ratio.  They are also relative, not absolute, probabilities that have been normalized such that the highest value is 1.  The probabilities have been calculated on a grid for each value of $f_0$, $\mu$, and $\sigma$.  The contour plots indicate the values of those hyperparameters that contribute the most to the odds ratio under the Nurture hypothesis and can also give some insight into why we do not see enough evidence for the Nurture hypothesis to favor it strongly over the Chance hypothesis.

The structure in the Nurture contour plots suggests a continuous distribution of solutions that fit the data with relatively high probability ($\gtrsim$ 70\% of the highest probability solution) under the Nurture hypothesis.  The highest probability region in the $f_0$ vs. $\mu$ plot indicates that under the Nurture explanation, somewhere between about 20\% and 50\% of systems would start out with a 2:1 resonance and would most likely get disrupted after a few Gyr, a timescale similar to the ages of the sample stars (see Figure \ref{fig:ageshist}).  The high probability region then stretches to the right across much of the range of $\mu$.  At the high-$\mu$ end, this corresponds to a disruption timescale much longer than the current age of the universe, and an initial fraction similar to the observed current fraction.  In this high-$\mu$ solution, roughly 20\% of systems start out with a 2:1 resonance, and these configurations are stable on very long timescales, which means we would see the vast majority of them still in resonance today. Thus this solution is essentially equivalent to the Chance hypothesis.

\begin{figure}[ht]
    \includegraphics[width=3.5 in]{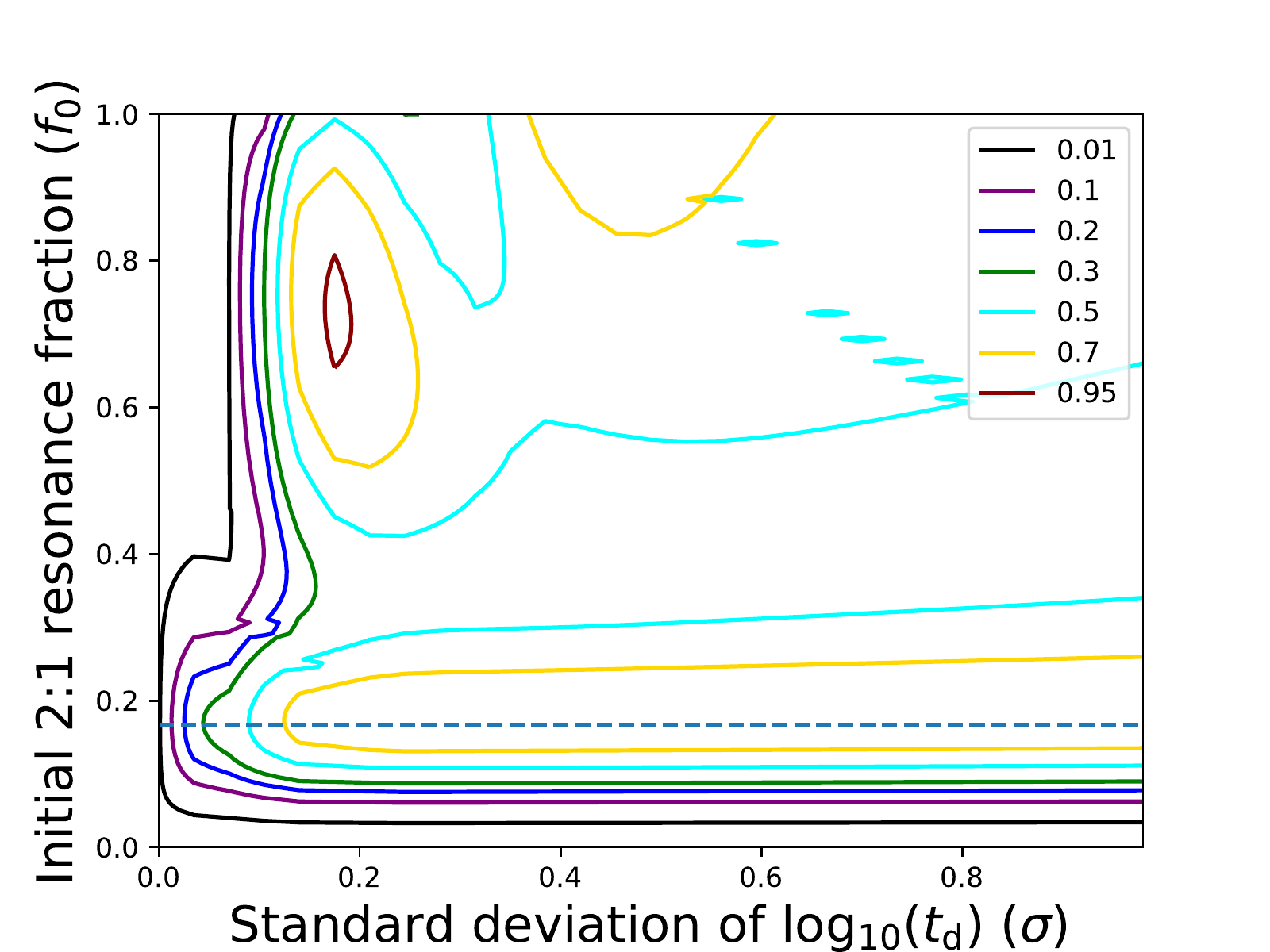}
    \includegraphics[width=3.5 in]{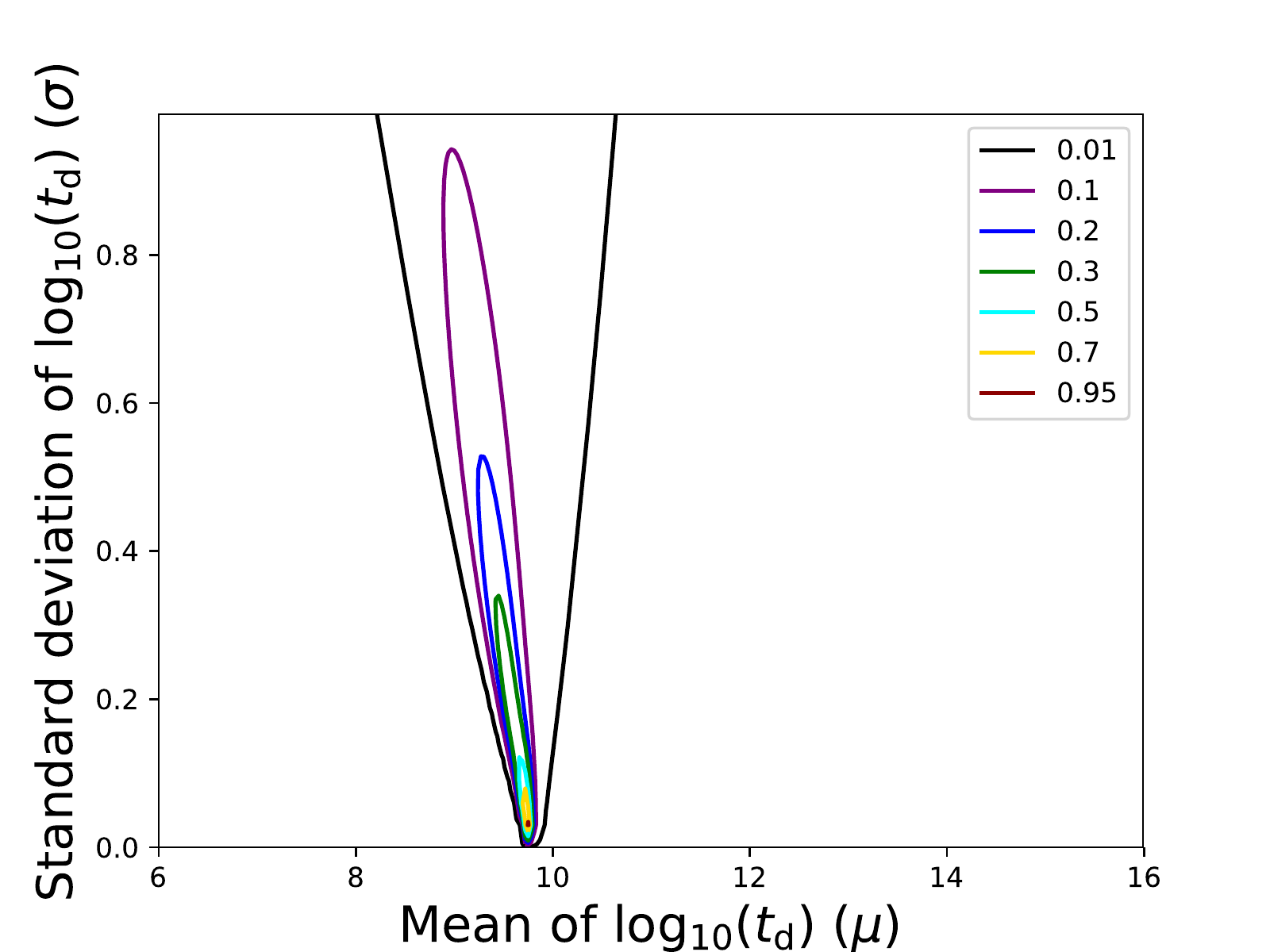}
    \caption{Close up views of the region from $\sigma=0$ to $\sigma=1$ in the $f_0$ vs. $\sigma$ (left) and $\sigma$ vs. $\mu$ (right) probability contour plots for the Nurture hypothesis for the 2:1 resonances application.  Contour levels are 0.01, 0.05, 0.2, 0.5, 0.6, 0.7, 0.8, and 0.95.  The observed fraction of systems with a 2:1 resonance (5/30) is shown with a dashed line.}
    \label{fig:rescontourplotszoomedin}
 \end{figure}

The $\sigma$ vs. $\mu$ plot indicates that there is very little variation in the disruption timescale of a few Gyr from system to system, as it exhibits a peak near $\sigma=0$.  This is supported by the very narrow high probability region at the far left of the $f_0$ vs. $\sigma$ plot.  Figure \ref{fig:rescontourplotszoomedin} shows close up views of the $\sigma=0$ to $\sigma=1$ region in both the $f_0$ vs. $\sigma$ and $\sigma$ vs. $\mu$ plots, which reveal that the peaks are near, but not exactly at, $\sigma=0$.  These peaks imply that in the solution where 2:1 resonance systems get disrupted after a few Gyr, there is very little variation in the timescale from system to system, so they would essentially get disrupted nearly all at the same age.  However, the $f_0$ vs. $\sigma$ plot also indicates a high probability for values of $\sigma$ from about 2.5 to 5, indicating a range of solutions with disruption timescale variations of several orders of magnitude.  This region also is at an initial 2:1 resonance fraction of roughly 20\%, suggesting it corresponds to the high-$\mu$ solution evident in the $f_0$ vs. $\mu$ plot.  This implies that in the solution where 2:1 resonances are broken on timescales much longer than the age of the universe (the solution equivalent to the Chance hypothesis), there is some notable variation in those timescales from system to system, such that we may see a few of the initially-resonant systems already disrupted today.  Most 2:1 resonances, though, will still be intact, so the observed 2:1 resonance fraction will be very close to the initial 2:1 resonance fraction.

The small islands in the upper right of the left panel of Figure \ref{fig:rescontourplotszoomedin} are low probability valleys caused by outlier systems.  However, while their effect is seen in the contour plot, the outliers do not bias the overall result; their removal results in an odds ratio of Nurture to Chance of 2.3, again a result only barely different from the original odds ratio of 2.2.

Overall, the area of the high-probability region is small compared to the entire hyperparameter space for the Nurture hypothesis.  Low-probability white space (i.e., outside the lowest probability contour) exists because of our lack of prior knowledge on resonance disruption timescales and is present in the upper right corners of the $f_0$ vs. $\mu$ and $f_0$ vs. $\sigma$ plots and in the entire upper half of the $\sigma$ vs. $\mu$ plot. This space is eventually marginalized over to compute the total probability for the Nurture hypothesis.  Since the low-probability regions take up a large portion of hyperparameter space compared to the high-probability regions, they diminish support for the Nurture hypothesis. In other words, the high probability hyper-parameters are too fine-tuned and are naturally penalized in this framework.

To investigate the effect of the priors on our results, we recompute the odds ratio using smaller prior ranges on $\mu$ and $\sigma$.  From the original priors of $p(\mu)=\text{U}(6,16)$ and $p(\sigma)=\text{U}(0,20)$, we first cut the ranges down to $p(\mu)=\text{U}(6,10)$ and $p(\sigma)=\text{U}(0,10)$, in log$_{10}$([yr]) space.  These alternative priors give an odds ratio of the Nurture to Chance hypothesis of 3.7.  This treatment lends more support to the Nurture hypothesis, but only a little bit; it is still not very strongly favored.  As an extreme case, we further cut down the priors to $p(\mu)=\text{U}(9,10)$ and $p(\sigma)=\text{U}(0,1)$.  Rather than being physically motivated, this cut is entirely based on the contour plots in Figure \ref{fig:rescontourplots} and designed to exclude the lower probability regions.  Even so, these priors only give an odds ratio of Nurture to Chance of 15.  This ratio is only moderately, but not decisively, in favor of the Nurture hypothesis.  Considering this particular calculation is contrived to result in high probability for the Nurture hypothesis anyway, we do not place much stock in this solution.  These results suggest that the odds ratio in this case is not strongly sensitive to the priors on $\mu$ and $\sigma$ used.

The bottom right panel of Figure \ref{fig:rescontourplots} shows a probability histogram of $f$ under the Chance hypothesis in blue, and the initial 2:1 resonance fraction ($f_0$) under the Nurture hypothesis with original hyperprior ranges on $\mu$ and $\sigma$ (orange), with the first reduction of $\mu$ and $\sigma$ hyperprior ranges (green, labeled as ``$H_{\rm nur}$ v. 2"), and with the second reduction of $\mu$ and $\sigma$ hyperprior ranges (red, labeled as ``$H_{\rm nur}$ v. 3").  The curves have been normalized to have the same area.  The observed 2:1 resonance fraction is shown with a dashed line.  The observational data of 5/30 systems having a 2:1 resonance is only slightly less than the best-fit initial 2:1 resonance fraction under the Nurture hypothesis, which is at about 25\%, and it is right at the peak of the best-fit 2:1 resonance fraction under the Chance hypothesis.

\subsection{Summary}
\label{subsec:ressummary}
In summary, we do not find enough evidence in the data we use to strongly support the hypothesis that 2:1 resonances are systematically disrupted.  This conclusion is robust against variations in the priors used for hyperparameters.

\section{Is Stellar Obliquity Realignment Primarily Driven by Age or by Stellar Temperature?}
\label{sec:alignment}

A star's obliquity angle, also referred to as the spin-orbit alignment angle, is the angle between the star's rotation axis and the orbital angular momentum vector of a planet around the star.  Our own Sun has a small -- though significant -- obliquity angle with the planets in the solar system \citep{beck2005,gomes2017}, and many systems have been observed in which the star has a much more severe misalignment (e.g., HAT-P-6b, \citealt{hebrard2011}; HAT-P-30b, \citealt{johnson2011}).  There are many factors and processes that may influence the excitation and damping of stellar obliquities, such as tidal interactions and Kozai-Lidov cycles, but the details as well as which factors play a dominant role remain unclear (e.g., Section 3.2 of \citealt{hotjupreview2018} and references therein).

\cite{winn2010} observed a trend of stellar obliquity with temperature for stars with hot Jupiters: hot stars are preferentially misaligned with their planets' orbits and cool stars are well aligned.  They set the division between aligned and misaligned groups at $T_{\star}=6250$ K (i.e. at the Kraft break; \citealt{kraft1967}) and noted that at approximately this temperature and below, stars have a sizable convective envelope.  The authors suggested that if this envelope is decoupled enough from the stellar core, it may aid the cooler stars in tidally realigning with their planets on relatively short timescales.  This observation of temperature dependence has also been supported by concurrent \citep{schlaufman2010} and follow up \citep{albrecht2012} studies.

However, \cite{triaud2011} looked at hot Jupiters orbiting A stars and found a trend with stellar age, in which younger stars have high obliquities and older stars are well aligned, with the separation at about 2.5 Gyr.  The ages of the stars were compiled from past studies and derived using stellar isochrones or evolutionary tracks, with the exception of WASP-17.  The age of WASP-17 was re-estimated for that paper using previously published data and evolutionary tracks.  Isochrone placement is a widely used age determination method and is based on well-understood physics, but uncertainties in temperature and other measured parameters can make precise age constraints difficult because the isochrones are often closely spaced \citep{soderblom2010}.  Most of the ages used in \cite{triaud2011} have uncertainties between 0.5 and 1 Gyr.  \cite{triaud2011} restricted their sample to stars with masses greater than $1.2 M_{\odot}$ and pointed out that such stars will cool by several hundred Kelvin over their main sequence lifetimes.  Thus, they proposed, it could be that stars are gradually realigned by tidal interactions as they age and also cool as they age, and that the cool aligned stars seen by \cite{winn2010} are the older versions of the hot misaligned stars that have had more time to realign with planetary orbits.  \citet{albrecht2012} later argued that the apparent trend with age may be due to a change in tidal realignment efficiency connected to temperature, rather than evolution of time.

\begin{figure}[ht]
\centering
    \includegraphics[width=3.0 in]{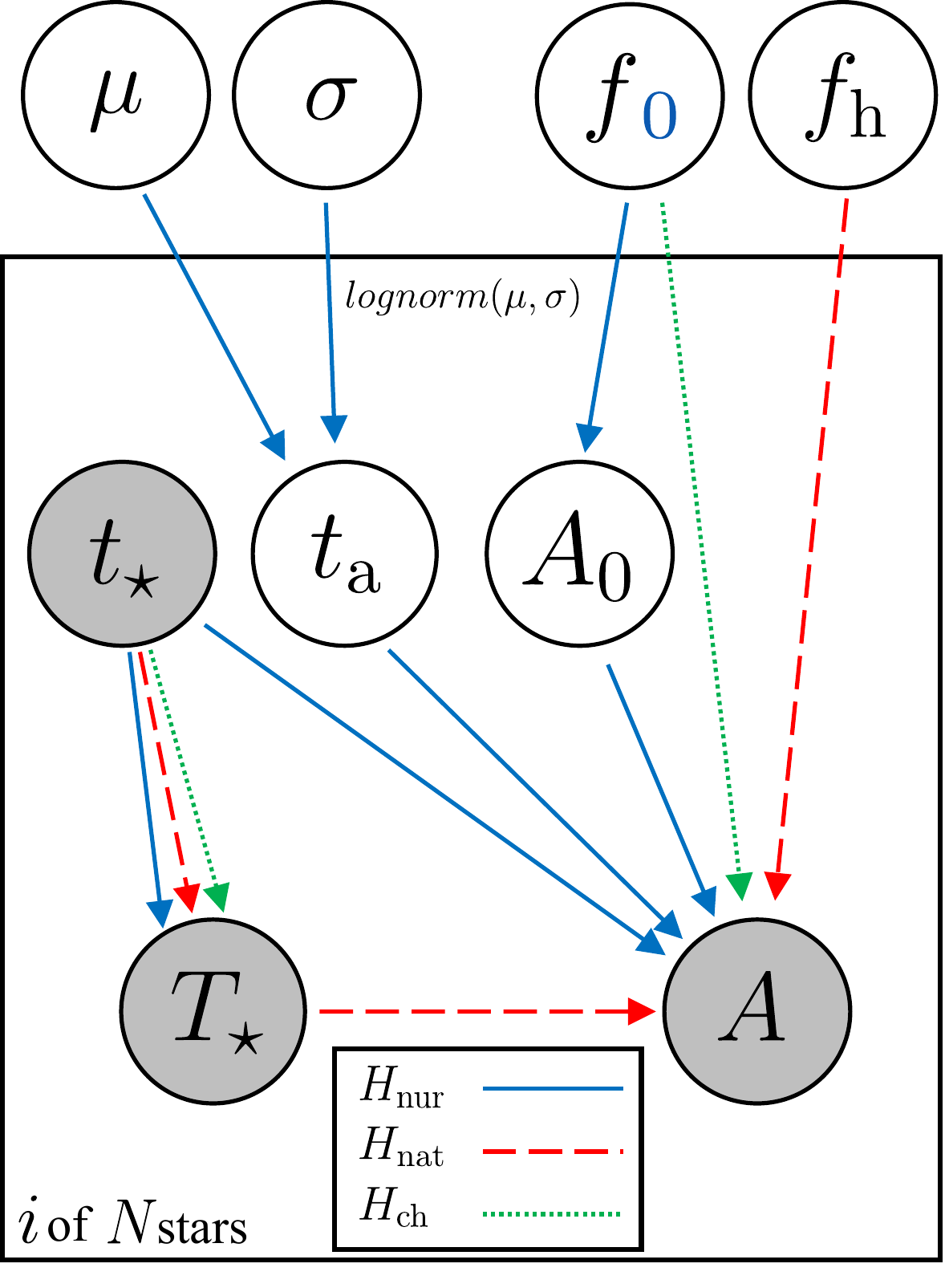}
    \caption{Graphical representation of the three hypotheses applied to the case of spin-orbit misalignment. Relations under the Nurture hypothesis are shown with a solid blue line, those under the Nature hypothesis with a dashed red line, and those under the Chance hypothesis with a dotted green line.  The ``$i$ of $N$ stars" indicates that the plate is iterated over each of the $N$ systems in the sample.  Gray circles represent observed parameters:  the current alignment state of system ($A$), the stellar age ($t_{\star}$), and the stellar temperature ($T_{\star}$).  White circles are unobserved individual parameters (on the plate) -- the initial alignment state ($A_0$) and the alignment timescale ($t_{\rm a}$) -- and hyperparameters -- the mean ($\mu$) and standard deviation ($\sigma$) of a log-normal prior for $t_{\rm a}$, the fraction of all systems observed to be misaligned ($f$), the fraction of all systems observed to start out misaligned ($f_0$), and the misaligned fraction of hot systems ($f_{\rm h}$).  The hyperparameters $f$ and $f_0$ are in the same circle because mathematically, they behave the same way in both the Nurture and Chance hypotheses; they are distinguished by the blue subscript, because $f_0$ occurs in the Nurture hypothesis while the Chance hypothesis uses $f$.}
    \label{fig:alignmentmodel}
\end{figure}

Because these stars' temperatures change as they age, it is unclear whether the true source of the trend is age or stellar temperature.  Here, we apply our framework to the \cite{triaud2011} sample, using the temperatures reported in the TEPCAT catalogue \citep{tepcat}, to see if the obliquities are better modeled by a relation with age or with temperature.  This dataset consists of 22 systems, 10 of which have misaligned stars  according to the criteria of \cite{triaud2011} that a star is misaligned if its measured obliquity angle is $>20^\circ$.  Uncertainties in the measured obliquity angle are typically between about $5^\circ$ and $10^\circ$.  There are 15 stars with temperature at or above 6250 K and seven stars with temperature below 6250 K.  All 10 misaligned stars have temperature at or above 6250 K.  Uncertainties in the stellar temperature are generally between about 50 K and 100 K.  We derive the relevant equations in Section \ref{subsec:aligneqns}, present the results we obtain in Section \ref{subsec:alignresults}, show and display contour plots of hyperparameters in Section \ref{subsec:aligncontours}, then summarize in Section \ref{subsec:alignsummary}.  We show a graphical representation of the model for this case in Figure \ref{fig:alignmentmodel}.

\subsection{Equations for each hypothesis}
\label{subsec:aligneqns}
\subsubsection{Hypothesis 1: Nurture (Age-driven)}
In this case, the property of interest is whether or not a planet's orbital axis is observed to be aligned with the spin axis of its host star.  For simplicity, we do not concern ourselves with the actual value of the measured obliquity angle, but rather merely whether that angle indicates that the system is aligned or misaligned.  We represent this with the parameter $A$, which is 1 if the star is misaligned and 0 if it is aligned, and use the criteria of \cite{triaud2011} that a star is misaligned if the observed obliquity is $>20^\circ$.  We designate the initial alignment state as $A_0$.  We also include an alignment timescale $t_{\rm a}$.  The additional observed data ${\bf X}_{\rm ob}$ is the stellar temperature for each system, $T_{\star}$.  Substituting the relevant parameters into the general Nurture hypothesis equations for a binary $X_{\rm p}$ that evolves in only one direction (Eqn. \ref{eqn:hyp1genbinaryfinal}) gives:
\begin{align}
    \label{eqn:alignnurfirst}
    p(A,t_{\star},T_{\star}|f_0)=
    \begin{cases}
    \text{\Large$\int$} \left[(1-f_0)+f_0\int_0^{t_{\star}}p(t_{\rm a}|A_0,T_{\star},{\bf X}_{\rm nob})dt_{\rm a}\right] p(t_{\star},T_{\star},{\bf X}_{\rm nob}) d{\bf X}_{\rm nob} &,A=0\\
    f_0\iint_{t_{\star}}^\infty p(t_{\rm a}|A_0,T_{\star},{\bf X}_{\rm nob}) p(t_{\star},T_{\star},{\bf X}_{\rm nob})dt_{\rm a} d{\bf X}_{\rm nob}&,A=1.
    \end{cases}
\end{align}
where the hyperparameter $f_{0}$ here represents the initial fraction of systems with stars observed to be misaligned from their planet's orbital axis.

In general, since realignment would occur via tidal interactions between the planet and the star, the alignment timescale would depend on properties like the stellar mass.  However, for the data we are considering here, the planetary systems are similar enough (hot Jupiters orbiting stars between about $1.2 M_\odot$ and $1.5 M_\odot$) that we can approximate $t_{\rm a}$ as being independent of other variables.  In order to cover a broad range of possible timescales spanning many orders of magnitude, we give $t_{\rm a}$ a log-normal distribution with unknown mean and standard deviation.  We give the mean $\mu$ a log-uniform hyperprior of U(6,16) and the standard deviation $\sigma$ a log-uniform hyperprior of U(0,20), both in log$_{10}$([yr]) space.  We again note that while the uniform hyperprior on $\mu$ is uninformative, the true uninformative hyperprior for $\sigma$ would be $1/\sigma^2$.  However, as we expect alignment timescales to span many orders of magnitude, a uniform hyperprior, which is weighted towards larger values of $\sigma$, is suitable for this case.

For simplicity, we treat $t_{\star}$ and $T_{\star}$ as being independent of other system parameters, both observed and not observed.  This means that the joint probability between them, $p(t_{\star},T_{\star})$ will cancel out when the odds ratios are calculated, as described in Section \ref{subsec:genoddsratios}.

With these assumptions, the equation becomes similar in form to that for the resonances case, with the inclusion of the temperature data.  We thus obtain a result very similar to Eqn. \ref{eqn:hyp1resfinal}:
\begin{align}
    \label{eqn:alignhyp1}
   p(A,t_{\star},T_{\star}|\mu,\sigma,f_{0})=
   \begin{cases}
   p(t_{\star},T_{\star})\left[(1-f_{0})+f_{0}\int_0^{t_{\star}} p(t_{\rm a}|\mu,\sigma) dt_{\rm a}\right] &, A=0 \\
   p(t_{\star},T_{\star})f_{0} \int_{t_{\star}}^\infty p(t_{\rm a}|\mu,\sigma) dt_{\rm a} &, A=1.
   \end{cases}
\end{align}

We give the initial fraction $f_0$ a uniform hyperprior from 0 to 1.  We note that this is not the uninformative hyperprior, since $f_0$ is a Bernoulli distribution parameter.  The uninformative hyperprior is Beta(1/2,1/2), and we will compare the odds ratios from the uniform hyperprior with those from the Beta hyperprior.

\subsubsection{Hypothesis 2: Nature (Driven by inherent system properties)}
In this case, the Nature hypothesis is that the stellar obliquity is driven by the stellar temperature. We do not concern ourselves with other unobserved system parameters here, so the integral over ${\bf X}_{\rm nob}$ disappears.  Then, adapting the general equation for the Nature hypothesis (Eqn. \ref{eqn:genhyp2final}) gives:
\begin{align}
\label{eqn:alignhyp2}
p(A,t_{\star},T_{\star})&=p(A|T_{\star})p(t_{\star},T_{\star}).
\end{align}

We separate the sample into hot and cool stars, with the dividing line at $T_{\star}=6250$ K (i.e., at the Kraft break), following \cite{winn2010}, and consider a system to be misaligned if its projected obliquity is $>20^\circ$, as did \cite{triaud2011}.  The idea behind the hypothesis of a trend driven by temperature is that stars start out with a range of obliquities and the cool stars are able to quickly realign with planetary orbits \citep{winn2010}. So although technically the obliquities change with time under this hypothesis too, we assume that hot stars realign so slowly that the change in obliquity is negligible and cool stars realign so quickly that we always observe them aligned. Indeed, all the cool stars in our sample are aligned.  So for cool stars, we use the following:
\begin{align}
    p(A|T_{\star}<6250\text{ K})=
    \begin{cases}
    1 &,A=0\\
    0 &,A=1.
    \end{cases}
\end{align}

Hot stars, on the other hand, retain the range of obliquities with which they started out.  Thus, we expect some of these stars to be well aligned, both as an expected occasional outcome of whatever process produces the obliquities and because what we observe is the projected, not true, obliquity angle.  Since we do not know the fraction of hot misaligned systems we would expect to see, we introduce the hyperparameter $f_{\rm h}$, the fraction of hot systems observed to be misaligned, which we give a uniform hyperprior from 0 to 1.  So we use the following for hot stars:
\begin{align}
    p(A|T_{\star}\geq6250\text{ K},f_{\rm h})=
    \begin{cases}
    1-f_{\rm h}&,A=0\\
    f_{\rm h}&,A=1.
    \end{cases}
\end{align}
Since the uniform hyperprior is not actually uninformative, we will also compare our results to the odds ratio obtained with assigning $f_{\rm h}$ the hyperprior Beta(1/2,1/2).  For comparison, in the dataset we use, there is an observed misaligned fraction of hot stars of 10/15.

\subsubsection{Hypothesis 3: Chance}
In the Chance hypothesis, as before, we are not concerned with any other unobserved parameters in ${\bf X}_{\rm nob}$.  So from the general equation for the Chance hypothesis with a binary planetary property of interest $X_{\rm p}$ (Eqn. \ref{eqn:genhyp3final}),
\begin{align}
    \label{eqn:alignhyp3}
    p(A,t_{\star},T_{\star}|f)=
    \begin{cases}
    p(t_{\star},T_{\star})(1-f) &, A=0\\
    p(t_{\star},T_{\star})f &, A=1
    \end{cases}
\end{align}
where $f$ is the fraction of systems observed to be misaligned.  We give $f$ a uniform hyperprior from 0 to 1, but will also compare to the results obtained with a hyperprior of Beta(1/2,1/2).

\subsection{Results}
\label{subsec:alignresults}
To obtain the odds ratios, we apply Eqn. \ref{eqn:alignhyp1} for the Nurture hypothesis, Eqn. \ref{eqn:alignhyp2} for the Nature hypothesis, and Eqn. \ref{eqn:alignhyp3} for the Chance hypothesis to our sample data to obtain individual system likelihoods.  For each hypothesis, we multiply those individual likelihoods and then marginalize over the hyperparameters to obtain the overall likelihood for each hypothesis, according to Eqn. \ref{eqn:oddsratio}.  The complete equations for each hypothesis are
\begin{align}
    p(H_{\rm nur})&=\iiint \prod_{A=0}\left[p(t_{\star},T_{\star})\left(1-f_{0}+f_{0}\int_0^{t_{\star}} p(t_{\rm a}|\mu,\sigma) dt_{\rm a}\right)\right]\nonumber\\&\qquad\quad\times\prod_{A=1}\left[p(t_{\star},T_{\star})f_{0} \int_{t_{\star}}^\infty p(t_{\rm a}|\mu,\sigma) dt_{\rm a}\right]p(\mu)p(\sigma)p(f_0)d\mu d\sigma df_0\\
    p(H_{\rm nat})&=\int\prod_{A=0}\left[p(A=0|T_{\star},f_{\rm h})p(t_{\star},T_{\star})\right]\prod_{A=1}\left[p(A=1|T_{\star},f_{\rm h})p(t_{\star},T_{\star})\right]p(f_{\rm h})df_{\rm h}\\
    p(H_{\rm ch})&=\int \prod_{A=0}\left[p(t_{\star},T_{\star})(1-f)\right]\prod_{A=1}\left[p(t_{\star},T_{\star})f\right]p(f)df.
\end{align}
The priors and probability functions of relevant parameters are given in Table \ref{tab:alignpriors}, where the ranges for $\mu$ and $\sigma$ are given in log$_{10}$([yr]).  Note that the joint prior of $t_{\star}$ and $T_{\star}$ will cancel out when we take the odds ratios, we do not include it in the table.

\begin{table}
	\centering
	\begin{tabular}{|ll|l|}
	    \hline
	    \multicolumn{2}{|l|}{Parameter} & Prior or Probability Function\\
		\hline
        \multicolumn{2}{|l|}{Alignment timescale (log$_{10}(t_{\rm a})|\mu,\sigma$)}&$\frac{1}{\sqrt{2\pi \sigma^2}} \text{exp}\left(\frac{-(\text{log}_{10}(t_{\rm a})-\mu)^2}{2\sigma^2}\right)$\\
        \hline
        \multicolumn{2}{|l|}{Mean of log$_{10}(t_{\rm a})$ ($\mu$)}&U(6,16)\\
        \hline
        \multicolumn{2}{|l|}{Standard deviation of log$_{10}(t_{\rm a})$ ($\sigma$)}&U(0,20)\\
        \hline    
        \multicolumn{2}{|l|}{Initial misaligned fraction ($f_0$)} & U(0,1)\\
        \hline
        \multicolumn{2}{|l|}{Misaligned fraction of hot stars ($f_{\rm h}$)} & U(0,1)\\
        \hline
        \multicolumn{2}{|l|}{Misaligned fraction ($f$)} & U(0,1)\\
        \hline
        \multirow{4}{*}{Alignment state}
        &$A=0|T_{\star}<6250\text{ K}$& 1\\
        &$A=1|T_{\star}<6250\text{ K}$& 0\\
        &$A=0|T_{\star}\geq6250\text{ K},f_{\rm h}$&$1-f_{\rm h}$\\
        &$A=1|T_{\star}\geq6250\text{ K},f_{\rm h}$&$f_{\rm h}$\\
        \hline
	\end{tabular}
	\caption{Priors and probability functions of relevant parameters for the obliquities case, with the parameters in the left column and the corresponding priors in the right column.  The ranges for $\mu$ and $\sigma$ are given in log$_{10}$([yr]).}
	\label{tab:alignpriors}
\end{table}

We obtain the following odds ratios:
\begin{align}
    &\frac{p(H_{\rm nat})}{p(H_{\rm nur})}=210\nonumber\\
    &\frac{p(H_{\rm nur})}{p(H_{\rm ch})}=1.4\nonumber\\
    &\frac{p(H_{\rm nat})}{p(H_{\rm ch})}=310\nonumber.
\end{align}
where $H_{\rm nur}$, $H_{\rm nat}$, and $H_{\rm ch}$ refer to the Nurture, Nature, and Chance hypotheses, respectively.  These odd ratios indicate that the Nature hypothesis, i.e., that stellar obliquity is caused by temperature, is strongly favored over the Nurture hypothesis, i.e., that stellar obliquity is caused by age.  The support for the Nature hypothesis compared to the Chance hypothesis is similarly strong.  Regarding the Nurture and Chance hypotheses, there is no preference for one over the other.

We also compute the odds ratios for when $f_0$, $f_{\rm h}$, and $f$ are each given the hyperprior Beta(1/2,1/2).  In this case, we find increased support for the Nurture hypothesis relative to Nature and Chance, and slightly increased support for the Nature hypothesis relative to Chance.  However, none of the odds ratios change by more than a factor of 1.5, and thus our general conclusions remain unchanged.

In Appendix \ref{sec:appendixc}, we perform additional variations in our calculations by removing systems whose alignment state is uncertain and by bootstrapping the data.  In both treatments, we find strong evidence for the Nature hypothesis over Nurture and Chance.

\subsection{Contour plots}
\label{subsec:aligncontours}
\begin{figure}[ht]
    \includegraphics[width=3.5 in]{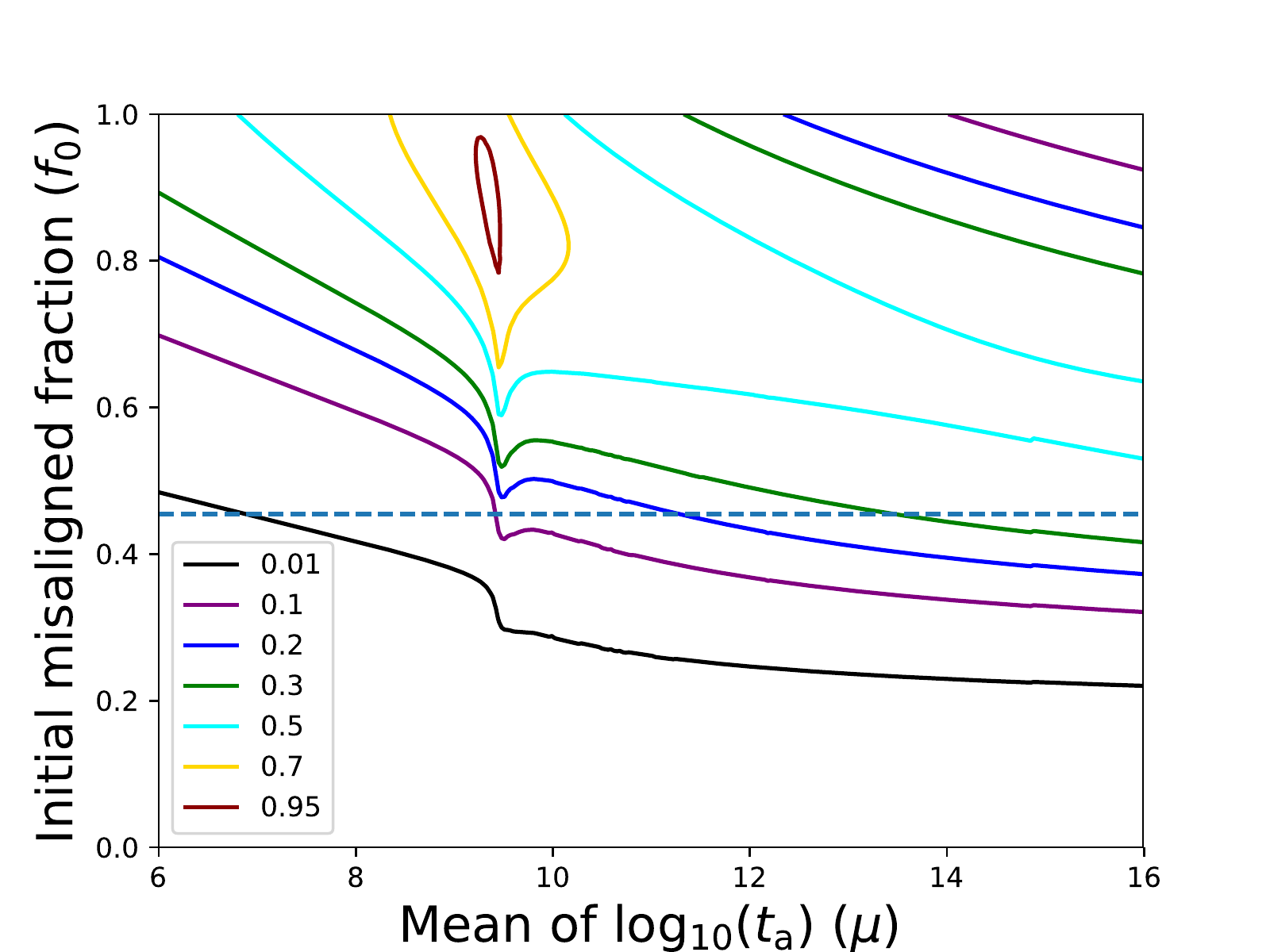}
    \includegraphics[width=3.5 in]{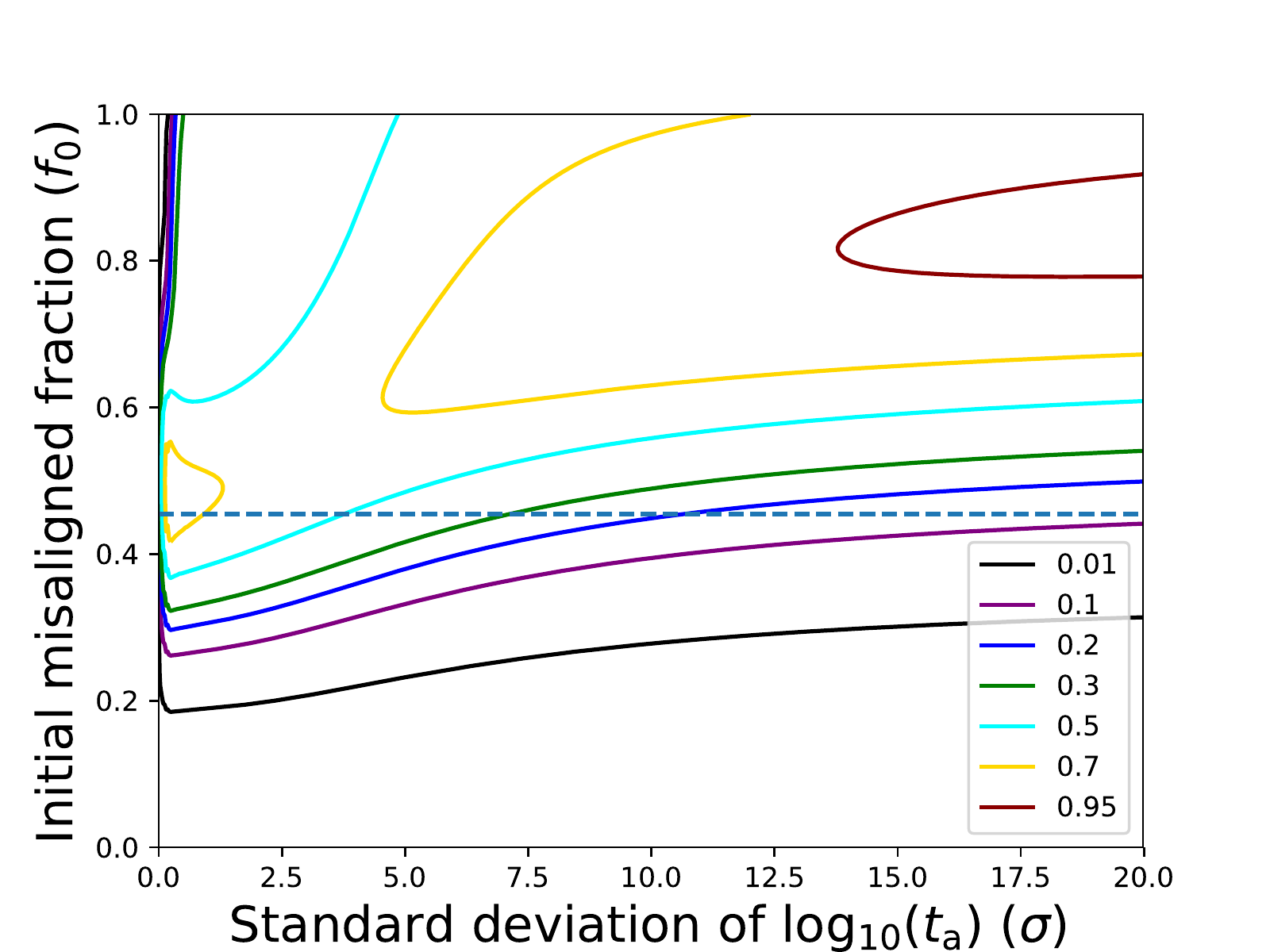}
    \includegraphics[width=3.5 in]{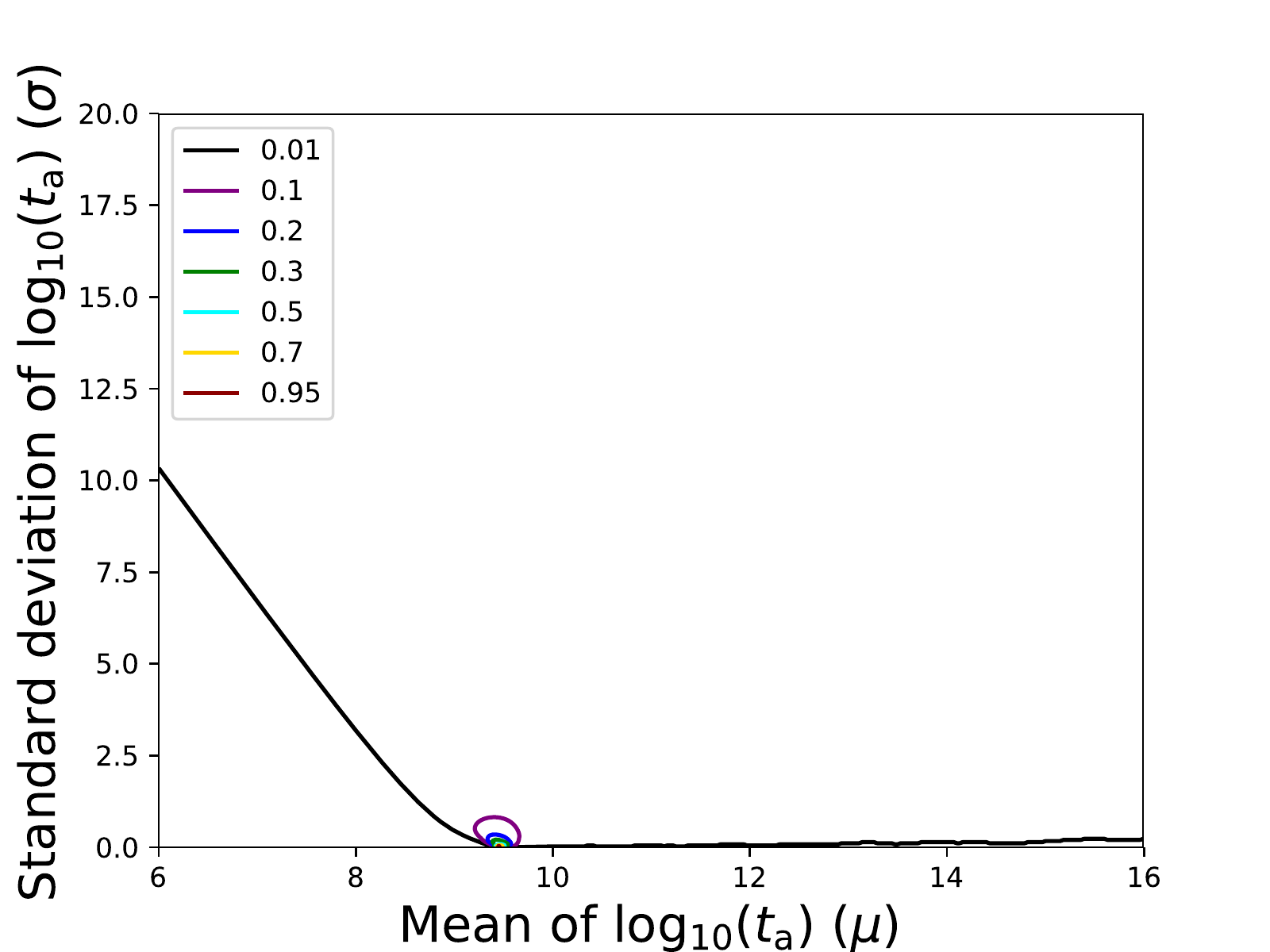}
    \caption{Probability contour plots for the Nurture hypothesis in the stellar alignment case, showing $\mu$ vs. $f_0$ (top left), $\sigma$ vs. $\mu$ (top right), and $\sigma$ vs. $f_0$ (bottom left).  Contour levels are 0.01, 0.05, 0.2, 0.5, 0.6, 0.7, 0.8, and 0.95.  The observed fraction of misaligned systems (10/22) is shown with a dashed line.  Note that because the Nurture hypothesis involves misaligned systems aligning over time, we do not expect $f_0$ to match up with the observed misaligned fraction.}
    \label{fig:aligncontourplots}
 \end{figure}

In Figure \ref{fig:aligncontourplots} we display probability contours between hyperparameters from the equations for the Nurture hypothesis.  These are are not plots of the posterior or the likelihood, but they are probabilities calculated from the part of the posterior that does not get cancelled in the odds ratios, and normalized relative to the highest value.  The probabilities have been calculated on a grid for each value of $f_0$, $\mu$, and $\sigma$.  The plots indicate the values of those hyperparameters that contribute the most to the odds ratio under the Nurture hypothesis and can also give some insight into why the Nurture hypothesis does not have enough evidence to favor it strongly over the Nature and Chance hypotheses.

The structure in the Nurture contour plots shows a continuous distribution of solutions that fit the data with high probability ($\gtrsim 50\%$) relative to the best solution.  The highest probability region in the $f_0$ vs. $\mu$ plot suggests that in the Nurture hypothesis, around 75\% to 95\% of systems start out with an observable misalignment and are realigned by a few Gyr, a timescale similar to the ages of stars in the sample.  The $f_0$ vs. $\mu$ plot also shows a region of relatively high probability extending to high values of $\mu$.  On the high-$\mu$ end, this corresponds to a solution in which $\sim 50$\% of systems start out misaligned and remain so on very long timescales, which means most of them would still be misaligned today (compare to the observed misaligned fraction of 10/22).  Such a solution is equivalent to the Chance hypothesis.

The peaks near $\sigma=0$ in the $\sigma$ vs. $\mu$ and $f_0$ vs $\sigma$ plots indicate a solution with little variation in alignment timescales from system to system.  There is also a high probability region in the upper right of the $f_0$ vs $\sigma$ plot, with high initial fraction and high $\sigma$, showing a solution where alignment timescales vary by many orders of magnitude; systems get aligned on average after a few Gyr, but there is a lot of variation, and some do not get aligned at all.

The high probability regions in these plots are relatively small compared to the low probability regions.  This fine-tuning serves to penalize the Nurture hypothesis when the hyperparameter space is marginalized over, so it does not win out over the Nature hypothesis and only barely has more support than the Chance hypothesis.

To see how much of an effect the large ranges for the hyperpriors of $\mu$ and $\sigma$ in the alignment timescale $t_{\rm a}$ have on the odds ratios, we cut down the priors of $\mu$ and $\sigma$ to $p(\mu)=$U(6,10) and $p(\sigma)=$U(0,10).  This reduction represents a scenario in which most stars are realigned by 10 Gyr or less and in which there is less variation in the alignment timescale from system to system.  The odds ratios then become $p(H_{\rm nat})/p(H_{\rm nur})=270$, $p(H_{\rm nur})/p(H_{\rm ch})=1.2$, and $p(H_{\rm nat})/p(H_{\rm ch})=310$.  So this adjustment actually decreases the support for the Nurture hypothesis, but not by much.  This suggests that other regions of the original hyperparameter space -- such as the region in the top right in the $f_0$ vs. $\sigma$ plot -- have relatively high-probability solutions that are important to the Nurture hypothesis, and the restricted ranges do not describe the data as well.

\begin{figure}[ht]
    \includegraphics[width=3.5 in]{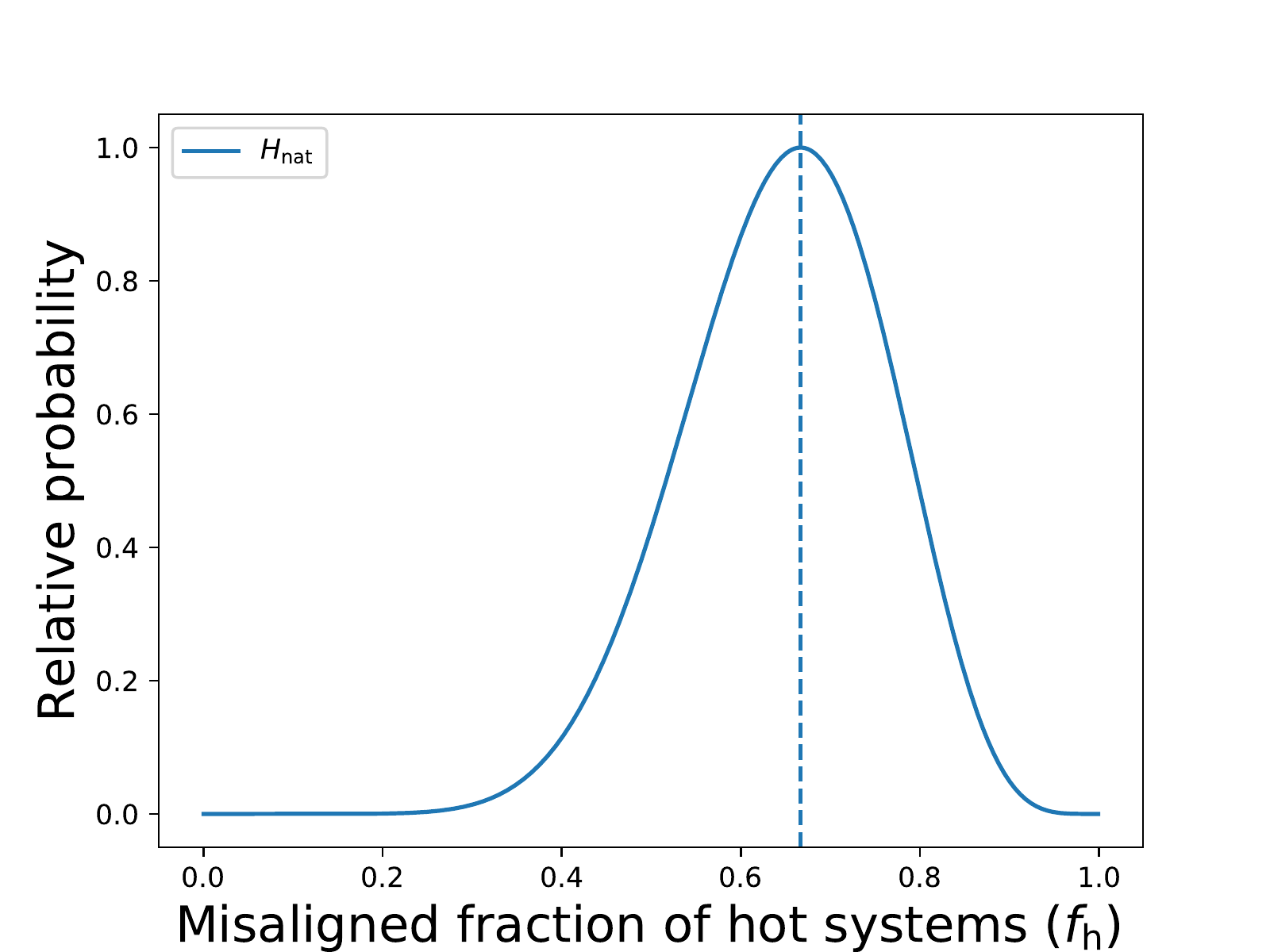}
    \includegraphics[width=3.5 in]{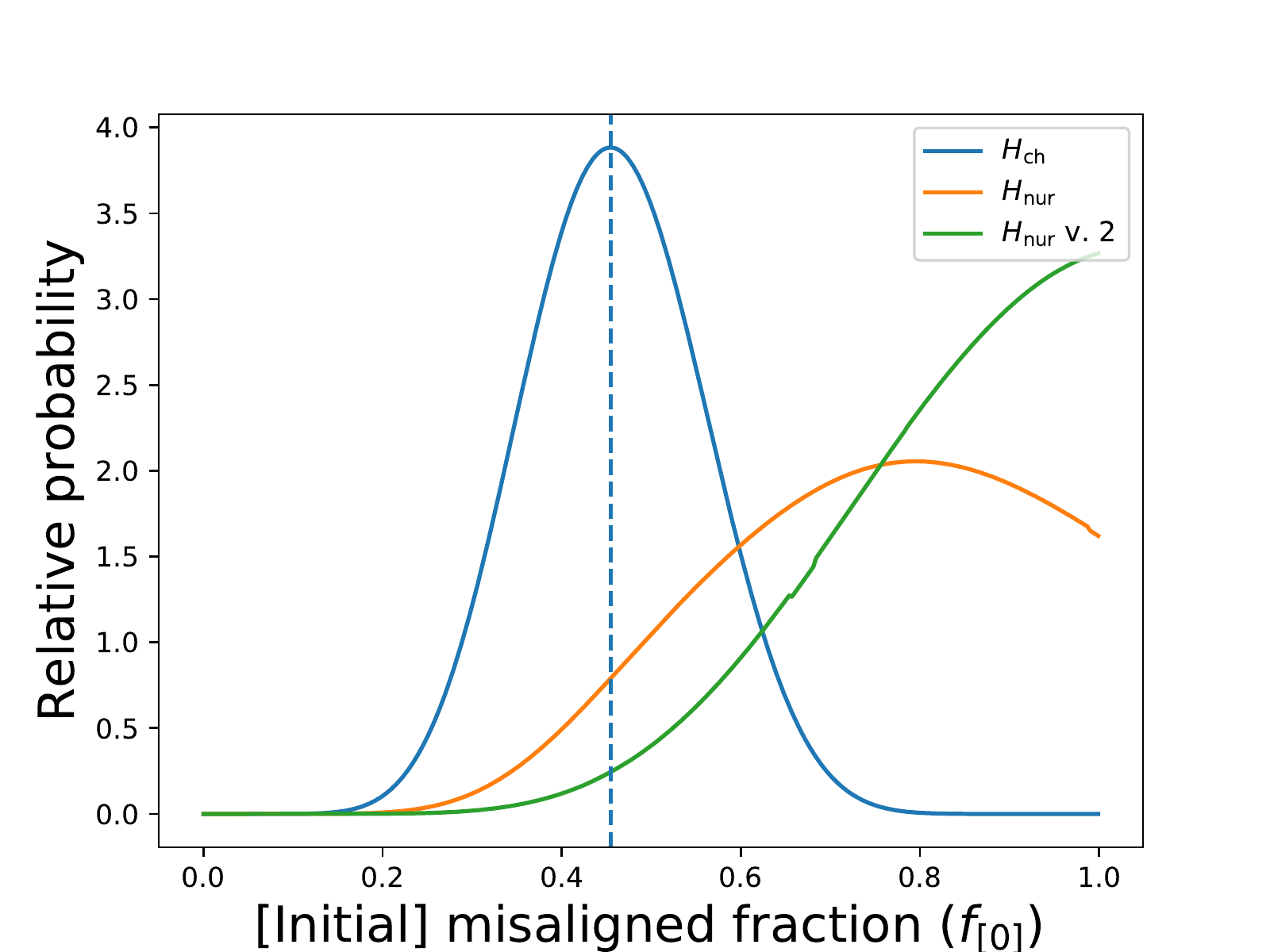}
    \caption{Left: Probability histogram of the fraction of hot systems observed to be misaligned under the Nature hypothesis ($f_{\rm h}$).  The observed misaligned fraction of hot systems (10/15) is shown with a dashed line.  Right:  Probability histogram of the fraction ($f$) of all systems, hot and cool, observed to be misaligned under the Chance hypothesis (blue) and the initial misaligned fraction ($f_0$) under the Nurture hypothesis with the original hyperprior ranges for $\mu$ and $\sigma$ (orange) and with the reduced hyperprior ranges (green, labeled as ``$H_{\rm nur}$ v. 2"), after marginalizing over $\mu$ and $\sigma$.  The curves have been normalized to have the same area.  The observed fraction of misaligned systems (10/22) is shown with a dashed line.  Because initially misaligned systems are aligned over time in the Nurture hypothesis, we do not expect $f_0$ to line up with the observed misaligned fraction.}
    \label{fig:alignf0hists}
 \end{figure}

The left panel of Figure \ref{fig:alignf0hists} shows the probability histogram for the fraction of hot systems ($f_{\rm h}$) observed to be misaligned under the Nature hypothesis.  The right panel of Figure \ref{fig:alignf0hists} shows the fraction ($f$) of all systems, hot and cool, observed to be misaligned under the Chance hypothesis (blue) and the initial misaligned fraction ($f_0$) under the Nurture hypothesis with the original hyperprior ranges for $\mu$ and $\sigma$ (orange) and with the reduced hyperprior ranges (green, labeled as ``$H_{\rm nur}$ v. 2"), after marginalizing over $\mu$ and $\sigma$.  The curves have been normalized to have the same area.  ~70\% of hot stars are seen to be misaligned with their planets, and ~45\% of all systems are observed to be misaligned.  Under the Nurture hypothesis, ~80\% of all systems start out misaligned.

\subsection{Summary}
\label{subsec:alignsummary}
Our calculations strongly favor the Nature hypothesis over both the Nurture hypothesis and the Chance hypothesis.  We conclude that of these three possibilities, the obliquities of stars in the data we use are best explained by temperature being the underlying driver.

\clearpage

\section{Do Hot Jupiters Show Evidence for Tidal Circularization?}

\label{sec:eccentricities}

To look for evidence of tidal circularization, \cite{quinn2014} analyzed the ages and eccentricities of hot Jupiters -- which they defined as having $M_{\rm P} > 0.3\text{ } M_{\rm J}$ and $P < 10$ days -- available from the literature at the time. They compared the ages with the expected orbital circularization timescale (see Eqn. \ref{eqn:tcir}) for each system, assuming a planetary tidal quality factor of $Q_{\rm P}=10^6$.  They found evidence that eccentric orbits circularize over time:  systems younger than their circularization timescales tend to be eccentric, and systems older than their circularization timescales tend to be circular.

\begin{figure}[ht]
    \includegraphics[width=7in]{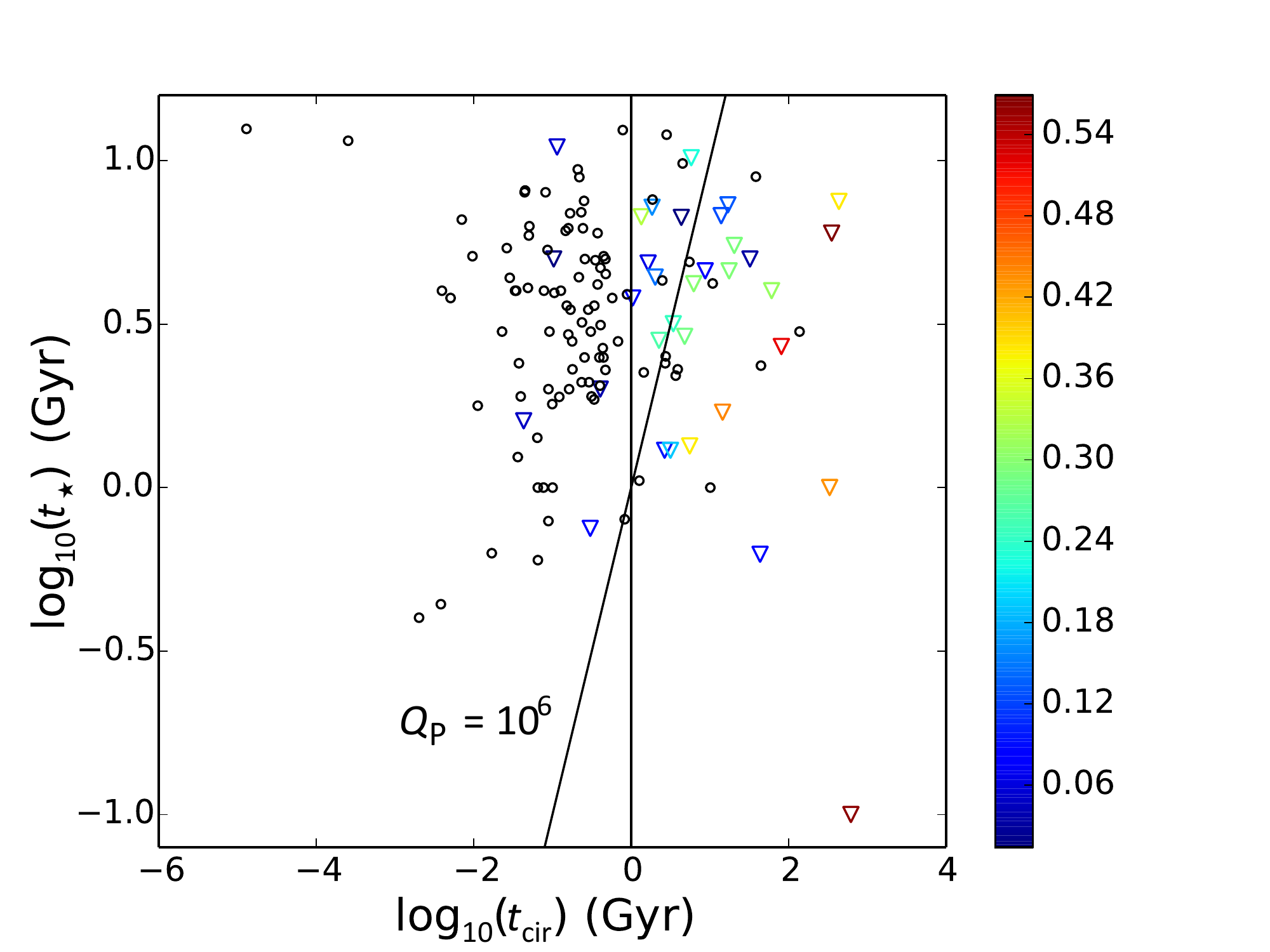}
    \caption{Stellar age ($t_{\star}$) versus circularization timescale ($t_{\rm cir}$) for known hot Jupiters, assuming $Q_{\rm P}=10^6$.  Planets with circular orbits are shown with black circles.  Planets on eccentric orbits are shown with triangles colored according to eccentricity, as shown in the color bar on the right.  The line of $t_{\star}=t_{\rm cir}$ is shown with a slanted solid black line.  A vertical line is shown as another potential way to divide the sample between primarily eccentric and primarily circular planets.}
    \label{fig:updatedquinnplot}
 \end{figure}

We present an updated version of \cite{quinn2014}'s Figure 4 as our Figure \ref{fig:updatedquinnplot}. Although \cite{quinn2014} did not explicitly list each of the individual systems they included in their sample, they stated that they pulled hot Jupiters and their host star ages from the Extrasolar Planets Encyclopaedia \citep{schn11}. Using their same mass and period filter with the additional constraint of $M_{\rm P}<13 M_{\rm J}$, we plot stellar age against circularization timescale, assuming $Q_{\rm P}=10^6$, for the hot Jupiters for which the stellar and planetary masses, semimajor axis, age, and measured eccentricity are available on the Extrasolar Planets Encyclopaedia, queried on May 28, 2019.  The sample contains 130 systems total.  The ages of the stars have been determined in a variety of ways, but the most common method is isochrone or evolutionary track fitting.  Uncertainties in stellar ages in the sample are typically $\sim0.5-2$ Gyr.  In this dataset, 11 systems do not have a measured planetary radius $R_{\rm P}$.  When this is the case, we follow \cite{quinn2014} and estimate $R_{\rm P}$ using the following empirical relation from \cite{weiss2013}:
\begin{equation}
    \frac{R_{\rm P}}{R_\oplus}=2.45\left(\frac{M_{\rm P}}{M_\oplus}\right)^{-0.039}\left(\frac{F}{\text{erg} \text{ s}^{-1} \text{ cm}^{-2}}\right)^{0.094}
\end{equation}
where $F$ represents the time-averaged incident flux the planet receives.

We divide the sample into eccentric systems and circular systems, classifying a given planet as eccentric if it has $e>0$ at the $3\sigma$ level.  This is a conservative criterion to only include systems with well measured eccentricities in the eccentric sample, and it means that the circular sample contains both truly low eccentricity planets as well as ambiguous cases. We exclude planets for which the reported eccentricity is $e=0$ with no error bars, as this indicates assumed circularity in the orbital solution rather than a measured value.  For planets for which only an upper limit on the eccentricity is given, we classify it as circular if $e<0.1$, and exclude it otherwise.  Typical uncertainties in eccentricity in the sample range up to 0.05.  Table \ref{tab:eccdata} gives the first few lines of our sample data; the full machine-readable table can be accessed at Table3.txt (see arXiv ancillary file).  In Figure \ref{fig:updatedquinnplot}, we over plot the $t_{{\star}}=t_{\rm cir}$ line \cite{quinn2014} used to divide the sample into circular and eccentric planets.

\begin{table}
	\centering
	\begin{tabular}{lccccccc}
		\hline
		Name & $e$ & $t_{\star}$ (Gyr) & $a$ (AU) & $M_{\rm P}$ ($M_{\rm J}$) & $R_{\rm P}$ ($R_{\rm J}$) & $M_{\star}$ ($M_{\odot}$) & Eccentric? (*)\\
		\hline
		CoRoT-16 b & 0.33$\pm{0.1}$ & 6.73 & 0.0618 & 0.535 & 1.17 & 1.098 & *\\
		CoRoT-20 b & 0.562$\pm{0.013}$ & 0.1 & 0.0902 & 4.24 & 0.84 & 1.14 & *\\
		CoRoT-23 b & 0.16$\pm{0.017}$ & 7.2 & 0.0477 & 2.8 & 1.08 & 1.14 & *\\
		...&&&&&&&\\
		\hline
	\end{tabular}
	\caption{Planetary and stellar data for the sample we use in the eccentricities case, giving the planet's name, eccentricity ($e$) with reported errors used to classify the planet as eccentric or circular, stellar age ($t_{\star}$) in Gyr, planetary semimajor axis ($a$) in AU, planetary mass ($M_{\rm P}$) in terms of $M_{\rm J}$, planetary radius ($R_{\rm P}$) in terms of $R_{\rm J}$, and stellar mass ($M_{\star}$) in terms of $M_{\odot}$.  The final column contains an asterisk (*) if the planet is classified as eccentric and no asterisk if the planet is classified as circular.  We consider a planet to have an eccentric orbit if it has $e>0$ at the $3\sigma$ level.  The full machine-readable version of this table can be accessed at Table3.txt (see arXiv ancillary file).}
	\label{tab:eccdata}
\end{table}

To investigate the difference between the sample subsets on each side of the $t_{{\star}}=t_{\rm cir}$ line, \cite{quinn2014} computed a Kolmogorov-Smirnov (KS) test and obtained $p = 3.3 \times 10^{-5}$, determining with $\sim$99.997\% confidence that they could reject the null hypothesis that the samples came from the same parent population.  They found that even when assuming generous errors in eccentricities of 0.1, the KS test rejected the null hypothesis with over $3\sigma$-level confidence. They concluded that planets younger than their cicularization timescales have higher eccentricities.

However, it is not immediately obvious that a $t_{{\star}}=t_{\rm cir}$ line better divides the sample than a vertical line would, as we show in Figure \ref{fig:updatedquinnplot}. For a sample of planets with similar masses and radii, a vertical line would represent a separation between eccentric vs. circular based only on semimajor axis and would not suggest a dependence on age. There are several possibilities for why we might see a dependence on semi-major axis with no significant age dependence. First, if the sample size has a relatively small age range and tidal circularization is at work, we would expect a noticeable division between planets that were circularized early on and planets that will not be able to circularize over the lifetimes of their stars. Larger radii while planets are young (making it even easier to circularize early on) could enhance this effect. Second, eccentricities are easier to excite at larger semi-major axes via planet-disk interactions (e.g., \citealt{duffell15}), planet-planet scattering (e.g., \citealt{petrovich14}), and/or forcing from an outer companion. Although many of the eccentricities are too high for these mechanisms or the planets lack the necessary nearby companions (see \citealt{hotjupreview2018} and references therein), the qualitative trend in e vs. a is consistent.

We perform a KS test on the planets to the right and left of the $t_{\star}=t_{\rm cir}$ line using the updated hot Jupiter data, and obtain a \textit{p}-value of $2.6\times10^{-6}$.  This means we can reject the null hypothesis that the subsamples come from the same parent distribution with 99.9997\% confidence, in agreement with \cite{quinn2014}'s results.  Next we perform a KS test for the subsamples on either side of the vertical line we added in Figure \ref{fig:updatedquinnplot}.  We obtain a \textit{p}-value of $1.3\times10^{-9}$ and can reject the null hypothesis with a confidence level of 99.9999999\%. The nature of the KS test is such that we cannot compare the results we have obtained here to determine which line does a better job of dividing the sample.  This is because, among other reasons, it is difficult to calculate accurate probabilities so far out on the tails of probability distributions because assumptions such as the lack of systematics break down; this renders the difference between our calculated $p$-values essentially meaningless.  Each test is to determine whether or not we can confidently reject the null hypothesis that the data on each side of the line come from the same parent distribution.  In both cases, we can reject the null hypothesis with a high level of confidence, demonstrating that there are multiple ways to effectively divide the data into primarily circular and primarily eccentric planets.

\begin{figure}[ht]
    \includegraphics[width=7in]{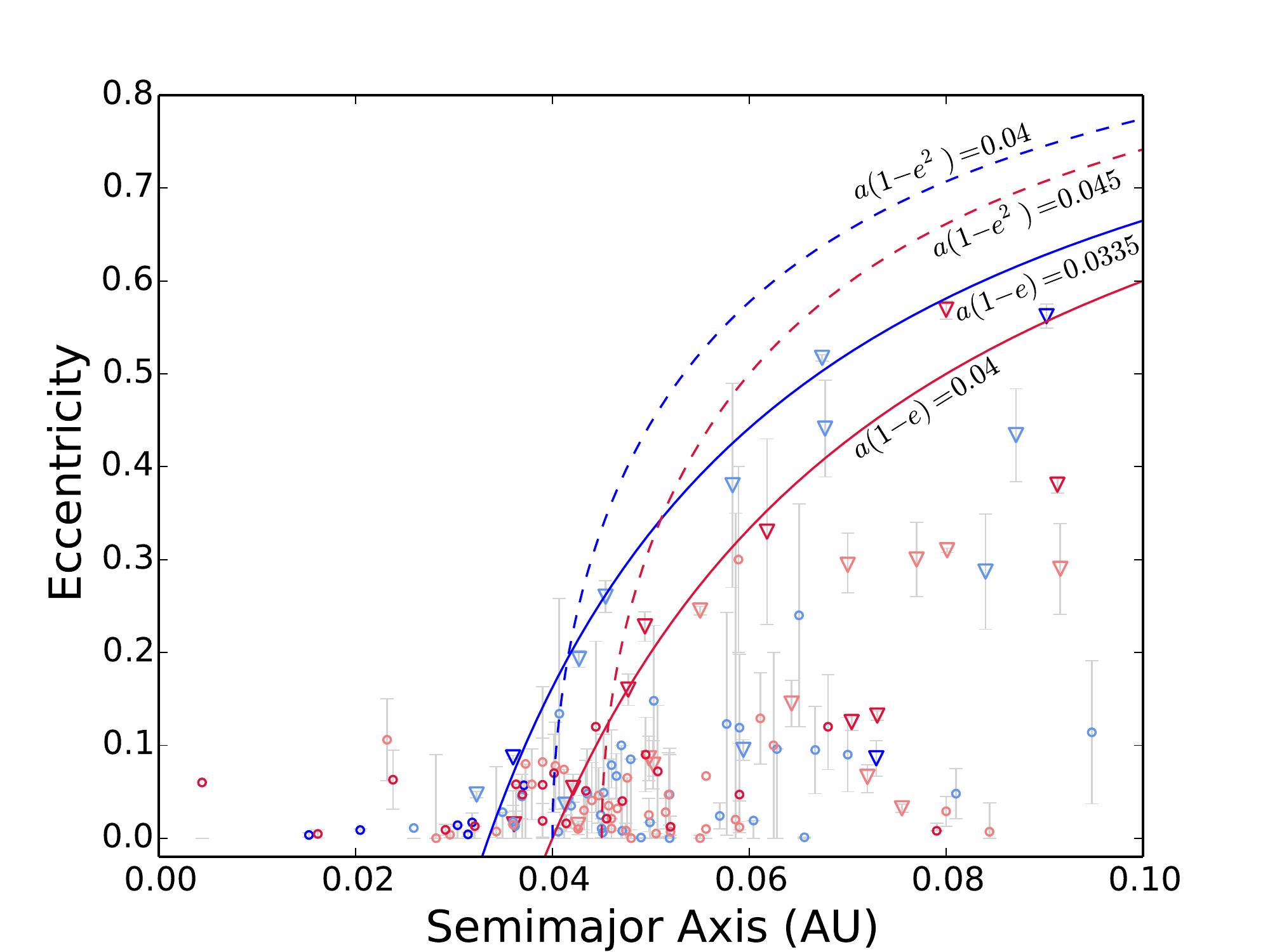}
    \caption{Eccentricity versus semimajor axis for known hot Jupiters.  Planets with circular orbits are shown as circles, and planets on eccentric orbits are shown as triangles.  The marker color indicates age:  blue is for $t_{\star}<1$ Gyr, light blue is for $1 \text{ Gyr} \leq t_{\star} <3 $ Gyr, light red is for $3 \text{ Gyr} \leq t_{\star} <6 $ Gyr, and red is for $6 \text{ Gyr} \leq t_{\star}$.  The dashed blue and red curves are lines of constant $a(1-e^2)$ for the blue and red points, respectively.  The solid blue and red curves are lines of constant $a(1-e)$ for the blue and red points, respectively.}
    \label{fig:evsa}
 \end{figure}

In Figure \ref{fig:evsa}, we plot eccentricity versus semimajor axis for the planets in the sample.  Eccentric planets are marked with a triangle and circular planets with a circle, but the marker colors now indicate the system's age.  Clearly, planets at larger semimajor axes tend to be more eccentric than those closer in.  However, the plot shows a tendency for younger systems to have higher eccentricities (we see blue triangles interior to the red triangles), supporting an additional dependence on age.  We also display lines of constant $a(1-e^2)$, which is the final semimajor axis the planets get circularized to, and of constant periapse\footnote{Contrary to expectation, the periapse lines seem to follow the outer envelope of data points better than the final semimajor axis lines, even though tidal circularization timescales are thought to scale with $a(1-e^2)$ at low eccentricities, e.g. \citealt{adamslaughlin2006,socrates2012}.} $a(1-e)$, for the blue and the red points.   It appears that the outer envelopes of red and blue points can be well traced by separate curves, indicating a possible age dependence.  Because stars of different masses have different lifetimes, this could instead be a trend in stellar mass or temperature,  but we do not see the same trend when we color the points according to stellar mass or temperature.

Here we seek to determine whether the available data truly exhibit a trend of eccentricity with stellar age or are better explained by a correlation with semimajor axis only.  We derive the relevant equations for each hypothesis in Section \ref{subsec:ecceqns}, present our results in Section \ref{subsec:eccresults}, show and discuss contour plots of hyperparameters in Section \ref{subsec:ecccontours}, then briefly summarize in Section \ref{subsec:eccsummary}.

\begin{figure}[ht]
\centering
    \includegraphics[width=3.0 in]{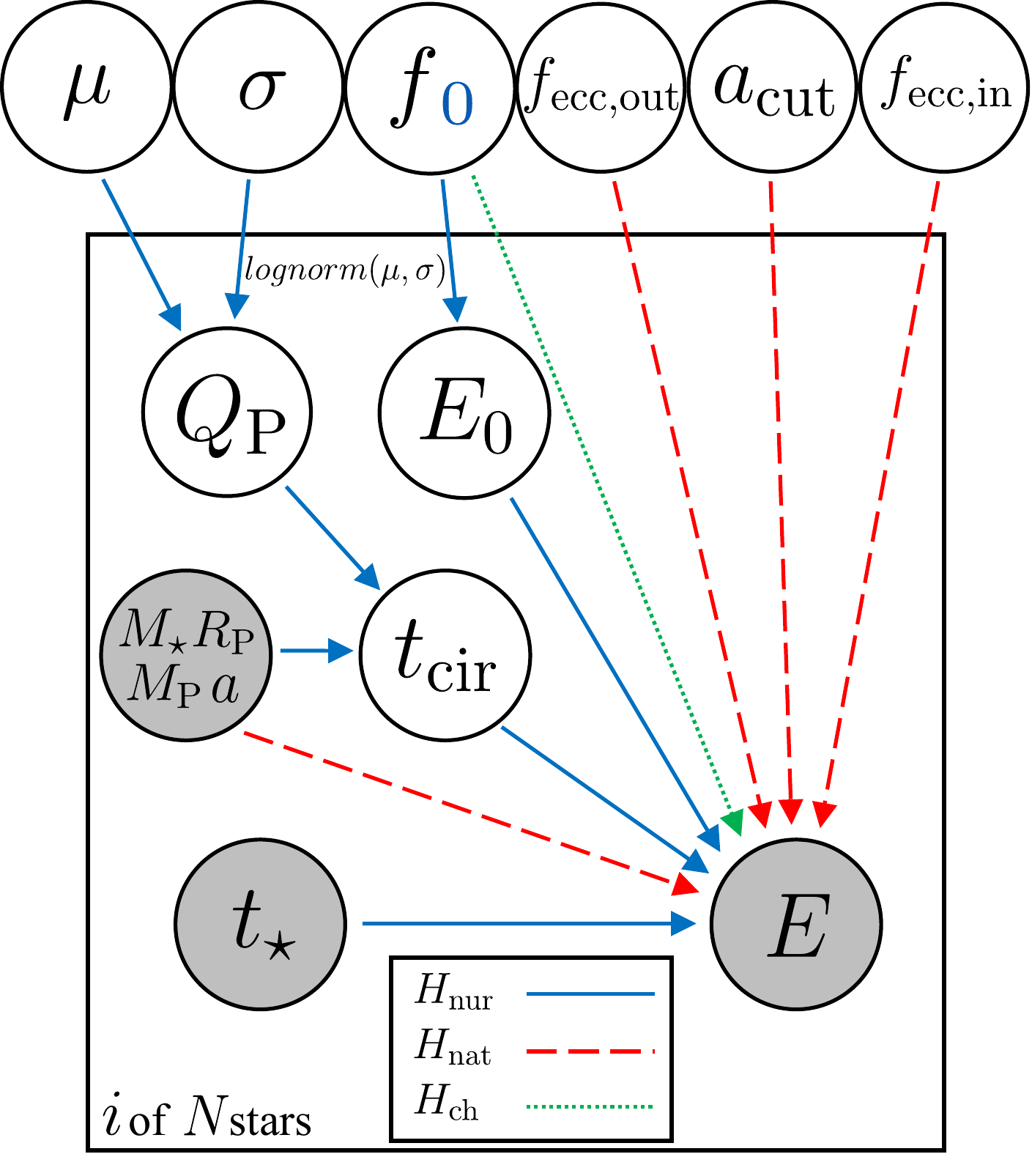}
    \caption{Graphical representation of the three hypotheses applied to the case of eccentricities. Relations under the Nurture hypothesis are shown with a solid blue line, those under the Nature hypothesis with a dashed red line, and those under the Chance hypothesis with a dotted green line.  The ``$i$ of $N$ stars" indicates that the plate is iterated over each of the $N$ systems in the sample.  Gray circles represent observed parameters:  the current eccentricity state of the system ($E$), the stellar age ($t_{\star}$), the stellar mass ($M_{\star}$), the planetary mass ($M_{\rm P}$), the planetary radius ($R_{\rm P}$), and the semimajor axis ($a$).  White circles are unobserved individual parameters (on the plate) -- the initial eccentricity state ($E_0$), the circularization timescale ($t_{\rm cir}$), and the planetary tidal quality factor ($Q_{\rm P}$) -- and hyperparameters -- the mean ($\mu$) and standard deviation ($\sigma$) of a log-normal prior for $Q_{\rm P}$, the fraction of eccentric systems ($f$), the initial fraction of eccentric systems ($f_0$), the semimajor axis cutoff ($a_{\rm cut}$), the fraction of eccentric systems beyond $a_{\rm cut}$ ($f_{\rm ecc,out}$), and the fraction of eccentric systems that remain eccentric within $a_{\rm cut}$ ($f_{\rm ecc,in}$).  The hyperparameters $f$ and $f_0$ are in the same circle because mathematically, they behave the same way in both the Nurture and Chance hypotheses; they are distinguished by the blue subscript, because $f_0$ occurs in the Nurture hypothesis while the Chance hypothesis uses $f$.}
    \label{fig:eccentricmodel}
\end{figure}

\subsection{Equations for each hypothesis}
\label{subsec:ecceqns}
\subsubsection{Hypothesis 1: Nurture (Age-driven)}
\label{subsec:eccnur}
In this application, the property of interest is whether the planet's orbit is eccentric.  Similar to the alignment case, we are not concerned with the specific value of the eccentricity, but instead classify a given planet as eccentric if it has $e>0$ at the $3\sigma$ level.  We represent this classification with $E$, which is 1 if the orbit is eccentric and 0 if it is circular.  The initial eccentricity state is designated as $E_0$ and behaves in the same way.

In the Nurture hypothesis, we compare the star's age to the planet's circularization timescale, $t_{\rm cir}$.  This timescale depends on several other system properties, most strongly the planetary radius and semimajor axis, and the systems we are looking at here are varied enough that we cannot discount the dependence on planet properties as we do in the alignment case.  The circularization timescale, i.e. the decay rate of orbital eccentricity, is given as follows:
\begin{equation}
\label{eqn:tcir}
    t_{\rm cir}=\frac{Q_{\rm P} M_{\rm P} a^{6.5}}{6\pi k_{\rm L} G^{0.5} M_{\star}^{1.5} R_{\rm P}^5}
\end{equation}
where $Q_{\rm P}$ is the tidal quality factor, $M_{\rm P}$ is the mass of the planet, $a$ is the semimajor axis, $M_{\star}$ is the mass of the star, $R_{\rm P}$ is the radius of the planet, and $k_{\rm L}$ is the Love number which is generally taken to be 0.38 \citep{socrates2012}.  This formulation is an approximation for low eccentricities, which is appropriate here because all of the systems have $e<0.6$ and most have $e<0.25$.

The tidal quality factor $Q_{\rm P}$ is related to planet's tidal dissipation efficiency; it is very poorly known and probably depends on the planet's composition, internal structure, rotation, and temperature \citep{socrates2012}.  For Jupiter, it has been constrained to $6\times10^4 < Q_{\rm J} < 2\times10^6$ \citep{yoderpeale1981}.  $Q_{\rm P}$ is a component of ${\bf X}_{\rm nob}$ and is the primary source of uncertainty in $t_{\rm cir}$ for a given planet.  To account for our lack of knowledge about this property, we assume the population of $Q_{\rm P}$ is a log-normal distribution with unknown mean $\mu$ and standard deviation $\sigma$.  These latter two variables are hyperparameters.  We give $\mu$ a log-uniform hyperprior of U(2,8), and we give $\sigma$ a log-uniform hyperprior of U(0,5), both in log$_{10}$ space.  Some planets in our sample may have tidal quality factors similar to that of Jupiter, but some may be quite different.  These hyperprior ranges for $\mu$ and $\sigma$ allow for both possibilities by comfortably enclosing and extending beyond the range of possible $Q_{\rm P}$ found for Jupiter.  Note that the hyperprior on $\mu$ is uninformative, while the hyperprior on $\sigma$ is not and is weighted towards large values of $\sigma$.  However, this is appropriate here since we expect a range of orders of magnitude in the tidal quality factor of hot Jupiters.

The other parameters in the calculation of $t_{\rm cir}$ -- $M_{\rm P}$, $M_{\star}$, $R_{\rm P}$, and $a$ -- are measured data and thus are contained in ${\bf X}_{\rm ob}$.  We assume the joint prior on $t_{\star}$, $M_{\rm P}$, $M_{\star}$, $R_{\rm P}$, and $a$ is independent of other parameters, both observed and not observed, as well as hyperparameters, and we do not consider any unobserved parameters other than $Q_{\rm P}$.  Then, from Eqn. \ref{eqn:hyp1genbinaryfinal}, substituting in the above parameters and rearranging gives
\begin{align}
    \label{eqn:ecchyp1}
    p(E,t_{{\star}},a,R_{\rm P},M_{\rm P},M_{\star}|f_0,\mu,\sigma)=
    \begin{cases}
    p(t_{\star},a,R_{\rm P},M_{\rm P},M_{\star})\\\quad\times\text{\Large$\int$} \left[1-f_0+f_0\int_0^{t_{\star}}p(t_{\rm cir}|a,R_{\rm P},M_{\rm P},M_{\star},Q_{\rm P},\mu,\sigma)dt_{\rm cir}\right] \\\quad\times p(Q_{\rm P}|\mu,\sigma) dQ_{\rm P} &,E=0\\
    p(t_{\star},a,R_{\rm P},M_{\rm P},M_{\star})\\\quad\times f_0\iint_{t_{\star}}^\infty p(t_{\rm cir}|a,R_{\rm P},M_{\rm P},M_{\star},Q_{\rm P},\mu,\sigma)dt_{\rm cir}\\\quad\times p(Q_{\rm P}|\mu,\sigma) dQ_{\rm P} &,E=1.
    \end{cases}
\end{align}
where $f_0$ is the fraction of systems that begin with detectable eccentric orbits.  We assign to $f_0$ a uniform hyperprior from 0 to 1.  We note that for $f_0$, the uninformative hyperprior is Beta(1/2,1/2); we will also compute odds ratios using this hyperprior, and compare to the results from the uniform hyperprior.  Since $p(t_{\star},a,R_{\rm P},M_{\rm P},M_{\star})$ is independent of unobserved system parameters, it will cancel out when the odds ratios are calculated.

Since the primary uncertainty in a given planet's $t_{\rm cir}$ comes from its tidal quality factor, we wish to reparameterize the above equation in terms of $Q_{\rm P}$.  We treat the prior of $t_{\rm cir}$ itself, $p(t_{\rm cir})$, as a Delta function centered at the value for $t_{\rm cir}$ given by Eqn. \ref{eqn:tcir}.  This means the integrals over $t_{\rm cir}$ in Eqn. \ref{eqn:ecchyp1} are either 1 or 0 depending on whether $t_{\rm cir}$ from Eqn. \ref{eqn:tcir} is greater than or less than $t_{\star}$.  We use Eqn. \ref{eqn:tcir} to write $Q_{\rm P}$ in terms of $t_{\rm cir}$ and introduce $Q_{\rm P,crit}$ as the value of $Q_{\rm P}$ such that $t_{\rm cir}=t_{\star}$ for a given planet.  We can then adjust the limits of integration over $Q_{\rm P}$ and obtain the following statements:
\begin{align}
    p(E,t_{{\star}},a,R_{\rm P},M_{\rm P},M_{\star}|f_0,\mu,\sigma)=
    \begin{cases}
    p(t_{\star},a,R_{\rm P},M_{\rm P},M_{\star})\\
    \times\left[1-f_0+f_0\int_0^{Q_{\rm P,crit}} p(Q_{\rm P}|\mu,\sigma)dQ_{\rm P}\right], &E=0\\
    p(t_{\star},a,R_{\rm P},M_{\rm P},M_{\star})f_0\int_{Q_{\rm P,crit}}^\infty p(Q_{\rm P}|\mu,\sigma)dQ_{\rm P}, &E=1.
    \end{cases}
\end{align}

\subsubsection{Hypothesis 2: Nature (Driven by inherent system properties)}
\label{subsec:eccnat}
In the Nature hypothesis, we test the idea that a dependence on semimajor axis -- with no contribution from  stellar age -- best explains the data, due to a relatively small span of stellar ages in the sample and/or eccentricities at larger semimajor axes being easier to excite and retain.

We introduce the parameter $a_{\rm cut}$, the cut-off semimajor axis, within which a planet has a circular orbit and beyond which a planet may acquire and/or retain an eccentric orbit.  We treat $a_{\rm cut}$ as a hyperparameter, i.e. a universal semimajor axis cutoff for eccentricity.  Thus $a_{\rm cut}$ behaves in much the same way as the 6250 K temperature division does in the obliquities case.  The primary difference is that we do not here have a well-defined cutoff value, so we must marginalize over $a_{\rm cut}$.  We give it a uniform prior from 0 to 0.1 AU, as this encompasses the range of semimajor axes in our sample.  Introducing $a_{\rm cut}$ and the other parameters into Eqn. \ref{eqn:genhyp2final} gives
\begin{align}
    \label{eqn:ecchyp2first}
    p(E,t_{\star},a,R_{\rm P},M_{\rm P},M_{\star}|a_{\rm cut})&= p(E|a,a_{\rm cut})p(t_{\star},a,R_{\rm P},M_{\rm P},M_{\star})
\end{align}
where we have noted that $a$ is the only component of ${\bf X}_{\rm ob}$ on which $E$ depends, and there are no ${\bf X}_{\rm nob}$.

If a planet is outside the semimajor axis cutoff, its orbit may be circular or eccentric.  We introduce the additional hyperparameter $f_{\rm ecc,out}$, which we give a hyperprior of U(0,1), to represent the fraction of planets with detectably eccentric orbits beyond $a_{\rm cut}$.  So for planets outside of $a_{\rm cut}$,
\begin{align}
    p(E|a\geq a_{\rm cut},f_{\rm ecc,out})=
    \begin{cases}
    1-f_{\rm ecc,out}&,E=0\\
    f_{\rm ecc,out}&,E=1.
    \end{cases}
\end{align}

Under the Nature hypothesis, a planet within the semimajor axis cutoff must have a circular orbit.  Thus for planets within $a_{\rm cut}$ we use:
\begin{align}
    p(E|a< a_{\rm cut})=
    \begin{cases}
    1&,E=0\\
    0&,E=1.
    \end{cases}
\end{align}

If a planet is within the semimajor axis cutoff, its orbit should be circular.  However, there may be some exceptions to this rule, such as planets whose orbits are influenced by an additional body that drives up their eccentricity.  Accordingly, we will also consider a special case of this hypothesis which includes a fraction of eccentric systems, $f_{\rm ecc,in}$, that remain eccentric within $a_{\rm cut}$.  Note that $f_{\rm ecc,in}$ is not a fraction of all the systems, but rather of all systems with detectably eccentric orbits, $f_{\rm ecc,out}$.  So for planets within $a_{\rm cut}$, in this special case,
\begin{align}
    p(E|a<a_{\rm cut},f_{\rm ecc,out},f_{\rm ecc,in})=
    \begin{cases}
    (1-f_{\rm ecc,in})f_{\rm ecc,out}+(1-f_{\rm ecc,out})&,E=0\\
    f_{\rm ecc,in}f_{\rm ecc,out}&,E=1.
    \end{cases}
\end{align}
We assign $f_{\rm ecc,in}$ a hyperprior of U(0,1).  Note that $f_{\rm ecc,in}$ is a hyperparameter.  We also note that for both $f_{\rm ecc,in}$ and $f_{\rm ecc,out}$, the true uninformative hyperprior is Beta(1/2,1/2); we will also compute odds ratios using this hyperprior on both $f_{\rm ecc,in}$ and $f_{\rm ecc,out}$ and compare to the results from the uniform hyperprior.

\subsubsection{Hypothesis 3: Chance}
\label{subsec:eccch}
From Eqn. \ref{eqn:genhyp3final},
\begin{align}
    \label{eqn:ecchyp3}
    p(E,t_{\star},a,R_{\rm P},M_{\rm P},M_{\star}|f)=
    \begin{cases}
    p(t_{\star},a,R_{\rm P},M_{\rm P},M_{\star})(1-f)&,E=0\\
    p(t_{\star},a,R_{\rm P},M_{\rm P},M_{\star})f&,E=1
    \end{cases}
\end{align}
as there are no components of ${\bf X}_{\rm nob}$ or other hyperparameters we consider here.  $f$ is the fraction of detectably eccentric systems, which we give a uniform hyperprior from 0 to 1.  We will also compare the results to those obtained using a hyperprior of Beta(1/2,1/2).

\subsection{Results}
\label{subsec:eccresults}
We apply Eqn. \ref{eqn:ecchyp1} for the Nurture hypothesis, Eqn. \ref{eqn:ecchyp2first} for the Nature hypothesis, and Eqn. \ref{eqn:ecchyp3} for the Chance hypothesis to the observed sample.  The complete equations for each hypothesis are
\begin{align}
    	p(H_{\rm nur})&=\iiint \prod_{E=0}[p(t_{\star},a,R_{\rm P},M_{\rm P},M_{\star})\left(1-f_{0}+f_{0}\int_0^{Q_{\rm P,crit}}p(Q_{\rm P}|\mu,\sigma) dQ_{\rm P}\right)]\nonumber\\&\times\prod_{E=1}[p(t_{\star},a,R_{\rm P},M_{\rm P},M_{\star})f_0\int_{Q_{\rm P,crit}}^\infty p(Q_{\rm P}|\mu,\sigma) dQ_{\rm P}]\nonumber\\ &\times p(\mu)p(\sigma)p(f_0)d\mu d\sigma df_0\\
    	p(H_{\rm nat})&=\iiint\prod_{E=0}\left[p(E=0|a,a_{\rm cut},f_{\rm ecc,out},f_{\rm ecc,in}) p(t_{\star},a,R_{\rm P},M_{\rm P},M_{\star})\right]\nonumber \\ &\times\prod_{E=1}\left[p(E=1|a,a_{\rm cut},f_{\rm ecc,out},f_{\rm ecc,in}) p(t_{\star},a,R_{\rm P},M_{\rm P},M_{\star})\right]\nonumber\\ &\times p(a_{\rm cut})p(f_{\rm ecc,out})p(f_{\rm ecc,in})da_{\rm cut} df_{\rm ecc,out} df_{\rm ecc,in}\\
    	p(H_{\rm ch})&=\int \prod_{E=0}\left[p(t_{\star},a,R_{\rm P},M_{\rm P},M_{\star})(1-f)\right]\prod_{E=1}\left[p(t_{\star},a,R_{\rm P},M_{\rm P},M_{\star})f\right]p(f)df.
\end{align}
Priors and probability functions of relevant parameters are summarized in Table \ref{tab:eccpriors}.  Note that the prior $p(t_{\star},a,R_{\rm P},M_{\rm P},M_{\star})$ will cancel out when we take the odds ratios, so we do not include it in the table.

\begin{table}
	\centering
	\begin{tabular}{|ll|l|}
		\hline
		\multicolumn{2}{|l|}{Parameter} & Prior or Probability Function \\
		\hline
        \multicolumn{2}{|l|}{Planetary tidal quality factor (log$_{10}(Q_{\rm P})|\mu,\sigma$)} & $\frac{1}{\sqrt{2\pi \sigma^2}} \text{exp}\left(\frac{-(\text{log}_{10}(Q_{\rm P})-\mu)^2}{2\sigma^2}\right)$\\
        \hline
        \multicolumn{2}{|l|}{Mean of log$_{10}(Q_{\rm P})$ ($\mu$)} & U(2,8)\\
        \hline
        \multicolumn{2}{|l|}{Standard deviation of log$_{10}(Q_{\rm P})$ ($\sigma$)} & U(0,5)\\
        \hline
        \multicolumn{2}{|l|}{Initial fraction of eccentric systems ($f_0$)} & U(0,1)\\
        \hline
        \multicolumn{2}{|l|}{Semimajor axis cutoff ($a_{\rm cut}$)} & U(0,0.1)\\
        \hline
        \multicolumn{2}{|l|}{Fraction of eccentric systems beyond $a_{\rm cut}$ ($f_{\rm ecc,out}$)} & U(0,1)\\
        \hline
        \multicolumn{2}{|l|}{Fraction of eccentric systems within $a_{\rm cut}$ ($f_{\rm ecc,in}$)} & U(0,1)\\
        \hline
        \multicolumn{2}{|l|}{Fraction of eccentric systems ($f$)} & U(0,1)\\
	   	\hline
	   	\multirow{4}{*}{Eccentricity state} & $E=0|a\geq a_{\rm cut},f_{\rm ecc,out}$ & $1-f_{\rm ecc,out}$\\
        & $E=1|a\geq a_{\rm cut},f_{\rm ecc,out}$ & $f_{\rm ecc,out}$\\
        & $E=0|a<a_{\rm cut}$ & 1\\
        & $E=1|a<a_{\rm cut}$ & 0\\
        \hline
	\end{tabular}
	\caption{Priors and probability functions of relevant parameters for the eccentricities case.}
	\label{tab:eccpriors}
\end{table}

We obtain the following odds ratios:
\begin{align}
    &\frac{p(H_{\rm nur})}{p(H_{\rm nat})} =1.3\times10^8\nonumber\\
    &\frac{p(H_{\rm nur})}{p(H_{\rm ch})} =1.5\times10^8\nonumber\\
    &\frac{p(H_{\rm nat})}{p(H_{\rm ch})}=1.1\nonumber.
\end{align}
These odd ratios indicate that the Nurture hypothesis, i.e. a correlation of eccentricity with age, is significantly favored over the Nature hypothesis, i.e. a correlation of eccentricity with semimajor axis.  The Nurture hypothesis is also very strongly favored over the Chance hypothesis.  There is nearly equal support for the Nature and Chance hypotheses; this is probably due to a few eccentric planets at small semimajor axes that essentially spoil the general trend of eccentricity with semimajor axis that can be seen in Figure \ref{fig:evsa}.  When we fix $a_{\rm{cut}}$ to be 0.03 AU -- just inside the eccentric planet with the smallest semimajor axis -- the odds ratio of Nature to Chance is 22, which is moderately in favor of the Nature hypothesis.

We also compute the odds ratios with $f_0$, $f_{\rm ecc,out}$, and $f$ assigned uninformative hyperpriors of Beta(1/2,1/2).  This results in an increase by roughly a factor of 2 in the support of the Nurture hypothesis relative to Nature and Chance.  The Nature vs. Chance ratio is essentially the same.  Thus the Nurture hypothesis is still the strong winner, and our overall conclusion is unchanged.

In the special case in which we include $f_{\rm ecc,in}$ in the Nature hypothesis, we obtain odds ratios of:
\begin{align}
    &\frac{p(H_{\rm nur})}{p(H_{\rm nat})} =5.5\times10^4\nonumber\\
    &\frac{p(H_{\rm nur})}{p(H_{\rm ch})} =1.5\times10^8\nonumber\\
    &\frac{p(H_{\rm nat})}{p(H_{\rm ch})}= 2.6\times10^3\nonumber.
\end{align}
Thus, the allowance of a few eccentric planets within the semimajor axis cutoff increases support for the Nature hypothesis by several orders of magnitude, but despite this, the Nurture hypothesis is still very strongly favored.  It may be expected that the value of $f_{\rm ecc,in}$ would be rather small; however, we found that assigning $f_{\rm ecc,in}$ a logarithmic prior to reflect the unknown scale of $f_{\rm ecc,in}$, while it increased support for Nature by a factor of a few, still did not allow the Nature hypothesis to be supported over Nurture.

We found that assigning $f_0$, $f_{\rm ecc,out}$, $f_{\rm ecc,in}$, and $f$ hyperpriors of Beta(1/2,1/2) decreased the support for the Nature hypothesis relative to Nurture and Chance by roughly a factor of 2, and decreased support for Nurture relative to Chance also by roughly a factor of 2.  This again leaves Nurture as the strong winner and does not change our overall conclusions.

In Appendix \ref{sec:appendixd}, we perform additional variations in our calculations by removing systems with highly uncertain ages or eccentricities and by bootstrapping the data.  We find that in both of these cases, the Nurture hypothesis is still very strongly favored over both Nature and Chance.

\subsection{Contour Plots}
\label{subsec:ecccontours}
\begin{figure}[ht]
    \includegraphics[width=3.5 in]{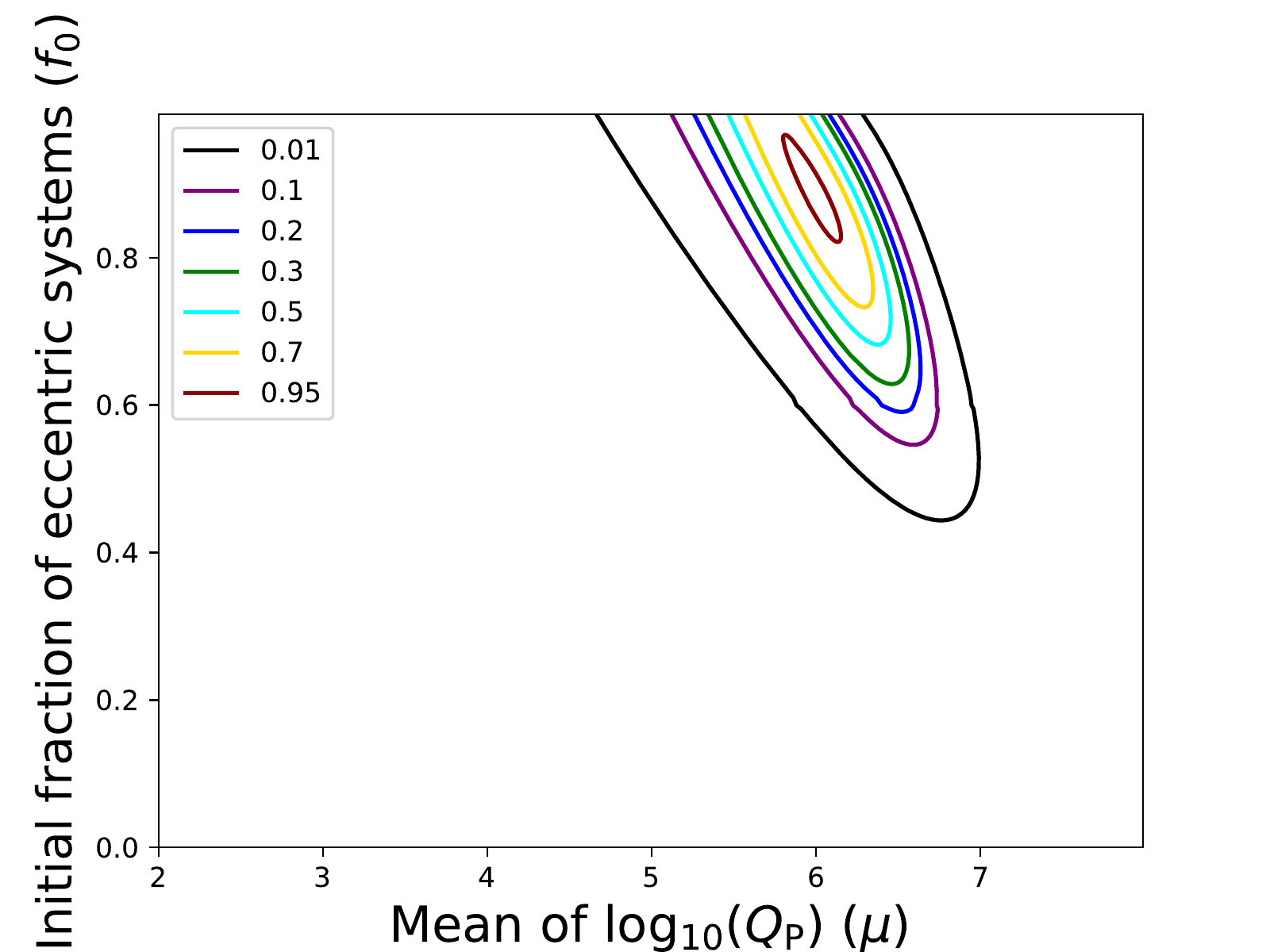}
    \includegraphics[width=3.5 in]{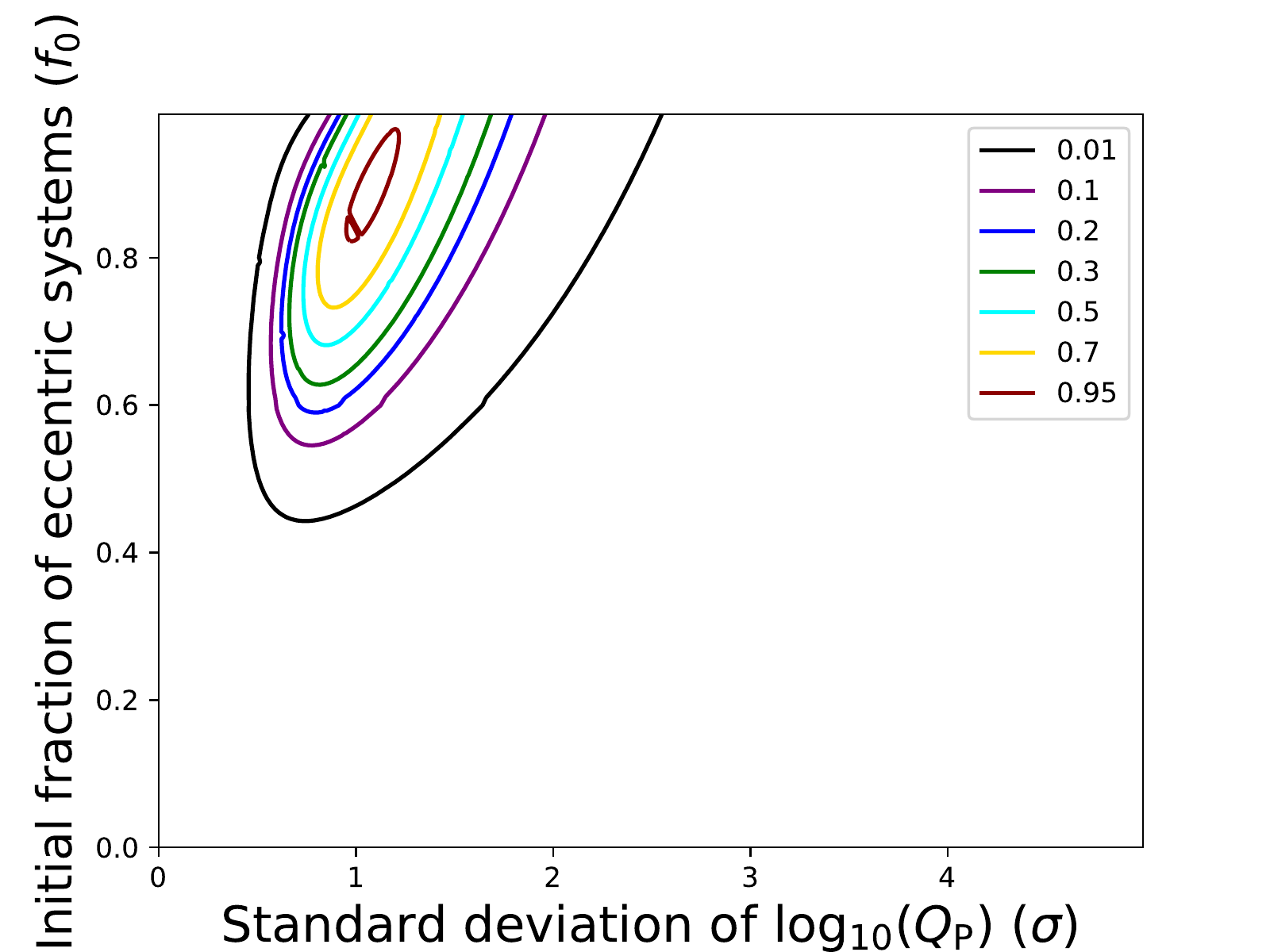}
    \includegraphics[width=3.5 in]{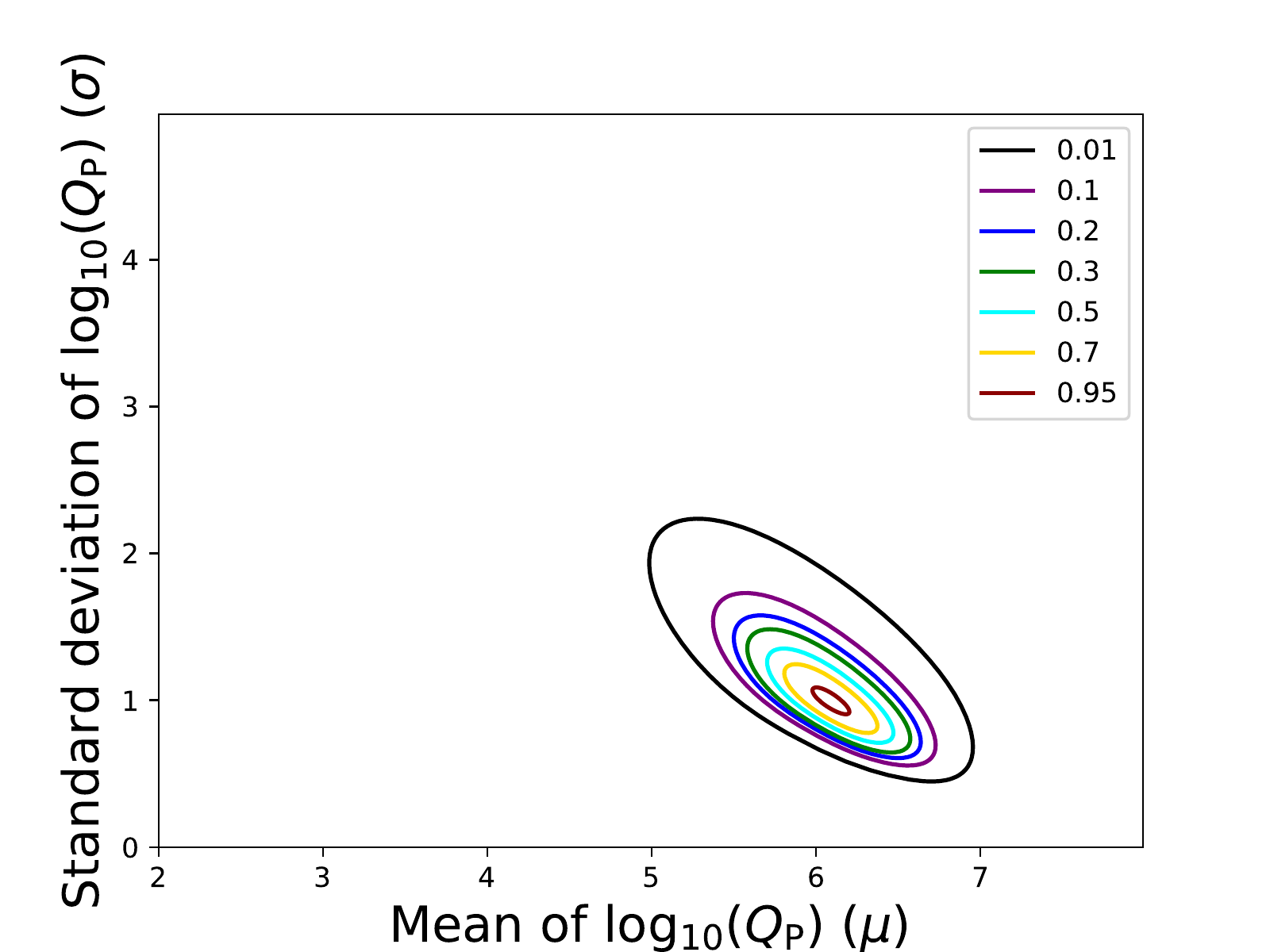}
    \caption{Probability contour plots for the Nurture hypothesis in the eccentricities case, showing $f_0$ vs. $\mu$ (top left), $f_0$ vs. $\sigma$ (top right), and $\sigma$ vs. $\mu$ (bottom left), where $\mu$ and $\sigma$ are the mean and standard deviation, respectively, of a log-normal prior on $Q_{\rm P}$.  Contour levels are 0.01, 0.05, 0.2, 0.5, 0.6, 0.7, 0.8, and 0.95.}
    \label{fig:eccqcontours}
 \end{figure}

In Figure \ref{fig:eccqcontours}, we display probability contours between the hyperparameters in the Nurture hypothesis, specifically, the initial fraction of eccentric systems $f_0$, and the mean $\mu$ and standard deviation $\sigma$ of the log-normal prior of the tidal quality factor $Q_{\rm P}$.  These plots indicate that most systems start out eccentric (high $f_0$) and have tidal quality factors of around $10^6$.  The peaks around $\sigma=1$ suggest there is a moderate amount of variation in tidal quality factors among the planets.

\begin{figure}[ht]
    \includegraphics[width=3.5 in]{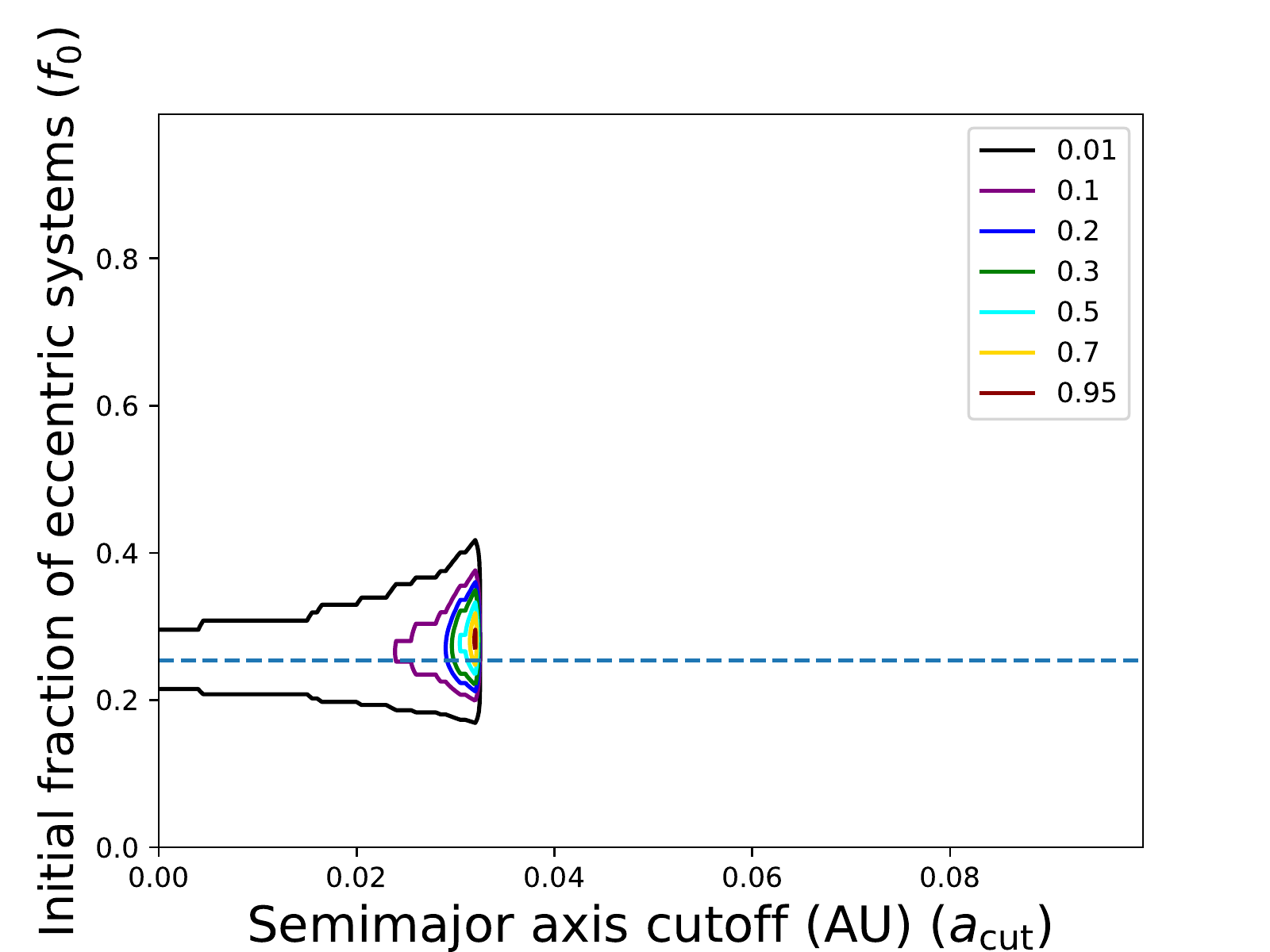}
    \caption{Probability contour plot for the Nature hypothesis in the eccentricities case showing $f_{\rm ecc,out}$ vs. $a_{\rm cut}$ in the case where we do not include the $f_{\rm ecc,in}$ parameter.  Contour levels are 0.01, 0.05, 0.2, 0.5, 0.6, 0.7, 0.8, and 0.95.  The observed fraction of detectably eccentric systems is shown as a blue dashed line.}
    \label{fig:eccnatcontours0fecc}
 \end{figure}

Figure \ref{fig:eccnatcontours0fecc} shows probability contours between the hyperparameters in the Nature hypothesis, the semimajor axis cutoff $a_{\rm cut}$ and the fraction of eccentric systems $f_{\rm ecc,out}$ outside of $a_{\rm cut}$ when we do not include $f_{\rm ecc,in}$.  This plot indicates a fraction of eccentric systems of around 30\%, very close to what is observed, and a semimajor axis cutoff of around 0.03 AU.

\begin{figure}[ht]
    \includegraphics[width=3.5 in]{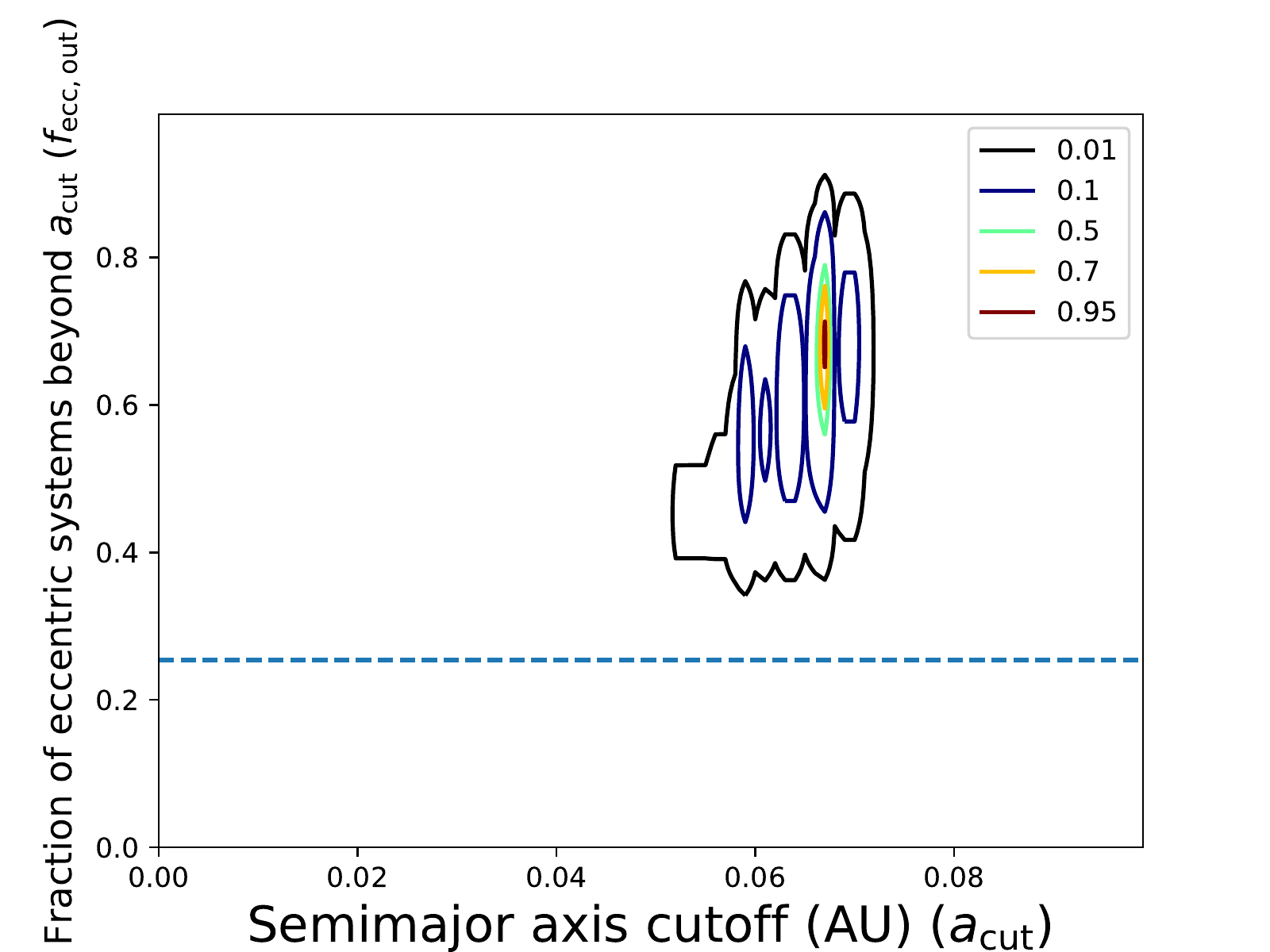}
    \includegraphics[width=3.5 in]{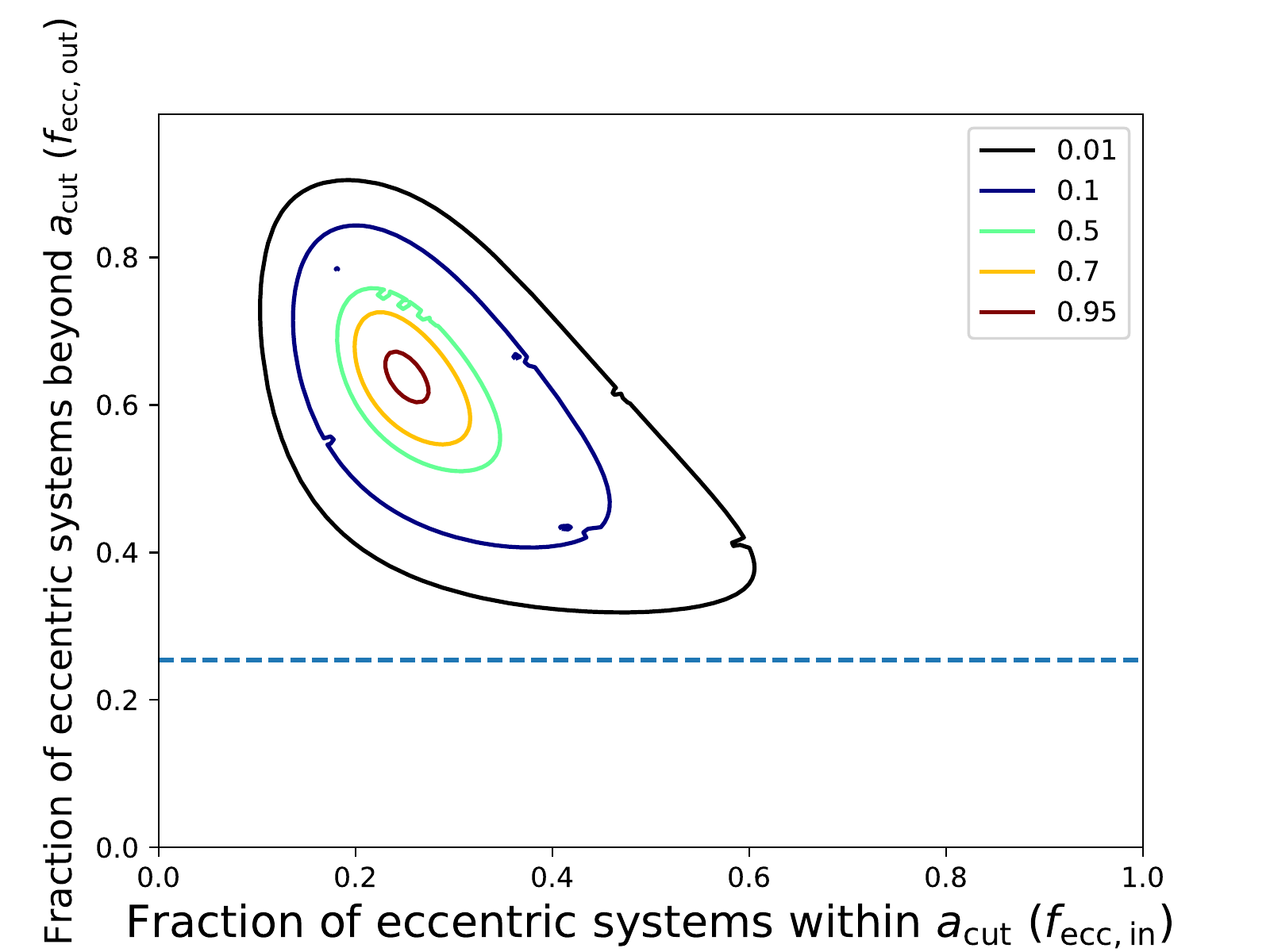}
    \includegraphics[width=3.5 in]{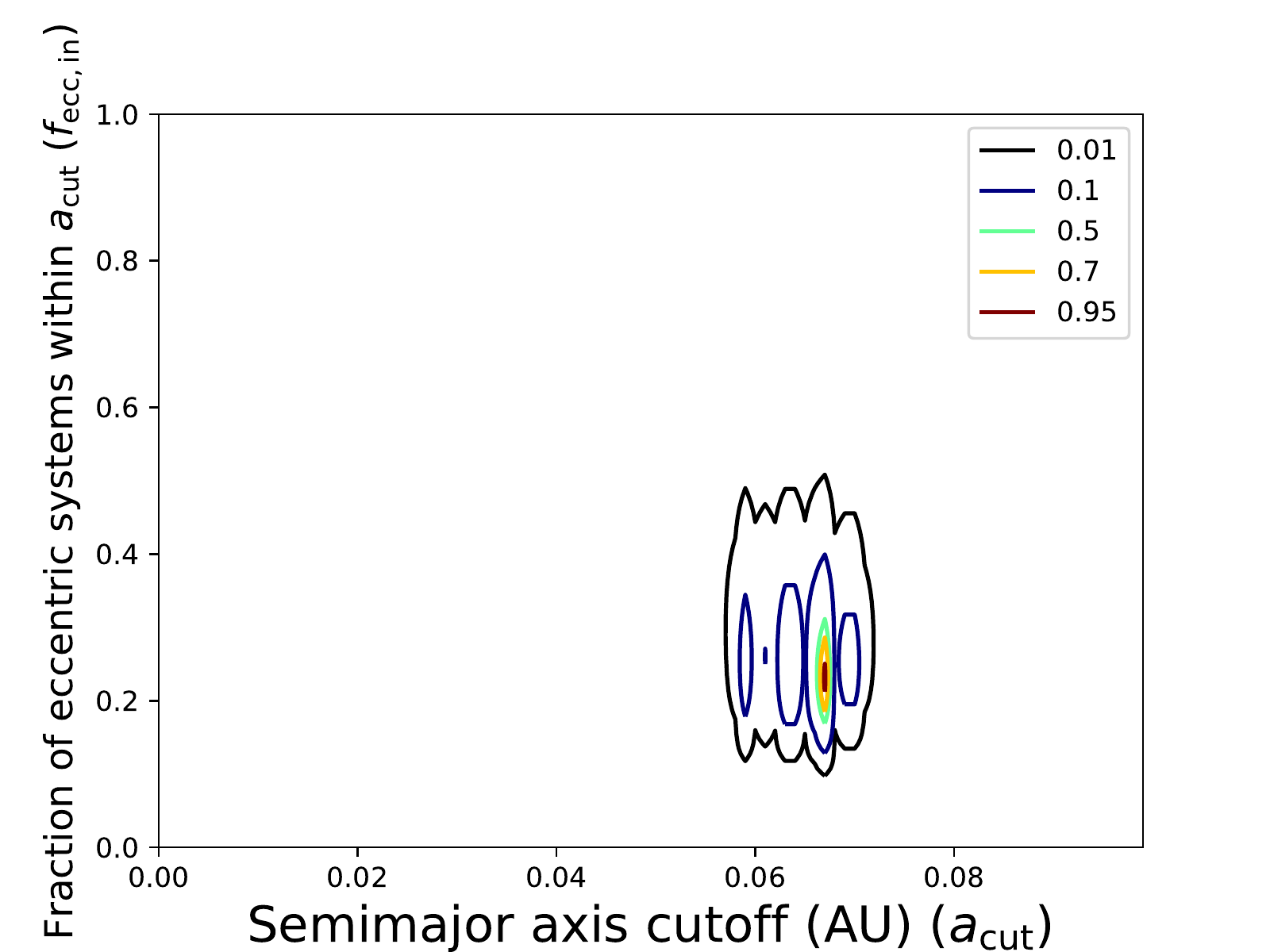}
    \caption{Probability contour plots for the Nature hypothesis in the eccentricities case where $f_{\rm ecc,in}$ is included, showing $f_{\rm ecc,out}$ vs. $a_{\rm cut}$ (top left) $f_{\rm ecc,out}$ vs. $f_{\rm ecc,in}$ (top right), and $f_{\rm ecc,in}$ vs. $a_{\rm cut}$ (bottom left).  Contour levels are 0.01, 0.05, 0.2, 0.5, 0.6, 0.7, 0.8, and 0.95.  The observed fraction of detectably eccentric systems is shown as a blue dashed line in the upper left and upper right plots.}
    \label{fig:eccnatcontours}
 \end{figure}

The panels in Figure \ref{fig:eccnatcontours} display contours between $f_{\rm ecc,out}$, $a_{\rm cut}$, and the fraction $f_{\rm ecc,in}$ of eccentric systems within $a_{\rm cut}$ for the special case in which $f_{\rm ecc,in}$ is included. The contours are not smooth because of the limited size of the observed sample.  These plots indicate that the best solution, under the Nature hypothesis in this special case, is a semimajor axis cutoff of $\sim 0.07$ AU, with $\sim 60\%$ of all planets starting out on eccentric orbits and $\sim 20\%$ remaining eccentric within $a_{\rm cut}$.

The small irregularities in the upper right panel of Figure \ref{fig:eccnatcontours} are low probability regions driven by close-in eccentric systems and more distant circular system, i.e. systems that do not fit the general pattern described by the Nature hypothesis.  Removing them from the sample decreases the Nurture vs. Nature odds ratio by a couple orders of magnitude, but the Nurture hypothesis still has strong support.

\begin{figure}[ht]
    \includegraphics[width=3.5 in]{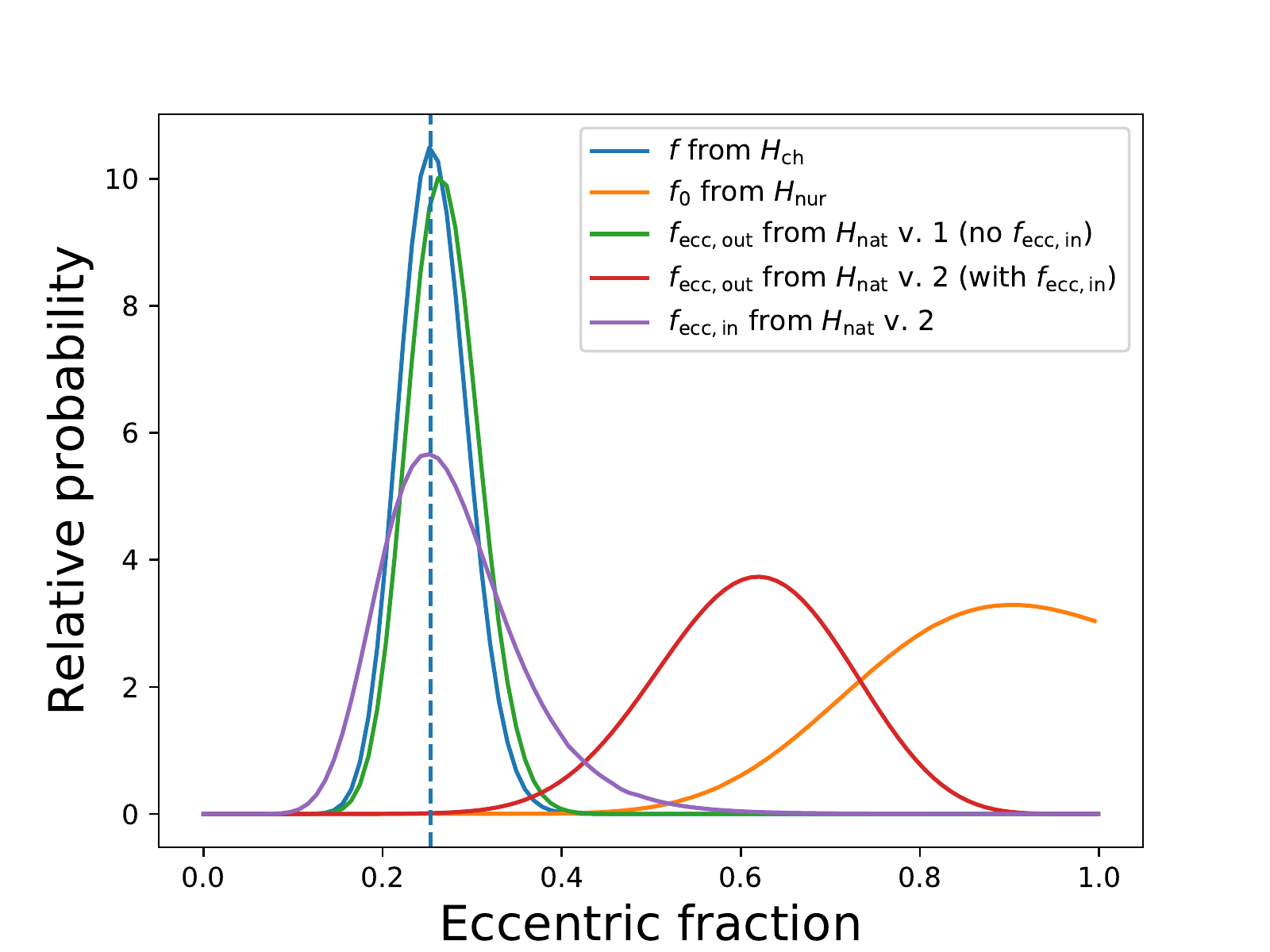}
    \caption{Probability histogram of the eccentric fraction, which under the Chance hypothesis is the eccentric fraction ($f$, blue), under Nurture is the initial eccentric fraction ($f_0$, orange), under Nature with no $f_{\rm ecc,in}$ is the eccentric fraction beyond $a_{\rm cut}$ ($f_{\rm ecc,out}$, green, labeled with ``$H_{\rm nat}$ v. 1"), and under Nature with $f_{\rm ecc,in}$ included is $f_{\rm ecc,out}$ (red, labeled with ``$H_{\rm nat}$ v. 2") and the eccentric fraction within $a_{\rm cut}$ ($f_{\rm ecc,in}$, purple, labeled with ``$H_{\rm nat}$ v. 2").  The observed fraction of eccentric systems (33/130) is shown with a blue dashed line. The curves have been normalized to have equal area.  Note that because the Nurture hypothesis describes eccentric systems circularizing over time, we do not expect $f_0$ to line up with the observed eccentric fraction.}
    \label{fig:eccfhists}
 \end{figure}

In Figure \ref{fig:eccfhists} we display probability histograms for $f$ in the Chance hypothesis, $f_0$ in the Nurture hypothesis, $f_{\rm ecc,out}$ in the Nature hypothesis, and $f_{\rm ecc,out}$ and $f_{\rm ecc,in}$ in the Nature hypothesis where we have included $f_{\rm ecc,in}$.  The curves have been normalized to have the same area, and the observed eccentric fraction (33/130) is shown with a dashed line.  Under the Nurture hypothesis, $\sim$90\% of planets form on eccentric orbits.  Under the Nature hypothesis with no $f_{\rm ecc,in}$, the fraction of eccentric systems outside the semimajor axis cutoff $a_{\rm cut}$ ($f_{\rm ecc,out}$) is only slightly more than the overall observed fraction of eccentric systems.  The contours in Figure \ref{fig:eccnatcontours0fecc} indicate that this version of the Nature hypothesis favors a value for $a_{\rm cut}$ of around 0.03 AU, just inside the closest-in eccentric planet.  There are only a few planets within 0.03 AU (see Figure \ref{fig:evsa}), so $f_{\rm ecc,out}$ is close to the observed fraction.  When $f_{\rm ecc,in}$ is included in the Nature hypothesis, a semimajor axis cutoff of around 0.07 AU is favored, as seen in Figure \ref{fig:eccnatcontours}.  Most of the sample is within this value, which leads to $f_{\rm ecc,in}$ being near the observed eccentric fraction, and the majority of planets in the sample beyond 0.07 AU are eccentric, giving $f_{\rm ecc,out}$ a higher value in this case.

\subsection{Summary}
\label{subsec:eccsummary}
In summary, we find very strong support for the hypothesis that eccentric hot Jupiters tend to be younger than circular hot Jupiters, compared to the hypothesis that eccentricity is due to semimajor axis or a chance relation.  This result supports the theory that hot Jupiter orbits are circularized over time, consistent with high-eccentricity migration.  In the future, we can improve our assessment by incorporating measurement uncertainties on the observed system parameters.  We note in particular that the strong dependence of $t_{\rm cir}$ on $a$ and $R_{\rm P}$ can inflate a measurement uncertainty of tens of percent into a factor of 2-10, which will be non-negligible in the calculation of $t_{\rm cir}$.

\clearpage

\section{Conclusions and Future Work}
\label{sec:conclusions}

We develop a Bayesian framework (Section \ref{sec:general}) for testing the strength of perceived correlations between planetary orbital and system properties.  We define hypotheses to describe possible relationships between planetary and system properties: the Nurture hypothesis, where the planetary property $X_{\rm p}$ is driven by age; the Nature hypothesis, where $X_{\rm p}$ is driven by an observed system property other than age; and the Chance hypothesis, where $X_{\rm p}$ is not driven by any observed system parameters.  We derive equations to use in calculating the odds ratios of pairs of these hypotheses.

The odds ratio itself is calculated using Eqn. \ref{eqn:oddsratio}.  The use of that equation requires knowledge of the likelihood $p(X_{\rm p},t_{\star},{\bf X}_{\rm ob})$ for each individual system under each hypothesis. This likelihood is given by Eqns. \ref{eqn:hyp1gen3v2}, \ref{eqn:genhyp2final}, and \ref{eqn:genhyp3cont} for the Nurture, Nature, and Chance hypotheses, respectively.  The equations for a binary parameter (i.e., when $X_{\rm p}$ can take on 0 or 1) -- used in the applications here -- are Eqns. \ref{eqn:hyp1gen3binary2}, \ref{eqn:genhyp2final}, and \ref{eqn:genhyp3final}.  If the binary parameter can only go from $X_{\rm p0}=1$ to $X_{\rm p}=0$ and not from $X_{\rm p0}=0$ to $X_{\rm p}=1$, the likelihood for the Nurture hypothesis can also be found with Eqn. \ref{eqn:hyp1genbinaryfinal}.

In the future, we can incorporate uncertainties in stellar ages into the assessment of the odds ratio, by adding a term to the likelihood that accounts for the probability of measuring $t_\star$ given the measurement uncertainties ($p(t_{\star,\rm obs}|t_\star)$) and marginalizing over $t_\star$ in the joint $X_{\rm p}, t_\star$ posterior.

We apply our equations to the question of 2:1 resonance disruption (Section \ref{sec:resonances}) and find a roughly equal amount of support for the Nurture hypothesis as for the Chance hypothesis.  One is not clearly favored over the other, so by our approach, it is uncertain whether the supposed trend is coincidental or not.  Going forward, though we are not aware of any such relation proposed in the literature, we will also investigate possible relations between 2:1 resonances and system properties other than age with the Nature hypothesis.  Additionally, we will be able to apply our framework to a larger dataset thanks to observations from the likes of TESS and Gaia that will not only reveal more exoplanets (and thus more systems with 2:1 resonances), but also more precise stellar ages.  TESS is expected to uncover many nearby planetary systems that will be relatively easy to follow up and characterize, which will make it easier to identify systems with 2:1 resonances.  It will also yield ages via gyrochronology.  The precise distances from Gaia will enable calibration of other stellar parameters and thus yield ages in that way.  A larger sample of 2:1 resonant systems with known ages may allow us to confidently favor one hypothesis in our framework over another.  This will help us understand planetary formation environments as well as the mechanisms that disrupt orbital resonances.

We also investigate the source of the observed stellar obliquity trend (Section \ref{sec:alignment}) and find that a correlation of alignment state with temperature is strongly favored over a correlation with age or a chance relation.  Thus our results favor the interpretation of \cite{winn2010} that hot stars tend to be misaligned with their planet's orbits and cool stars are aligned, rather than the interpretation that stars are gradually realigned as they age and cool down or that there is no relation between alignment state and other system properties.  This indicates that whatever process realigns stellar spin and planetary orbits occurs predominantly in cooler stars (possibly due to the size of the convective zone) and happens on timescales much shorter than the lifetime of the stars \citep{winn2010}.  The processes that produce misalignments and govern their subsequent evolution are not well understood (see Albrecht et al. in prep, for a review), and the correlation seen here could also be an indication that different misaligning mechanisms operate at different stellar temperatures or, probably more directly, stellar masses.

Finally, we apply our framework to the set of known hot Jupiters with measured eccentricities and stellar ages to determine if it truly exhibits an evolutionary trend or is better described by a relation between eccentricity and semimajor axis (Section \ref{sec:eccentricities}).  We find very strong support for the Nurture hypothesis compared to the Nature and Chance hypotheses.  Thus our results agree with those of \cite{quinn2014} that the data show hot Jupiter orbits circularizing over time.  This supports the high-eccentricity migration mechanism for hot Jupiter formation.

There are other planet-star correlations suggested in the literature to which we can apply this framework.  \citet{bj2016} looked at the Kepler dichotomy around M-dwarfs.  They suggested that the observed excess of singly-transiting planets may be a result either of different formation processes producing some planetary systems with high mutual inclinations, or of the dynamical scattering of planetary inclinations over longer timescales.  \cite{puwu2015} also suggested that many planetary systems may start out tightly packed and then experience dynamical instability that leads to the loss of some of the planets, producing the high number of singly- and doubly-transiting systems.  \cite{veras2015} have shown that the high-precision stellar ages expected from the PLATO mission will enable the detection of a trend of exoplanet frequency with age.  We will investigate this case in our future work and look forward to the increased insight afforded by the high-quality data from upcoming missions.
 
 \acknowledgments We thank the referee for the helpful comments on this paper.  We gratefully acknowledge support from the Alfred P. Sloan Foundation's Sloan Research Fellowship and NASA XRP 80NSSC18K0355. We thank Tom Loredo, Darin Ragozzine, and Leslie Rogers for helpful conversations. The Center for Exoplanets and Habitable Worlds is supported by the Pennsylvania State University, the Eberly College of Science, and the Pennsylvania Space Grant Consortium. This research was supported in part by the National Science Foundation under Grant No. NSF PHY-1748958.

\bibliography{bibliography} \bibliographystyle{aasjournal}

\appendices
\def\x{\ref{eqn:hyp1genbinaryfinal} }
\def\y{\ref{eqn:hyp1gen3binary2}}
\section{Derivation of Equation \x from Equation \y}
\label{sec:appendixa}
To obtain Eqn. \ref{eqn:hyp1genbinaryfinal} from Eqn. \ref{eqn:hyp1gen3binary2}, for when $X_{\rm p}$ can go from 1 to 0 but not from 0 to 1, we separate out Eqn. \ref{eqn:hyp1gen3binary2} for each combination of $X_{\rm p}$ and $X_{\rm p0}$ and adjust the limits of integration over $t_{\rm e}$ accordingly:

\begin{align}
    p(X_{\rm p},t_{{\star}},{\bf X}_{\rm ob}|f_0,{\bf Y})&=\iint p(X_{\rm p}=0|t_{{\star}},t_{\rm e},X_{\rm p0}=0,{\bf Y})\nonumber\\&\times p(t_{\rm e}|X_{\rm p0},{\bf X}_{\rm ob},{\bf X}_{\rm nob},{\bf Y}) p(t_{{\star}},{\bf X}_{\rm ob},{\bf X}_{\rm nob}|{\bf Y})\nonumber\\&\times p(X_{\rm p0}=0|f_0)dt_{\rm e} d{\bf X}_{\rm nob}&&\nonumber\\
    &=\iint (1) \times\delta(t_{\rm e}-\infty) p(t_{{\star}},{\bf X}_{\rm ob},{\bf X}_{\rm nob}|{\bf Y})\nonumber\\&\times (1-f_0)dt_{\rm e} d{\bf X}_{\rm nob}&&\nonumber\\
    \label{eqn:appeq1}
    &=(1-f_0)\int p(t_{{\star}},{\bf X}_{\rm ob},{\bf X}_{\rm nob}|{\bf Y}) d{\bf X}_{\rm nob}&&,X_{\rm p}=0,X_{\rm p0}=0
\end{align}
\begin{align}
    p(X_{\rm p},t_{{\star}},{\bf X}_{\rm ob}|f_0,{\bf Y})&=\iint p(X_{\rm p}=0|t_{{\star}},t_{\rm e},X_{\rm p0}=1,{\bf Y})\nonumber\\&\times p(t_{\rm e}|X_{\rm p0},{\bf X}_{\rm ob},{\bf X}_{\rm nob},{\bf Y}) p(t_{{\star}},{\bf X}_{\rm ob},{\bf X}_{\rm nob}|{\bf Y})\nonumber\\&\times p(X_{\rm p0}=1|f_0)dt_{\rm e} d{\bf X}_{\rm nob}&&\nonumber\\
    &=\iint_0^{t_{\star}} (1)\times p(t_{\rm e}|X_{\rm p0},{\bf X}_{\rm ob},{\bf X}_{\rm nob},{\bf Y}) \nonumber\\&\times p(t_{{\star}},{\bf X}_{\rm ob},{\bf X}_{\rm nob}|{\bf Y}) (f_0)dt_{\rm e} d{\bf X}_{\rm nob}&&\nonumber\\
    &+ \iint_{t_{\star}}^{\infty} (0)\times p(t_{\rm e}|X_{\rm p0},{\bf X}_{\rm ob},{\bf X}_{\rm nob},{\bf Y}) \nonumber\\&\times p(t_{{\star}},{\bf X}_{\rm ob},{\bf X}_{\rm nob}|{\bf Y}) (f_0)dt_{\rm e} d{\bf X}_{\rm nob}&&\nonumber\\
    \label{eqn:appeq2}
    &=f_0\iint_0^{t_{\star}}p(t_{\rm e}|X_{\rm p0},{\bf X}_{\rm ob},{\bf X}_{\rm nob},{\bf Y})dt_{\rm e}\nonumber\\&\times p(t_{{\star}},{\bf X}_{\rm ob},{\bf X}_{\rm nob}|{\bf Y}) d{\bf X}_{\rm nob}&&,X_{\rm p}=0,X_{\rm p0}=1
\end{align}
\begin{align}
    \label{eqn:appeq3}
    p(X_{\rm p},t_{{\star}},{\bf X}_{\rm ob}|f_0,{\bf Y})&=0&&,X_{\rm p}=1,X_{\rm p0}=0\end{align}
\begin{align}
    p(X_{\rm p},t_{{\star}},{\bf X}_{\rm ob}|f_0,{\bf Y})&=\iint p(X_{\rm p}=1|t_{{\star}},t_{\rm e},X_{\rm p0}=1,{\bf Y})\nonumber\\&\times p(t_{\rm e}|X_{\rm p0},{\bf X}_{\rm ob},{\bf X}_{\rm nob},{\bf Y})p(t_{{\star}},{\bf X}_{\rm ob},{\bf X}_{\rm nob}|{\bf Y})\nonumber\\&\times p(X_{\rm p0}=1|f_0)dt_{\rm e} d{\bf X}_{\rm nob}&&\nonumber\\
    &=\iint_0^{t_{\star}} (0)\times p(t_{\rm e}|X_{\rm p0},{\bf X}_{\rm ob},{\bf X}_{\rm nob},{\bf Y})\nonumber\\&\times p(t_{{\star}},{\bf X}_{\rm ob},{\bf X}_{\rm nob}|{\bf Y}) (f_0)dt_{\rm e} d{\bf X}_{\rm nob}&&\nonumber\\
    &+\iint_{t_{\star}}^\infty (1)\times p(t_{\rm e}|X_{\rm p0},{\bf X}_{\rm ob},{\bf X}_{\rm nob},{\bf Y})\nonumber\\&\times p(t_{{\star}},{\bf X}_{\rm ob},{\bf X}_{\rm nob}|{\bf Y}) (f_0)dt_{\rm e} d{\bf X}_{\rm nob}&&\nonumber\\
    \label{eqn:appeq4}
    &=f_0\iint_{t_{\star}}^\infty p(t_{\rm e}|X_{\rm p0},{\bf X}_{\rm ob},{\bf X}_{\rm nob},{\bf Y})dt_{\rm e}\nonumber\\&\times p(t_{{\star}},{\bf X}_{\rm ob},{\bf X}_{\rm nob}|{\bf Y}) d{\bf X}_{\rm nob}&&,X_{\rm p}=1,X_{\rm p0}=1.
    \end{align}
  Combining Eqns. \ref{eqn:appeq1} and \ref{eqn:appeq2} and Eqns. \ref{eqn:appeq3} and \ref{eqn:appeq4} gives the desired result:
  \begin{align}
    p(X_{\rm p},t_{{\star}},{\bf X}_{\rm ob}|f_0,{\bf Y})&=\int \left[1-f_0+f_0\int_0^{t_{\star}}p(t_{\rm e}|X_{\rm p0},{\bf X}_{\rm ob},{\bf X}_{\rm nob},{\bf Y})dt_{\rm e}\right] \nonumber\\&\qquad\times p(t_{{\star}},{\bf X}_{\rm ob},{\bf X}_{\rm nob}|{\bf Y}) d{\bf X}_{\rm nob} &&,X_{\rm p}=0\nonumber\\
    p(X_{\rm p},t_{{\star}},{\bf X}_{\rm ob}|f_0,{\bf Y})&=f_0\iint_{t_{\star}}^\infty p(t_{\rm e}|X_{\rm p0},{\bf X}_{\rm ob},{\bf X}_{\rm nob},{\bf Y})dt_{\rm e}\nonumber\\&\qquad\times p(t_{{\star}},{\bf X}_{\rm ob},{\bf X}_{\rm nob}|{\bf Y}) d{\bf X}_{\rm nob}&&,X_{\rm p}=1.
\end{align}
Again, this is only valid for the situation of a binary $X_{\rm p}$ that can go from $X_{\rm p0}=1$ to $X_{\rm p}=0$ but cannot go from $X_{\rm p0}=0$ to $X_{\rm p}=1$.

\section{Additional exploration of resonances case}
\label{sec:appendixb}
Our investigation into whether 2:1 resonances get disrupted over time yielded an odds ratio for Nurture vs. Chance of 2.2, a result which does not favor one hypothesis over the other.  Here we further explore the case of 2:1 resonance disruption by performing several variations on our original odds ratio calculation.

We first examine the effect of high stellar age uncertainties on the results of the resonances case.  \cite{kz2011} retrieved ages for their sample from large chromospheric activity surveys, and the surveys themselves do not report uncertainties for individual stars.  However, most of the systems have multiple reported ages.  We remove systems which have a difference of $\gtrsim 3$ Gyr between the highest and lowest reported ages, which includes one 2:1 resonant system and four systems without 2:1 resonances.  This gives an odds ratio of Nurture to Chance of 2.2, the same as our original result, suggesting that in this application, the stellar age uncertainties do not bias the overall result.

As a further step, we bootstrap the data 300 times -- i.e. redraw from the original data a sample of the same size as the original sample, with replacement -- and recompute the odds ratio for each iteration.  By so doing, we obtain a distribution of odds ratios.  This distribution is primarily concentrated between about 1.6 and 3, but has a few high-value outliers as well.  We take the log$_{10}$ of the odds ratios.  The median value of the log$_{10}$ distribution of Nurture vs. Chance ratios is 0.34 and the standard deviation is 0.15.  In only 0.67\% of cases is the odds ratio greater than 10.  Thus there is still not enough evidence to favor either the Nurture or the Chance hypothesis.

Finally, the threshold for a system having a 2:1 resonance used by \cite{kz2011} -- a normalized commensurability proximity value (see Eqn. \ref{eqn:ncp}) of $\delta<0.1$ -- and which we have adopted here is somewhat arbitrary.  As another variation in our calculations, we treat the 2:1 resonance threshold as a hyperparameter.  Eqn. 14 in \cite{malhotra2012} estimates the fractional range of orbital motion over which a resonance angle could librate (without very fast precession of the periapse); for a Jupiter-mass planet, this is about 1\%.  We marginalize over a range of 1\% to 20\% of the deviation of the period ratio from commensurability.  With this treatment, we obtain a Nurture to Chance odds ratio of 2.0, which is similar to our original value of 2.2.

\section{Additional exploration of obliquities case}
\label{sec:appendixc}
The application of our framework to the stellar obliquity case strongly favored the Nature hypothesis, which says that stellar obliquity is driven by stellar temperature, rather than age.  Here we present further exploration of this case by performing several variations on our original calculations.

First, we investigate how uncertainties in measured obliquity angle and stellar age impact the results of the obliquities case.  Rather than simply removing those systems with highly uncertain obliquity angles, we are interested in those with a level of uncertainty that could easily place them on either side of the $20^\circ$ threshold.  We remove from our sample systems with measured obliquity angle within $1\sigma$ of $20^\circ$; this results in the exclusion of three aligned systems.  We also exclude any systems with an uncertainty in stellar age $>1$ Gyr, though the only system for which this is the case is already excluded by the angle criterion.  This modified sample results in reduced support for the Nature hypothesis, relative to Nurture and Chance, by about a factor of 2, while the support of Nurture relative to Chance remains essentially unchanged.  However, the support for the Nature hypothesis compared to Nurture and Chance remains very strong, and we still conclude that it is the best explanation of the three for the data in question.

Additionally, we bootstrap the data 300 times -- i.e. redraw from the original data, with replacement, a sample of the same size as the original sample -- and recompute the odds ratios each time to obtain distributions of the odds ratios.  We then take the log$_{10}$ of the ratios.  The log$_{10}$ distribution of Nurture vs. Chance ratios has a median of 0.17 and a standard deviation of 0.16.  The log$_{10}$ distribution of Nature vs. Nurture ratios has a median of 2.2 and standard deviation of 0.91, and the log$_{10}$ distribution of Nature vs. Chance ratios has a median of 2.4 and standard deviation of 0.90.  These values correspond to odds ratios that are slightly reduced from our original calculations, but not enough to change our overall conclusions.  In only 4\% of cases is the Nature vs. Nurture ratio less than 10, and in only 2\% of cases is the Nature vs. Chance ratio less than 10.  Thus the Nature hypothesis is still strongly favored.

Finally, the $20^\circ$ threshold for stellar misalignment is somewhat arbitrary and does not have a strong physical basis.  We have used this in our calculations to follow \cite{triaud2011}.  An alternative criterion is that of \cite{winn2010}, who consider a system to be misaligned if the measured obliquity angle is $>10^\circ$ at $>3\sigma$ level.  However, when we apply this alternative criterion to the dataset we use (that of \citealt{triaud2011}), we obtain the exact same subsamples of aligned and misaligned stars as with the $20^\circ$ threshold.  Thus, the criterion we use provides a solid sample division, and we do not marginalize over the threshold in this case.

\section{Additional exploration of eccentricities case}
\label{sec:appendixd}
For the question of whether hot Jupiters show evidence of tidal circularization, we found very strong evidence that hot Jupiter eccentricities are driven by age rather than semimajor axis.  Here we further explore this case by performing several variations on our original odds ratio calculations.

To examine the effect of highly uncertain data points on the results in the eccentricities case, we remove systems with age uncertainties $\geq 3$ Gyr or eccentricity uncertainties $\geq 0.1$.  This excludes 10 eccentric systems and 19 circular systems.  With this reduced sample, the Nurture vs. Nature ratio goes from $1.3\times10^8$ to $5.0\times10^5$.  The Nurture vs. Chance ratio experiences a similar drop, from $1.5\times10^8$ to $2.6\times10^5$.  The Nature vs. Chance ratio drops by about a factor of 2.  Despite orders of magnitude less relative support for the Nurture hypothesis, it is still clearly the most favorable of the three options.

As an additional test, we bootstrap the data 300 times, i.e. redraw from the original data, with replacement, a sample of the same size as the original sample, to obtain distributions of each of the odds ratios.  We then look at the log$_{10}$ of those distributions.  The log$_{10}$ distribution of Nature vs. Chance ratios has a median value of -0.39 and a standard deviation of 0.65.  The log$_{10}$ Nurture vs. Nature distribution has a median of 9.4 and a standard deviation of 2.5.  Finally, the Nurture vs. Chance distribution has a median of 9.3 and a standard deviation of 2.6.  Additionally, both the Nurture vs. Nature and Nurture vs. Chance ratios are greater than 100 in every case.  Thus a correlation driven by age is still overwhelmingly favored.

\end{document}